\documentclass[11pt,english,reqno]{bourbaki}
\usepackage{amsmath}
\usepackage{amsthm}
\usepackage{amsfonts}
\usepackage{amssymb}
\usepackage{epic}
\usepackage{eepic}
\usepackage{times}
\usepackage{graphics}

\def\bc{\bar\gamma} 
\def\c{\gamma} 
\def\pl{\partial} 
\def\bpl{\bar \partial} 
\def\H3p{H_3^+}
\def\QR{\mathbb{R}} 
\def\QC{\mathbb{C}} 
\def\a{\alpha}
\def\b{\beta} 
\def\c{\gamma} 
\def\d{\delta} 
\def\e{\epsilon} 

\newcommand{\SLC}{{\rm SL(2,\BC \rm)}}
\newcommand{\SLR}{{\rm SL(2,\BR \rm)}}
\newcommand{\SU}{{\rm SU(2)}}
\def\vph{\varphi} 
\def\raa{{r}} 
\def\ras{{s}}

\theoremstyle{plain}
\newtheorem{thm}{Theorem}
\newtheorem{claim}{Claim}

\newtheorem{lem}{Lemma}

\theoremstyle{remark}
\newtheorem{rem}{Remark}

\newcommand{\beq}{\begin{equation}}
\newcommand{\eeq}{\end{equation}}

\newcommand{\id}{\mbox{id}}

\newcommand{\pa}{\partial}
\newcommand{\ot}{\otimes}
\newcommand{\ra}{\rightarrow}

\newcommand{\ti}{\times}

\newcommand{\fr}[2]{{\textstyle \frac{#1}{#2} }}

\newcommand{\fsl}{{\mathfrak s}{\mathfrak l}}
\newcommand{\hfsl}{\widehat{\fsl}}

\newcommand{\bra}{\langle}

\newcommand{\ket}{\rangle}
\newcommand{\Id}{{\rm Id}}
\newcommand{\al}{\alpha}

\newcommand{\be}{\beta}
\newcommand{\ga}{\gamma}
\newcommand{\Ga}{\Gamma}
\newcommand{\de}{\delta}
\newcommand{\De}{\Delta}
\newcommand{\ep}{\epsilon}
\newcommand{\la}{\lambda}
\newcommand{\om}{\omega}
\newcommand{\Om}{\Omega}
\newcommand{\si}{\sigma}

\newcommand{\vf}{\varphi}

\newcommand{\bJ}{\bar{J}}

\newcommand{\bu}{\bar{u}}

\newcommand{\bw}{\bar{w}}
\newcommand{\bx}{\bar{x}}
\newcommand{\bm}{\bar{m}}
\newcommand{\bz}{\bar{z}}

\newcommand{\CA}{{\mathcal A}}

\newcommand{\CC}{{\mathcal C}}
\newcommand{\CD}{{\mathcal D}}

\newcommand{\CF}{{\mathcal F}}
\newcommand{\CG}{{\mathcal G}}
\newcommand{\CH}{{\mathcal H}}
\newcommand{\CI}{{\mathcal I}}
\newcommand{\CJ}{{\mathcal J}}

\newcommand{\CO}{{\mathcal O}}  
\newcommand{\CP}{{\mathcal P}}  
  
\newcommand{\CR}{{\mathcal R}}
\newcommand{\CS}{{\mathcal S}}

\newcommand{\SH}{{\mathsf H}}

\newcommand{\SL}{{\mathsf L}}
\newcommand{\bL}{\bar{\SL}}

\newcommand{\SV}{{\mathsf V}}

\newcommand{\BR}{{\mathbb R}}

\newcommand{\BC}{{\mathbb C}}

\newcommand{\BS}{{\mathbb S}}

\newcommand{\BZ}{{\mathbb Z}}

\newcommand{\Fus}[6]{F_{{\scriptstyle #1}{\scriptstyle #2}}\bigl[
\begin{smallmatrix} #3 & #4 \\ #5 & #6 \end{smallmatrix}\bigr]}

\DeclareMathOperator{\sgn}{sgn}

\DeclareMathOperator{\Tr}{Tr}

\newcommand{\rf}[1]{(\ref{#1})}

\newcommand{\aufz}
{\begin{list}{$\bullet$}{\topsep0cm \itemsep0cm \parsep0cm}}
\newcommand{\eaufz}{\end{list}}
\newcounter{num}
\newcommand{\remlst}{\begin{list}
{(\arabic{num})}{\usecounter{num}\topsep0cm \itemsep0cm \parsep0cm}}
\begin{document}
\bigskip
\hfill\begin{minipage}{3cm}
SfB 288 Preprint\\ AEI 2001-146\\ LPTENS 01/47
\end{minipage}
\vspace{0.8cm}

\begin{center}
{\Large \textbf{BRANES\, IN\, THE\, EUCLIDEAN\,  AdS$_3$}}

\vspace{0.8cm}

{\bf B.\ Ponsot\footnote{Max-Planck-Institut f\"ur Gravitationsphysik, 
Albert Einstein Institut, Am M\"uhlenberg 1, 14476 Golm, 
Germany, {\it bponsot@aei.mpg.de}}, \ 
V.\ Schomerus\footnotemark[1] \footnote{Laboratoire de Physique 
Th\'eorique de l'\'Ecole Normale Sup\'erieure, 24 rue Lhomond, 
F-75231 Paris Cedex 05, France, {\it vschomer@aei.mpg.de}},\ \  
J.\ Teschner\footnote{Institut f\"ur theoretische Physik, 
Freie Universit\"at Berlin, Arnimallee 14, 14195 Berlin, 
Germany, {\it teschner@physik.fu-berlin.de}}}
\vspace{2cm}
\end{center}
\begin{quote}\small{ 
In this work we propose an exact microscopic description
of maximally symmetric branes in a Euclidean $AdS_3$
background. As shown by Bachas and Petropoulos, the most 
important such branes are localized along a Euclidean $AdS_2 
\subset AdS_3$. We provide explicit formulas for the coupling 
of closed strings to such branes (boundary states) and for the 
spectral density of open strings. The latter is computed in two 
different ways first in terms of the open string reflection 
amplitude and then also from the boundary states by world-sheet 
duality. This gives rise to an important Cardy type consistency 
check. All the results are compared in detail with the geometrical 
picture. We also discuss a second class of branes with spherical 
symmetry and finally comment on some implications for D-branes in 
a 2D back hole geometry.}
\end{quote}

 \setcounter{equation}{0} 
\section{Introduction}
\def\tr{{\rm tr}}

String theories on Anti-deSitter ($AdS$) spaces have received 
enormous attention over the last years because of their 
conjectured duality with gauge theories on the boundary 
of the $AdS$ space (see \cite{AGMOO} and references therein). 
Unfortunately, strings moving in $AdS_p$ are rather difficult 
to study and therefore most of the tests and uses of the 
duality have been restricted to a super-gravity limit in which the 
$AdS$ space is only weakly curved. For $p=3$ the situation 
is much better because the string equations of motion for 
$AdS_3$ can be solved with a non-vanishing NSNS 3-form 
field strength so that there is no need for non-zero RR
background fields. This allows to study the $AdS/CFT$ 
correspondence in a truly stringy regime. 
\smallskip

In this paper we shall work with the Euclidean counterpart 
$\H3p$ of $AdS_3$. Let us be a bit more specific and describe 
the model for $\H3p$ we will be using. To this end, we 
identify $AdS_3$ with the group manifolds of $SL(2,\BR)$. 
In fact, 
\begin{equation} \label{matX} 
\left( \begin{array}{cc} X_0+X_1 & X_2 + X_3 \\
       X_2-  X_3 & X_0 - X_1 \end{array} \right)
 \in \SLR 
\end{equation} 
implies that $X_0^2 - X_1^2 - X_2^2 + X_3^2 = 1$ which is 
the defining equation of $AdS_3 \subset \QR^4$. One can 
imagine this space as an infinite solid cylinder which is 
parametrized by the global coordinates $(\rho,\theta,\tau)$ 
such that (see Figure)   
\begin{figure}[ht]
\includegraphics{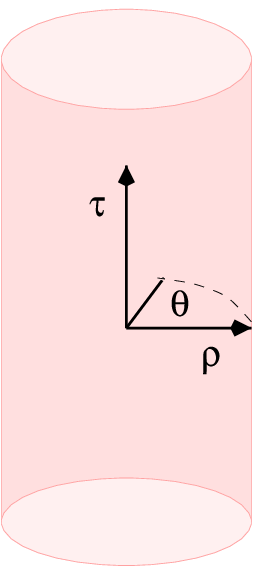}\hspace*{7cm} \vspace*{-7cm}  
\end{figure} 

\hfill \begin{minipage}[t]{7cm} 
\begin{eqnarray} 
  X_0 + i X_3  & = & e^{i \tau } \cosh \rho \ \ \ , \\[2mm]   
  X_1 + i X_2  & = & e^{i \theta} \sinh \rho \ \ .
\end{eqnarray}
Upon rotation to a Euclidean time $\tau \rightarrow i \tau$, 
the coordinate $X_3$ gets replaced by $i X_3$. When we make 
this substitution in the matrices (\ref{matX}) above then 
we end up the the space $H_3$ of hermitian $2\times 2$ 
matrices $h$ with $\det h = 1$. It consists of two 
components and the component of the identity matrix 
is given by \vspace*{2mm}
$$ \H3p = \{ h \in \SLC\, |\,  h^\dagger\, = \, h\, ,\,  \tr h > 0 \} 
\ . $$ \vspace*{-2mm} 
\end{minipage}   

This is the space on which we want to study string theory. 
We have used the $AdS/CFT$  correspondence as our main motivation.
Let us note, however, that there are various other good reasons
to be interested in $\H3p$. Part of them are related to the fact 
that one can descend from $\H3p$ to the coset $\H3p/ \QR_\tau$ 
describing a 2D Euclidean black hole \cite{Wit1}. The relevant 
action of $\QR$ on $\H3p$ is given by constant shifts in the 
Euclidean time $\tau$. The black hole geometry appears as part 
of many interesting string backgrounds. One example is the near 
horizon geometry of non-extremal NS5-branes \cite{HoSt,MaSt}. 
Furthermore, it can emerge as a factor when Calabi-Yau spaces 
develop an isolated singularity \cite{OoVa}. 
\smallskip

Recently, there has been considerable progress towards the 
construction of perturbative closed string theory on $AdS_3$,
see \cite{MaOo1,MaOo2,MaOo3} and references therein. These works 
show that partition function
and scattering amplitudes of string theory on $AdS_3$ can 
be constructed with the help of the $H_3^+$ gauged WZNW model. 
The procedure of constructing amplitudes for string theory on 
$AdS_3$ from correlation functions associated to a Euclidean 
target may be seen as some analog of the usual Wick-rotation. 
It is therefore crucial for the success of such a procedure 
to have sufficient control over the $H_3^+$ model. The first 
important step was the calculation of the partition function 
\cite{Ga} which allows to determine the spectrum. Crossing 
symmetric correlation functions on the sphere were constructed 
in \cite{Te2,Te3} from the three-point functions of the model. 
The latter were first obtained in \cite{Te1,FZZH+,Te2}.
\medskip

In the present paper we want to study D-branes on backgrounds 
containing $AdS_3$. For the Lorentzian models some possible 
brane geometries were first analyzed by \cite{Sta} using the 
relation between $AdS_3$ and the group $\SLR$ along with 
results from \cite{AlSc1} which show that branes on group 
manifolds can wrap conjugacy classes. It was later shown by 
Bachas and Petropoulos \cite{BaPe} that the most interesting 
branes 
on $AdS_3$ are associated with twined conjugacy classes 
in the sense of \cite{FFFS1}. These can be localized along 
\ $AdS_2 \subset AdS_3$ \ (see
\begin{figure}[ht]
\includegraphics{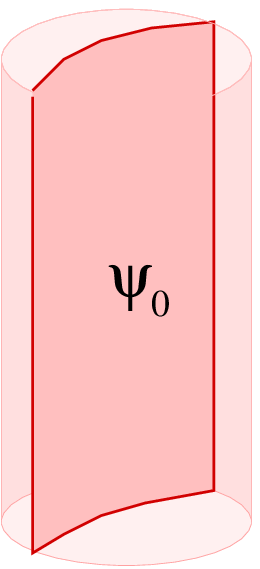}\hspace{7cm} \vspace*{-6.7cm}
\end{figure}   

\hfill \begin{minipage}[t]{7cm}  
 Figure) and they are 
parametrized by a single real parameter $\Psi_0$. In 
addition one can have branes localized along $H_2$, $dS_2$,
the light cone, as well as point-like branes. Not all of these
geometric possibilities correspond to physical brane configurations, 
though: The branes localized along $dS_2$, for example, were 
found to have a supercritical electric field on their 
world-volume \cite{BaPe}. 

In view of the above-mentioned possibility to construct perturbative 
closed string theory on $AdS_3$ via the $H_3^+$ WZNW model 
it is natural to expect that at least the physical 
branes in $AdS_3$ can also be described by means of the  
corresponding Euclidean model.
\end{minipage} 
\bigskip

Our main focus in this paper will be on the Euclidean counterparts 
of the $AdS_2$ branes. We shall also find analogues for the point-like 
branes as well as branes along the two connected components 
$H^\pm_2$ of $H_2$. In addition, it is possible to localize branes 
along 2-spheres, though the exact solution will tell us later that 
these spheres have an imaginary radius. Let us note, however, that the 
branes on $H^\pm_2$ are related to the Euclidean $AdS_2$ branes by 
a symmetry transformation on $\H3p$ so that is suffices to study 
the latter. 
\medskip 

In this paper we will analyse branes using microscopic 
techniques of boundary conformal field theory (BCFT).
As we mentioned already, we shall concentrate on the $AdS_2$ 
branes since their analysis is much more difficult 
than for the point-like and spherical branes. This is related 
to the fact that the former are non-compact and extend
to the boundary of $AdS_3$. Hence, in the exact solution we will 
find a continuous open string spectrum with a rather non-trivial 
spectral density. For completeness, we shall also spell out all 
the relevant formulas that are needed to treat the point-like
and spherical branes. 
\smallskip

Let us now explain our main results in more detail. In string theory, 
D-branes can be characterized by their couplings to closed string 
states. In the case of maximally symmetric branes on $\H3p$, the 
relevant states are associated with bulk fields $\Phi^j(u|z)$ where 
$j \in - \frac12 + \QR^+$ and  $u \in \QC$. These fields live on 
the upper half plane $\Im z \geq 0$. Couplings of closed string 
modes to the brane are encoded in the one-point functions of the 
bulk fields. 
\begin{equation} \label{intro1} 
 \langle \Phi^j(u|z)\rangle_{BC} \ = \ 
\ \frac{A^j(u)^{BC}}{|z-\bz|^{2\De_j}}\ \ .  
\end{equation}  
Here, $\Delta_j$ are the conformal dimensions of the fields $\Phi^j$ 
and the label $BC$ refers to the choice of the boundary condition. 
The form of the 1-point functions is fixed by conformal invariance 
up to some constants $A^j(u)^{BC}$. The latter contain the same information 
as the boundary state. 
\smallskip

Another interesting quantity in boundary conformal field theory is 
the partition function. It encodes information on spectrum of open 
strings that are living on the brane. For maximally symmetric branes
the partition function can be expanded in terms of characters $\chi^j$
of the chiral algebra, i.e.\  very schematically one has 
\begin{equation} \label{intro2} 
Z_{BC}(q) \ = \ \int_{\BS_{BC}} dj  \ \rho^{BC}(j) \ \chi_j(q) \ \ .  
\end{equation}  
Here, the integration extends over a set that might depend on the 
branes and it might be either continuous or discrete. In the latter 
case, the integral gets replaced by a sum. When $\BS_{BC}$ is 
continuous, the partition function involves a non-trivial spectral 
density function $\rho_{BC}$ which describes the density of open 
string modes with `momentum' $j$. Following Cardy, the partition 
function may be computed from the boundary state of the brane by 
world-sheet duality. 
\smallskip

But there is another way of obtaining $\rho_{BC}$. It involves one 
more interesting quantity to study in case of non-compact branes: 
the so-called {\em reflection amplitude} of open strings. Open strings
states can be created by boundary operators $\Psi^j_{BC}(u,x)$. 
Here $u$ is a real variable which carries an action 
of the space-time symmetry that is left unbroken by the brane and $x$ 
is the usual coordinate for the boundary of the world-sheet. One 
can then study the amplitude that describes the scattering of an open 
string that is sent in with momentum $j_1$ from the boundary of 
$AdS_3$ into an outgoing open string with momentum $j_2$. This  
defines the reflection amplitude,
\begin{equation} \label{intro3} 
\langle \Psi^{j_1} (u_1|x_1) \, \Psi^{j_2}(u_2|x_2) \rangle_{BC}   
\ \sim \ \delta(j_1 - j_2)  \ R(j)^{BC} \, 
\frac{1}{|x_1 - x_2|^{\Delta_j}} \ \ . 
\end{equation}
Here, $j_i \in  -1/2 + i \QR^+$ and we omitted some $(j_i,u_i)$ 
dependent factor that is determined by the unbroken symmetry. 
It is one of the fundamental observations in scattering theory that 
one can often recover the spectral density $\rho_{BC}$ from such a 
reflection amplitude. For the reader's convenience we have included 
a review of this relation in Appendix B. Comparison between the two 
ways of obtaining $\rho_{BC}(j)$ is an important consistency check. 
\medskip     
 
The aim of this work is to determine the one-point functions 
(\ref{intro1}), the open string reflection amplitude (\ref{intro3}) 
and the open string spectral density (\ref{intro2}) for all 
maximally symmetric branes of the model. As we have explained, 
these branes split into two classes. The first one consists of 
branes which break the $\SLC$ symmetry of the background to a 
subgroup $\SLR$. All branes from this class are related by a 
symmetry transformation to a Euclidean $AdS_2$-brane. Their 
solution is given by eqs.\ (\ref{res1},\ref{res2},\ref{Zrel}) 
below. Analogous results are also provided for a second class 
of D-branes in $\H3p$ which possess an $\SU$ symmetry (see eqs.\
(\ref{res1a},\ref{res3a})). They behave as if they were localized 
along a discrete set of 2-spheres with an {\em imaginary} radius.  

In our exposition we shall begin with a discussion of the 
semi-classical limit of the model (Section 2) where the stringy 
corrections are turned off. This allows us to introduce all the 
relevant objects in a rather familiar and simple setup. It is
also reassuring to see later that the results we obtain by very 
different methods in the full string theory do indeed possess 
the expected semi-classical behavior. The one-point functions are 
then constructed in Section 3 by solving certain factorization
constraints. Similar techniques are also employed in Section 4
in order to find the reflection amplitude for open strings. The
consistency between these data is then discussed in Section 5 
where we show that they are related by world-sheet duality. 
\medskip
 
Several recent publications have addressed the problem of 
constructing branes in $AdS_3$ \cite{GKS,PaSa,RaRo,LOPT,HiSu,Raj}. 
However, it seems to us that even the most basic of the relevant 
data, namely the one-point function, 
is not available so far.\footnote{When
this paper was nearly completed, we were informed by H.\ Ooguri
that P.\ Lee, H.\ Ooguri and J.\ Park have also found the exact
expression for the one-point function of $AdS_2$-branes. We thank 
H.\ Ooguri for kindly sending us a draft of their paper. It 
has some overlap with the discussion in Sections 2 and 3.} 
The discussion in \cite{GKS,PaSa} focuses mainly on a series of 
boundary conformal field theories that includes the point-like
brane and the ones that we called `spherical' above. As we shall 
show below, however, the dependence of the one-point functions 
on the coordinate $u$ of the CFT on the boundary of $AdS_3$ has
not been stated correctly in those papers. The authors of 
\cite{GKS} did observe that their one-point functions produced 
some puzzling singularities at the boundary of $AdS_3$. The 
correct formulas turn out to be regular, as one would have 
expected. The $AdS_2$ branes in $AdS_3$ were even less well 
understood. Semi-classical expressions for the one-point 
functions have been proposed in \cite{RaRo} and it was also 
suggested that these formulas might hold true in the string 
regime. Our analysis shows that this is not the case. We will
comment more on the discrepancies with the existing literature 
as we proceed.       

\setcounter{equation}{0} 
\section{Strings on $\H3p$ -- the semi-classical limit} 
\subsection{Bulk geometry and the closed string action} 

\paragraph{\it Geometry of $\H3p$.} 
As we have explained and motivated in the introduction, we are 
interested in studying string theory on the space $\H3p$ of
Hermitian $2 \times 2$ matrices $h$ with determinant $\det h = 
1$ and positive trace. It is convenient to parametrize this 
space through coordinates $(\phi, \c,\bc)$ such that 
\begin{equation}\label{pggpar} 
 h \ = \ \left( \begin{array}{cc} e^\phi \ & \ e^\phi \bc \\
                   e^\phi \c \ & \  e^\phi \c \bc + e^{-\phi} 
                  \end{array} \right) \ \ . 
\end{equation}
Here, $\phi$ runs through the real numbers and $\c$ is 
a complex coordinate with conjugate $\bc$. We can visualize the 
geometric content of these coordinates most easily by expressing 
them in terms of the more familiar global coordinates $(\rho,
\tau,\theta)$ that we also used in the introduction, 
$$ \c \ = \ e^{\tau + i \theta} \, \tanh \rho \ \ \ \ \ 
   \mbox{ and } \ \ \ \ \ e^\phi \ = \ e^{-\tau} \cosh \rho 
   \ \ \ . 
$$  
At fixed $\ga,\bc$, the boundary of $\H3p$ is 
reached in the limit of infinite $\phi$. The boundary is now represented
as the complex plane with coordinates $\c,\bc$, which 
are related to the coordinates $(\tau,\theta)$ via the usual 
conformal mapping from the cylinder to the complex plane,
$\c \ = \ e^{\tau + i \theta}$. 
\smallskip

Let us note in passing that $\H3p$ admits an action of the 
group $\SLC$ which is defined as follows
\begin{equation} \label{SL2Cact}
 h \ \longrightarrow \ g \, h \, g^\dagger \ \ \ 
  \mbox{ for } \ \ \ g \in SL(2,\QC)\ \ . 
\end{equation}
Since the stabilizer of this action is isomorphic to the 
subgroup $\SU \subset \SLC$ we can identify $\H3p$ with 
the coset $\H3p = \SLC/\SU$.      
\medskip 

The space $\H3p$ comes equipped with the following metric and
$H$-field, 
\begin{eqnarray}
ds^2 & = & \, d\phi^2 + e^{2 \phi} \, d\c d \bc \ \ \ , \\[2mm]
H & = & 2 \, e^{2 \phi} \, d\phi \wedge d \bc \wedge d \c \ \ . 
\end{eqnarray} 
We shall introduce 2-form potentials $B$ 
for the 3-form $H$ later on. 
\medskip

\paragraph{\it The string action.} 
To write down the action functional, we need to choose some 
2-form potential $B'$ for $H$. Note that the space $\H3p$ is 
topologically trivial which implies that such a potential always 
exists and, moreover, that the resulting action for closed strings
does not depend on the particular choice we make. For the 
moment, we shall work with 
$$ B' \ = \ e^{2 \phi} \, d \c \wedge d \bc \ \ . $$ 
Putting all this information together, we arrive at the 
following action functional for closed strings moving on 
$\H3p$, 
\begin{equation} \label{Saction} 
 S(\phi,\gamma,\bc) \ = \ \frac{k}{\pi} \, \int \ dz \, 
    d\bar z \ \left( 
    \pl \phi \bpl \phi + e^{2 \phi} \, \pl \c \bpl \bc 
     \right) \ \ . 
\end{equation} 
One should note that this model has some obvious 
defect, namely it has an imaginary B-field 
that causes the theory to be non-unitary. The problem 
is quite easy to understand. Recall that the string 
equations of motion relate the curvature $R$ of the 
background to the square of the $H$-field (provided 
that the dilaton is constant). Now it is also well known 
that strings on a 3-sphere have a perfectly unitary 
description. When we pass to $\H3p$, the curvature 
changes its sign and we have to multiply the 3-form 
$H$ with $\sqrt{-1}$ to be consistent with the string 
equations of motion. This factor $\sqrt{-1}$ is then 
certainly passed on to the potential $B'$. Such problems
disappear when we descend to the black hole geometry 
$\H3p/ \QR$ since the latter has vanishing $H$-field 
for purely dimensional reasons.  
\medskip

\paragraph{\it The currents.} We want to conclude this 
subsection with a few remarks on the chiral currents 
of the model. Let us introduce the following matrices  
\begin{equation}
T_+ \ =\ \left(
\begin{matrix} 0 & -1\\ 0 & 0 \end{matrix}
\right),\quad
T_- \ =\ \left(
\begin{matrix} 0 & 0\\ 1 & 0 \end{matrix}
\right),\quad
T_0 \  = \ \frac{1}{2}\left(
\begin{matrix} 1 & 0\\ 0 & -1 \end{matrix}
\right). 
\end{equation}
These are matrix representatives of the Lie algebra $\SLR$, 
i.e.\ they obey the relations $[T_0,T_\pm] = \pm T_\pm$ and 
$[T_-,T_+] = 2 T_0$. For the chiral currents we use 
$$ J(z) \ := \ k \, h^{-1} \bpl h \ \ \quad \ \ \ 
   \bJ(\bz) \: = \ - k \, \pl h\, h^{-1} \ . $$
When we expand them according to $J (z) = T_+ J^+ + T_- J^- 
+ 2 T_0 J^0$, we obtain expressions for the components  
\begin{eqnarray}
J^-(z) & := &  k\, e^{2 \phi} \, \bpl \c \\[2mm]
J^0(z) & := &  k\, \left(\bpl \phi -  e^{2\phi}\, \bc \, 
              \bpl \c \right) \\[2mm] 
J^+(z) & := & k\, \left( \bc^2 \, e^{2 \phi}\, \bpl \c - 
           \bpl \bc - 2 \, \bc\, \bpl \phi \right) \ \ .   
\end{eqnarray} 
The components of the anti-holomorphic currents are 
constructed in an analogous way. Both sets of currents are 
related by complex conjugation $ (J^\pm)^* = (\bar J)^{\mp}$ 
and $ (J^0)^* = - \bar J^0$.

\subsection{Brane geometry and the boundary conditions} 
\def\tr{\mbox{\rm tr}}  
\def\bx{\bar x}
\def\a{\alpha} 
\def\diag{{\mbox{\rm diag}}}

\paragraph{\it General results.} 
In this section we want to present the possible geometries for
branes in $\H3p$ which preserve half of the $\SLC$ symmetry
(\ref{SL2Cact}). Let us recall that $\SLC$ contains two 
important 3-parameter subgroups, namely the groups $\SLR$ 
and $\SU$. We shall analyse equations of the form 
\begin{equation} \tr(C h) \ = \ c \ \ ,\label{eq} 
\end{equation}   
where $C$ is a $2 \times 2$-matrix and $h \in \H3p$. It turns 
out that there are two important cases to distinguish. If the 
matrix $C$ is of the form  \footnote{One could admit matrices 
$C$ of a slightly more general form and with $U \in$ GL(2,$\BC$) 
but this extra freedom can be absorbed in a rescaling of the 
constant $c$.}
$$ C \ = \ U^\dagger U \ \ \ \ \mbox{where } \ \ \ \ U 
   \in \SLC \ \ . $$ 
then the equation (\ref{eq}) preserves a subgroup of $\SLC$ that 
is conjugate to SU(2). More precisely, for every $g \in$ SU(2) 
the element $U^{-1} g U \in  \SLC $ is a symmetry of the 
equation. A second possibility is to impose eq.\ (\ref{eq}) with 
a matrix $C$ of the following form:
$$ C \ = \ U^\dagger \, \om 
\, U \ \ \  \mbox{ where } \ \ \ \om:=\left( \begin{array}{cc} 
   0 & 1 \\ 1 & 0  \end{array} \right)\ \  \mbox{ and }
   \ \ U \in \SLC\ \ . $$
In this case, the equation preserves a subgroup of $\SLC$ 
that is conjugate to $\SLR$, i.e.\ it is left invariant by 
the action with $U^{-1} g U \in \SLC$ for all $g\in\SLC$ that satisfy 
$\om(g^\dagger)=g^{-1}$, where 
\[
\om(g):=\om\, g\, \om^{-1}\;\, .
\]
It is easy to see that this condition implies that the 
matrices $g$ must be of the 
form $g=\bigl(\begin{smallmatrix} \al & i \be \\ i\ga & \de
\end  {smallmatrix}\bigr)$ and therefore generate a subgroup of $\SLC$ 
that is conjugate to $\SLR$. 
\medskip

For each of these two cases it suffices to consider the the 
special choice $U = 1$. In fact, the submanifold defined by an 
equation with nontrivial $U$ is obtained from the one corresponding 
to $U = 1$ through the symmetry transformation $U \in$ 
SL(2,$\BC$). This means that it suffices to consider just 
two different choices of $C$. These will be described in more 
detail now.   
\medskip

\paragraph{\it The $AdS_2$ branes.} The first case corresponds to 
surfaces which are characterized by the equations
$$  \tr (\,\om\,h\,)   = \ c \ . $$
Solutions of these equations 
form Euclidean $AdS_2$-planes ending at $\theta = \pm \pi/2$ on the 
boundary. In terms of the coordinates introduced above one gets the 
equations
$$  e^\phi \, ( \gamma  + \bc ) \ = \ c \ \ \ 
 \mbox{ or } \ \ \ \ 2 \, \sinh(\rho) \cos(\theta) 
 \ = \ c   \ \ . $$ 
It is convenient to introduce a new set of coordinates $(\psi, 
\nu,\chi)$ on $\H3p$ in which these branes are coordinate planes
$\psi =\raa$. We can achieve this by setting 
\begin{equation}\label{param1}
h\;=\;  c(\nu,\chi)\cdot h_{\psi} \cdot
c^{\dagger}(\nu,\chi) \ \ , 
\end{equation}
where 
\begin{equation} \nonumber
h_{\psi}\;\equiv\;\left( \begin{matrix} \cosh \psi & \sinh \psi \\
\sinh \psi  & \cosh \psi \end{matrix}\right),
\;\;{\rm and}\;\;
c(\nu,\chi)\;\equiv\; 
\left( \begin{matrix} e^{\frac{\chi}{2}} & 0 \\
i\nu e^{\frac{\chi}{2}}  & e^{-\frac{\chi}{2}} \end{matrix}\right).
\end{equation}
Definition \rf{param1} is equivalent to:
\begin{equation} 
h \;\equiv\;
 \left( \begin{matrix} e^{\chi}\cosh\psi & 
\sinh \psi +i\nu e^{\chi}\cosh\psi \\
\sinh \psi -i\nu e^{\chi}\cosh\psi  &  (e^{-\chi}+\nu^2 e^{\chi})\cosh \psi
\end{matrix}\right) \ \ . 
\end{equation}
In these coordinates the metric $ds^2$ and the B-field have the from 
\begin{eqnarray} 
ds^2 & = & d\psi^2 + \cosh^2 \psi \ 
( \, e^{2\chi} \, d\nu^2 + d\chi^2 \, ) 
\\[2mm] 
B & = & 2 i \, \left( \frac{1}{2} \sinh 2\psi \ + \psi \right) 
   \ e^\chi \, d\nu \wedge d \chi \ \ . 
\end{eqnarray} 
The $AdS_2$ branes are described by the equation $\psi = \raa$. 
Note that in the coordinates $(\psi,\chi,\nu)$, the boundary of an 
$AdS_2$-brane is at $\chi = \infty$. It is parametrized by $\nu = 
\pm e^\tau$. The $SL(2,\BC)$-invariant measure has the form 
$$ dh\ = \ 2d\nu\,d\chi e^{\chi} \,d\psi\cosh^2\psi\ \ . $$
Given the expressions for the metric and the  B-field, it is 
straightforward to write down the open string action for the 
fields $\psi,\nu,\chi$. Vanishing of the boundary terms in the 
variation of this action is equivalent to the following simple 
boundary conditions for currents 
\begin{equation}\label{glue1}   
 J^{\pm}(z) \ = \ \bar J^{\pm}(\bz)  \ \ \ , \ \ \ 
   J^0(z) \ = \ \bar J^0 (\bz) \ \ .
\end{equation} 
holding all along the boundary $z = \bz$. Note that these 
gluing conditions are consistent with the *-operation and they 
imply that the boundary current obeys $(J^\pm)^* = J^\mp$ and
$(J^0)^* = - J^0$.   
\medskip

It seems worth noting that the $\SLC$-translates 
of the Euclidean $AdS_3$-branes include branes which correspond to
$H_2^{\pm}$ in the Minkowskian picture: These branes are characterized by
the equation  
$$  \tr \left( \begin{array}{cc} -\frac{i}{4} & 0 \\ 
    0 & \frac{i}{4} \end{array}
     \right) \ h \ = \ c' \ \ . $$
As in the previous cases we display this equation in our coordinates:
$$ e^{\phi} (\gamma \bar \gamma - 1) + e^{- \phi} \ = \ c
\ \ \ \ \mbox{ or } \ \ \ \ \ 2 \cosh \rho \sinh \tau \ = \ c \ \ .
$$
It is now easy to see  that these solutions are extended 
along $H_2^-$ for $c<0$ and along $H_2^+$ for $c>0$. For $c=0$ one 
gets a disc at $\tau = 0$. Although they look quite differently from 
the $AdS_2$ branes, they are related to the latter by an SL(2,$\BC$)
transformation $U$ of the form 
$$ U \ = \ \frac{1}{\sqrt{2}}\ \left( \begin{array}{cc} 
 e^{i\frac{\pi}{4}} & e^{i\frac{\pi}{4}} \\ -e^{-i\frac{\pi}{4}}  
 & e^{-i\frac{\pi}{4}} \end{array} \right) \ \ .$$ 
\medskip

\paragraph{\it Spherical branes.} To get an idea about the subsets
that preserve an $\SU$-symmetry let us study equations of the form 
$$   \tr \left( \begin{array}{cc} 1 & 0 \\ 0 & 1 \end{array}
     \right) \ h \ = \ c \ \ .$$
When rewritten in terms of the coordinates $(\phi,\gamma,\bar \gamma)$ 
or the global coordinates $(\rho,\tau,\theta)$ these equations read   
$$ e^{\phi} (\gamma \bar \gamma + 1) + e^{- \phi} \ = \ c 
\ \ \ \ \mbox{ or } \ \ \ \ \ 2 \cosh \rho \cosh \tau \ = \ c \ \ . 
$$ 
Solutions exist for $c \geq 2$ and they are point-like when $c=2$ 
and spherical otherwise. None of them extends to the boundary because 
near to the boundary the equation would become $\gamma \bar \gamma + 1 = 0$. 
\smallskip

It is convenient to introduce a new set of coordinates $(\Lambda, 
\phi,\mu)$ on $\H3p$ in which these branes are coordinate planes
$\Lambda =\Lambda_0$. We can achieve this by setting 
\begin{equation}\label{param2}
h\;=\;  c(\mu,\vph)\cdot h_{\Lambda} \cdot
c^{\dagger}(\mu,\vph) 
\end{equation}
where 
\begin{equation} \nonumber
h_{\Lambda}\;\equiv\;\left( \begin{matrix} \cosh \Lambda & \sinh 
\Lambda \\
\sinh \Lambda  & \cosh \Lambda \end{matrix}\right),
\;\;{\rm and}\;\;
c(\mu,\vph)\;\equiv\; 
\left( \begin{matrix} e^{i \frac{\mu}{2}} \cos \frac{\vph}{2}  &  
  e^{-i \frac{\mu}{2}} \sin \frac{\vph}{2}  \\
 -e^{i \frac{\mu}{2}} \sin \frac{\vph}{2}  &  
e^{-i \frac{\mu}{2}} \cos \frac{\vph}{2}  
\end{matrix}\right).
\end{equation}
Definition \rf{param2} is equivalent to:
\begin{equation}
h \;\equiv\;
\cosh \Lambda \ {\bf 1}_2 + \sinh \Lambda \ 
 \left( \begin{matrix} \cos\mu \sin \phi & 
   \cos\mu \cos \phi + i \sin \mu  \\
 \cos \mu \cos \phi - i \sin \mu & - \cos \mu \sin \phi
\end{matrix}\right)\cdot 
\end{equation}
In the new coordinates the metric $ds^2$ and the B-field have the from 
\begin{eqnarray} 
ds^2 & = & d\Lambda^2 + \sinh^2 \Lambda \ 
( \, \cos^2 \mu \, d\vph^2 + d\mu^2 \, ) 
\\[2mm] 
B' & = & 2 i \, \left( \frac{1}{2} \sinh 2\Lambda \ - \Lambda \right) 
   \ \cos\mu \, d\vph \wedge d \mu \ \ . 
\end{eqnarray} 
The spherical branes are described by the equation $\Lambda = 
\Lambda_0 \geq 0$. Note that in the coordinates $(\Lambda,\vph,\mu)$, 
the boundary of $AdS_3$ is at $\Lambda = \infty$. The 
$SL(2,\BC)$-invariant measure is given by $dh= 2 d\vph \,d\mu 
\cos\mu  \,d\Lambda\sinh^2\Lambda$. A straightforward computation 
shows that the currents must satisfy 
\begin{equation} \label{glue2}  
 J^{\pm} \ = \ \bar J^{\mp} \ \ \ , \ \ \ J^0 \ = \ - \bar J^0 \ \ .
\end{equation} 
along the boundary $z = \bar z$ in order for the boundary terms
in the variation of the action to vanish. Once more this is 
consistent with the $*$-structure but this time the induced
action on the boundary currents is $(J^\pm)^* = J^\mp$ and 
$(J^0)^* = J^0$, i.e.\ we have an su(2) current algebra on 
the boundary of the world-sheet.

\subsection{Semi-classical limit of closed string couplings} 

Our aim in this subsection is to study the semi-classical limit 
of the closed string couplings to the brane. Following \cite{MMS1},
we will read them off by expanding the $\delta$-functions
that describe localization on the branes in a 
basis of eigen-functions for the Laplace operator on $\H3p$. 
The latter are in one-to-one correspondence with the primary 
fields of the bulk theory. We will explain this in more detail 
after a short review of the harmonic analysis on 
$\H3p$ \cite{GGV,Te0}. 
\medskip

\paragraph{\it Harmonic analysis on $\H3p$.} Any wave function on 
$\H3p$ can be expanded in terms of eigen-functions of the 
Laplace operator on $\H3p$. We recall that there exists an 
action of $\SLC$ on $\H3p$ which commutes with the Laplace 
operator. This implies that each eigen-space must carry some
representation of $\SLC$. It is not difficult to show that 
the possible eigenvalues are given $j(j+1), j = - \frac12 + 
iP,$ where $P$ is a non-negative real number and that the 
associated eigen-spaces carry the irreducible representation 
$D_j$ from the principal continuous series. Explicitly, the 
eigen-functions are given by the following formula 
\begin{eqnarray} 
\Phi^j(u| \phi,\c,\bc)  & = & - \frac{2j+1}{\pi} \  
  (v_u \, h \, v_u^\dagger)^{2j} \label{phief} \\[2mm] 
& = & - \frac{2j+1}{\pi} \left( |u - \c|^2 e^\phi + e^{-\phi} 
      \right)^{2j} \nonumber 
\end{eqnarray}  
Here, $u$ is a complex coordinate and $v_u = (-u,1)$. In the 
second line we inserted the parametrization (\ref{pggpar}) 
of $\H3p$. The transformation law of the functions (\ref{phief})
under the action of $\SLC$ is easily worked out, 
\begin{equation} \label{trafo1} 
   \Phi^j (u|g h g^\dagger) \ = \ 
   |\beta u + \delta|^{4j}\ \Phi^j({g \cdot u}|h) 
 \ \  \mbox{ where } \ \ g \cdot u \, = \, \frac{\a u + \c}{\b u + \d} \;\, ,
\end{equation}   
and $\a,\b,\c,\d$ are the four matrix elements of $g \in 
\SLC$. For later use we shall also spell out the 
asymptotics of the eigen-functions near the boundary 
of $\H3p$,  
\begin{eqnarray} \label{phiasym}
\Phi^j(u| \phi,\c,\bc) &
\stackrel{\phi 
 \rightarrow \infty}{\sim} &  - I^j(u|\c) \ e^{2 j \phi } 
  + \, \delta(\c - u) \, e^{- 2 (j+1) \phi } \\[2mm]
\mbox{ where }  & & I^j(u|\c) \ := \  \frac{2j+1}{\pi} 
 |\c - u |^{4j} \, \ \ \  
\end{eqnarray} 
is the integral kernel of the unitary intertwiner that implements 
the equivalence between the representations $D_j$ and $D_{-j-1}$ 
of $\SLC$ \cite{GGV}. The functions $\Phi^j(u|h)$ 
can be considered as the wave-function of some particle that was 
created with `radial momentum' $j$ at the boundary point with 
coordinates $u,\bu$ \cite{BORT}. They are in 
one-to-one correspondence with the ground states of the bulk 
conformal field theory on $\H3p$ and form a basis in the space 
of square integrable functions on $\H3p$. 
\medskip

\paragraph{\it The $AdS_2$-branes.} We now want to determine the 
semi-classical one-point function $\langle\Phi^j\rangle_{\raa}$ 
which is supposed to describe the amplitude for absorption/emission of 
closed string modes with asymptotic radial momentum $j$ by the 
brane. These amplitudes can be regarded as Fourier-transforms 
of the amplitudes $\bra\Phi^{h'}\ket_{\raa}$ for the 
absorption/emission of point-like localized closed string modes 
$\Phi^{h'}$ with wave-functions $\Phi^{h'}(h)=\de(h-h')$. 
The latter must of course vanish away from the surface 
$\psi=\raa$. Moreover, homogeneity of the brane world-volume
(equivalent to its $\SLR$-symmetry) imply that the amplitude 
can only depend on the transverse coordinate $\psi$. Hence, we  
conclude that $\bra\Phi^{h'}\ket_{\raa}\propto \de(\psi-\raa)$ 
up to a constant. 
 
Altogether this means that the one-point function $\langle\Phi^j
\rangle_{\raa}$ can be read off from the Fourier-expansion of 
$\de(\psi-\raa)$ w.r.t.\ the basis formed by $\Phi^j$. This 
expansion takes the form   
\begin{eqnarray} 
 \delta(\psi - \raa) & = & \kappa_1 \, \int_{\BS} dj \int du^2 
 \ (\Phi^j(u|\psi,\chi,\nu))^* \ \left( d_0^j(u) \cosh \raa 
   (2j+1) \right. \nonumber \\[0mm] 
  & & \label{deltapsi}
 \hspace{5cm} \left. -  d_1^j(u) \sinh \raa (2j+1)\right) \\[3mm] 
  \mbox{ where } \ & & d_\e^j(u) \ = \ |u+ \bar u|^{2j} 
     \sgn^\e (u + \bar u) \  \ , \nonumber 
\end{eqnarray} 
and $\kappa_1$ is defined as $\kappa_1 = (2/\pi i) \cosh \raa$. 
To prove this statement we make use of the following auxiliary formula 
\begin{equation} \label{aux1} 
\cosh \psi \, \int d^2 u \;(\Phi^j(u|\psi,\chi,\nu))^* \, d_\e^j(u) \ = \ 
 \left\{ \begin{array}{ll} \cosh\psi(2j+1) & \mbox{ for }  \e = 0 
 \\[1mm]  \sinh\psi(2j+1) & \ \mbox{ for } \e = 1 
\end{array} \right.  
\end{equation}
which is derived in Appendix A. The main idea is to show first that 
the integral is constant along the surfaces of constant $\psi$, i.e.\ 
the orbits of the $\SLR$ action on $\H3p$. At this point one makes 
use of the transformation law (\ref{trafo1}) of the functions 
$\Phi^j$ together with the fact that the functions $d^j_\e$ 
depend only on the sum $u + \bar u$. Then one exploits that $\Phi^j$ 
are eigen-functions of the Laplace operator and derives a second order 
differential equation for the $\psi$-dependence of the integrals. The 
latter has two independent coefficients which can finally be determined 
by studying the integral near the boundary of $\H3p$, i.e.\ in the
limit $\phi \rightarrow \infty$ where $\Phi^j$ is known to behave 
according to formula (\ref{phiasym}). 

Once the auxiliary formula is established, it is straightforward 
to obtain eq.\ (\ref{deltapsi}). In fact, one has 
\begin{eqnarray*}
 & &  \int_{\BS} dj \left( \cosh \psi (2j+1) \cosh \raa (2j+1) 
   - \sinh \psi (2j+1) \sinh \raa (2j+1) \right)  \\[2mm]  
 & &  \hspace*{1cm} = \  \frac{i}{2} \int_{- \infty}^\infty dP 
   \  e^{2i(\psi-\raa)P} = \frac{\pi i}{2} \delta(\psi - \raa) 
\ \ .
\end{eqnarray*} 
The reason we have gone through these technical steps here 
was to show how the $\delta$-functions arises from the two terms 
involving $d_0$ and $d_1$. One single term alone would give an 
answer that is either symmetric or anti-symmetric under the 
reflection $\psi \rightarrow - \psi$. Only if the two terms work 
together, we can obtain a $\delta-$ function that is localized 
at a single point $\raa$ on the real line. 
\medskip
    
\paragraph{\it The spherical branes.} 
A similar analysis can be performed for the 2-spheres in 
$\H3p$. The formula for the decomposition of the $\delta$ 
function of a 2-sphere characterized by $\Lambda = \Lambda_0$ 
is given by 
\begin{eqnarray} 
 \delta(\Lambda - \Lambda_0) \ = \ \kappa_2 \, \int_{\BS} dj \int du^2 
 \ (\Phi^j(u|\psi,\chi,\nu))^* \ \sinh \Lambda_0 (2j+1) \  
   (u \bar u + 1)^{2j}  \ \ 
\end{eqnarray}  
where $\kappa_2 = (4i/\pi) \sinh \lambda_0$. Note that $\sgn(u \bar u + 1) = 
1$. Correspondingly, there appears only one term in the 
expansion of the spherically symmetric $\delta$-function in 
contrast to what we found for the $AdS_2$-branes above. 
The proof of this formula follows the same ideas as described in the 
previous paragraph, but of course one now has to use the SU(2) action 
on $\H3p$. The counterpart of the auxiliary formula \rf{aux1} turns out to be
$$ \sinh \Lambda \, \int d^2 u \; (\Phi^j(u|\Lambda,\vph,\mu))^* 
   (u \bar u +1)^{2j} \ = \ \sinh \Lambda (2j+1) \ \ . $$ 
In the end one must recall that the coordinate $\Lambda$ is 
restricted to non-negative values so that 
$$ \int_{\BS} dj \ \sinh \Lambda (2j+1) \sinh \Lambda_0 (2j+1) 
 \ = \ \frac{\pi}{4i} \delta(\Lambda - \Lambda_0) \ \ . $$ 
It is the restriction $\Lambda \geq 0$ that really allows us
to decompose the spherically symmetric $\delta$-functions in 
the way we have described with only a single $u$-dependent 
term appearing at each momentum $j$.   
\medskip 

In conclusion, the expansions of the $\delta$-functions on the 
Euclidean $AdS_2$- and the $S^2$-branes lead us to expect that 
the semiclassical limits of the one-point functions are given by
\begin{eqnarray} \label{1pfexp1} 
 \langle \Phi^j(u|z) \rangle^{AdS_2}_{\raa} \ 
  & \stackrel{k \rightarrow \infty}{\sim} &  
   |u+\bar u|^{2j} \exp ( - \sgn(u+\bar u)\raa (2j+1) ) \\[2mm]
 \langle \Phi^j(u|z) \rangle^{S^2}_{\Lambda_0} \ 
 \ \ \  & \stackrel{k \rightarrow \infty}{\sim} &  
   (u\bar u + 1)^{2j} \sinh \Lambda_0 (2j+1) \label{1pfexp2} 
\end{eqnarray} 
Let us anticipate that we shall indeed find an expression 
for $\langle \Phi^j(u|z) \rangle^{AdS_2}_{\raa}$ with
the expected semi-classical  
behavior. For the spherical branes, however, 
the 1-point functions will have the form (\ref{1pfexp2}) with an 
imaginary parameter $\Lambda_0$ corresponding to an `imaginary
radius' of the 2-spheres.    
  
\subsection{Semi-classical limit of open string spectra}

Finally, we would like to understand the semi-classical 
(point-particle) limit of 
the open string theory on the brane. The wave functions of open 
strings can be expanded in eigen-functions of the Laplace operator 
on the brane. In order to get an idea of the spectrum of open 
strings on our branes we have to understand the spectrum of the
Laplace operator. We will discuss this for the two different 
cases separately. 
\medskip

\paragraph{\it The $AdS_2$-brane.} With the data and notations provided
in Subsection 2.2. we can easily write down the Laplace operator on 
the brane $AdS_2^{\raa}$, 
\begin{equation} \label{Casads2}  
      Q_{\raa} \ \sim \ \pl^2_\chi + \pl_\chi + e^{-2 \chi} 
      \pl^2_\nu \ \ .   
\end{equation} 
Once more, it is easy to write down an explicit 
formula for these eigen-functions
\begin{eqnarray} 
\Xi^j(u|\raa;\nu,\chi)  & = &   \  
  (v'_u \, h \, {v'}_u^\dagger)^{j}|_{\psi = \raa} \label{psief} \\[2mm] 
& = &    \ 
   \cosh^j \raa \  \left( (u - \nu)^2 e^\chi + e^{-\chi} 
      \right)^{j} \nonumber 
\end{eqnarray}  
Here, $u$ is a real coordinate and $v'_u = (iu,1)$. In the 
second line we inserted the parametrization (\ref{pggpar}) 
of $\H3p$.

Recall that the $AdS_2$ branes admit an action of $\SLR$ which 
commutes with $Q_{\raa}$ so that the eigen-functions of the 
Laplace operator form representations for $\SLR$. It turns out 
that eigen-functions for a given eigenvalue $j(j+1), j = - \frac12 
+ iP,$ carry an irreducible representation $\CP_j$ from the principal 
continuous series.  The transformation law of the functions (\ref{psief})
under the action of $\SLR$ is easily worked out, 
\begin{equation} \label{trafo2} 
   \Xi^j(u|g h g^\dagger) \ = \ 
   |\beta u + \delta|^{2j}\ \Xi^j_{g \cdot u} (h) 
 \ \  \mbox{ where } \ \ g \cdot u \, = \, \frac{\a u + \c}{\b u + \d}\;\, , 
\end{equation}   
and $g=(\begin{smallmatrix} \de & -i\be \\ i\ga & \de 
\end{smallmatrix})$ is a $\SLC$-matrix conjugate to an element of the
$\SLR$ subgroup that preserves
the Euclidean $AdS_2$. The asymptotics
of these solutions are given by 
\begin{eqnarray} \label{psiasym}
c^{-1}(j)\;\Xi^j(u| \raa; \nu ,\phi) \ 
 & \stackrel{\phi
 \rightarrow \infty}{\sim} & \  J^j(u|\nu) 
   \ e^{j \phi } + \nonumber \\[1mm] 
 & & \hspace*{-2mm}  +  \cosh^{2j+1} \raa \frac{c(-j-1)}{c(j)} 
   \ \delta(\nu - u) \, e^{- (j+1) \phi }, 
\  \\[2mm]
{} \mbox{ where}\qquad\qquad & & \hspace*{-22mm} 
J^j(u|\nu) \ = \ |u-\nu|^{2j}\, c^{-1}(j)\ \ ,\quad 
  c(j)\ := \ \sqrt{\pi}\, \frac{\Gamma(j+ \frac12)}{\Gamma(j+1)}\ \ .
\label{Icdef}\end{eqnarray}    
$J^j(u|\nu)$ 
is the integral kernel of the unitary transformation that 
implements the isomorphism between the two representations 
$\CP_j$ and $\CP_{-j-1}$ of $\SLR$ \cite{GGV}. The normalizing 
factor $c(j)$ is the so-called Harish-Chandra c-function which 
plays a central role in the harmonic analysis of non-compact 
groups.
 
The set of functions $\{\Xi^j;j\in-1/2+i\BR^+\}$ 
forms a basis for the space of wave-functions 
of a particle on the Euclidean $AdS_2$. We shall see that they 
are in one-to-one correspondence with the ground states of the 
boundary conformal field theory. An important piece of information  
is the prefactor of the second term in (\ref{psiasym}):
The first term describes a 
plane wave that is injected with some
momentum parametrized by $j$ at the boundary 
of the $AdS_2$-brane. Accordingly, the second term gives the 
outgoing signal, leading us to interpret
the non-trivial coefficient in front 
of the second term as a semi-classical 
reflection amplitude, 
\begin{equation}
R_{\rm c}(P)\ \equiv\ R_{\rm c} (\raa;P) \ = \  -(\cosh\raa)^{2iP}
\frac{\Ga(1-iP)}{\Ga(1+iP)} \frac{\Ga\bigl(\frac{1}{2}+iP\bigr)}
{\Ga\bigl(\frac{1}{2}-iP\bigr)}.
\end{equation}
We will determine the rather nontrivial stringy corrections to
this formula in Section 4. 
\smallskip

Let us note that some
important motivation to be interested in such reflection 
amplitudes derives from their relation with relative spectral 
densities. This relation is reviewed in Appendix B. 
It allows to predict how the density $\rho(P)$ of states
in a quantum mechanical system changes when the scattering 
potential is varied. In our case, we shall fix one 
$AdS_2$-brane with parameter $\raa_*$ and use it as 
a reference to compare with the spectral densities of the 
other branes. The precise relation is 
$$  \rho_{\rm rel}(P|\raa,\raa_\ast) \ = \ 
  \frac{1}{2 \pi i} 
 \ \frac{\pl}{\pl P} \ \log \frac{ R_{\rm c}(\raa;P)} 
 {R_{\rm c}(\raa_\ast;P)} \ = \ \frac{1}{\pi}\ 
 \log \frac{ \cosh \raa}{\cosh \raa_\ast }\ \ . 
$$     
Informally one may think of $\rho_{\rm rel}(P|\raa,\raa_\ast)$
as $\rho^{\rm c}_0(P) - \rho^{\rm c}_\ast (P)$.
In the semiclassical limit, this quantity is is completely 
unrelated to the closed string couplings we described in the 
previous subsection. But this changes when we turn to the 
stringy analogue. In fact, in string theory the couplings 
of closed strings to the brane allow to compute the open 
string spectral density by using world sheet duality 
(``Cardy computation''). As we shall see below, the stringy 
couplings to the brane do indeed provide a formula for the 
spectral density that reduces to the semi-classical 
expression when the stringy corrections are turned off. 
\medskip

\paragraph{\it Spherical branes.} For the spherical branes our 
discussion of the semi-classical limit of open string theory 
can be rather short as this is a lot simpler than for the 
$AdS_2$ branes. In addition, the following remarks can at best 
serve as some kind of guiding ideas since it will turn out 
that the conformal field theory does not seem to allow one to 
construct branes that are localized along a finite 2-sphere.
 
A priori, we would expect the 
following picture to emerge. As is well known, the space of 
functions on a 2-sphere is spanned by spherical harmonics 
$\Psi^j_m(\phi,\mu), j = 0,1, \dots; |m| < j $. They are 
eigen-functions of the standard Laplace operator on $S^2$ 
with eigen-value $j(j+1)$ and they transform according to 
the $2j+1$-dimensional representation of SU(2). Now let us
take into account that the spherical branes come equipped 
with a non-vanishing B-field. By standard arguments, 
this implies that the space of wave functions must be finite 
dimensional with a dimension that grows as we increase the 
parameter $\Lambda_0$ of the 2-sphere. Since the number of 
states is an integer, we conclude that $\Lambda_0$ must be
quantized too, i.e.\ boundary theories will only exist for
a discrete set of $\Lambda_0$. All these expectations are 
essentially copied from the findings for spherical branes 
in $S^3$ \cite{AlReSc1,AlReSc2} and they do give rise to a 
rather accurate picture of the open string sector for the 
SU(2)-symmetric branes on $\H3p$. But let us stress one more 
that the open string couplings will not quite fit into this
geometric framework.

\setcounter{equation}{0} 
\section{The closed string sector}

We shall now look for quantum corrections to the expressions 
for the semi-classical closed string couplings that we constructed 
in the previous section. In other words, our aim here is to obtain 
the exact 1-point functions $\langle \Phi^j(u|z)\rangle^{\rm BC}$
that describe maximally symmetric branes on $\H3p$.\footnote{Note 
that these one-point functions contain the same information 
as the boundary state $|BC\rangle$ of the corresponding BCFT.}  
These 1-point functions are strongly constrained by the gluing 
conditions (\ref{glue1}) or (\ref{glue2}) for chiral 
currents. The latter are certainly required for conformal 
invariance of the boundary conformal field theory, as usual. 
Throughout most of this section we shall concentrate on the 
$AdS_2$ branes which are characterized by   
\begin{equation}\label{gluecond}
J^{\pm}\;=\;\bJ^{\pm}, \qquad J^0\;=\;\bJ^0 \ \ \ .
\end{equation}
A short subsection on the case of spherical branes appears at the 
end of the section. Obviously, the gluing condition (\ref{gluecond}) 
is not sufficient for the construction of a consistent boundary
conformal field theory. In addition, one has to satisfy consistency conditions 
that arise from the factorization properties of correlation 
functions. The most important condition arises from the factorization
of 2-point functions of bulk operators in the presence of a boundary
\cite{CL,FZZ}. Together, the gluing condition and the factorization 
constraints can be expected to determine the 1-point functions 
completely. Let us emphasize that in this approach there is 
really no need for any geometric intuition of the type we have 
gained in the previous section. 

The analysis of gluing and factorization constraints will lead us to 
a rather plausible candidate for the quantum corrections to the 
semi-classical boundary state. There is a caveat, though, which comes
from the fact that we can evaluate only one special factorization 
constraint that arises from considering 2-point functions in which  
one of the two bulk fields corresponds to a degenerate current 
algebra representation. It turns out that this condition is not 
sufficient to fully determine the form of the 1-point function. 
Additional requirements have to be imposed in order to narrow down 
the remaining freedom. We shall later use a non-rational analogue 
of the Cardy condition for that purpose.

\subsection{Primary bulk fields}

\paragraph{\it Some basic facts.}  Let us collect some basic properties of 
the primary bulk fields that will be used in the present paper (see 
\cite{Te1,Te2} for more details). We are interested in bulk fields 
$\Phi^j(u|z)$, $\Im z \geq 0,$ which obey the following operator 
product expansion (OPE) with respect to the currents, 
\begin{equation}\label{prim2}
  J^a(z)\,\Phi^j(u|w)\, = \, \frac{1}{z-w}      
  \, \CD_{j,u}^a \Phi^j(u|w)\ \ , \ \  
\bJ^a(\bz)\, \Phi^j(u|w)\, =\, \frac{1}{\bz-\bw}\bar{\CD}_{j,u}^a 
\Phi^j(u|w)\ ,
\end{equation}
where the differential operators $\CD_{j,u}^a$ are defined by
\begin{equation}
\CD_{j,u}^+\ =\ -u^2\pa_{u}+2ju \qquad \CD_{j,u}^0\ =\ -u\pa_{u}+j 
\qquad \CD_{j,u}^-\ =\ -\pa_{u}
\end{equation}
and the same expressions with $\bar u$ instead of $u$ are used to define 
$\bar{\CD}^{a}_{j,\bu}$. The fields $\Phi^j(u|z)$ are primary also 
w.r.t.\ the Sugawara Virasoro algebra with conformal dimensions 
\begin{equation}
\De_j\ =\ -\frac{1}{k-2}j(j+1) \ = \ - b^2 j(j+1)  \ \ . 
\end{equation}
In this expression and throughout most of our text, we parametrize
$k$ through $b^2\equiv (k-2)^{-1}$. 

Semi-classically one may think of 
the fields $\Phi^j(u|z)$ as being related to the functions $\Phi^j(u|h)$
that were defined
in eq.\ (\ref{phief})
by identifying $h$ with the field $h(z)$ that appears in the 
action of the $H_3^+$ WZNW model,
\[
 \Phi^j(u|z)\;=\;\Phi^j(u|h(z))\;\, .
\]
In terms of the coordinates
$(\phi,\c,\bc)$ this amounts to
\begin{equation}\label{classfield}
\Phi^j(u|z)\ =\ \frac{2j+1}{\pi}\Bigl((\ga(z)-u)({\bar \ga(z)}-\bar{u})
 e^{\phi(z)}+e^{-\phi(z)}
\Bigr)^{2j}\ \ .
\end{equation}
\medskip

\paragraph{\it Normalization.} \label{bulk_norm} A useful way to fix the 
normalization of these primary fields is to specify their asymptotic 
behavior near the boundary of $H_3^+$ \cite{Te2}. 
The fields $\Phi^j(u,z)$ are 
normalized such that    
\begin{equation}\label{opasym}
\Phi^j(u|z) \;\sim \;\;:e^{2(-j-1)\phi(z)}:\de^2(\ga(z)-u)
 + B(j) :e^{2j\phi(z)}: |\ga(z)-u|^{4j}\ \ .
\end{equation}
This should be compared with the asymptotic behavior (\ref{phiasym}) 
of the functions $\Phi^j(u|h)$. The only difference is that the 
the coefficient function $B(j)$ which appears for the fields 
$\Phi^j(u|z)$ is now given by\footnote{The following expression 
differs by a factor of $\pi$ in the expression for $\nu_b$
from the one given in \cite{Te2}. This means that the primary fields
denoted $\Phi^j(x|z)$ in \cite{Te2} differ by a factor $\pi^j$ from 
our primary fields $\Phi^j(u|z)$.} 
\begin{equation}\label{Bcoeff}
 B(j)\ = \ -\nu_b^{2j+1}\frac{2j+1}{\pi}
\frac{\Ga(1+b^2(2j+1)\bigr)}{\Ga(1-b^2(2j+1)\bigr)}, 
\quad \nu_b \ =\ \frac{\Ga(1-b^2)}{\Ga(1+b^2)} \ \ . 
\end{equation}
This expression reduces to the corresponding coefficient in rel.\ 
(\ref{phiasym}) in the limit $b \rightarrow 0$. Hence, the asymptotic 
behavior (\ref{opasym}) is a deformation of rel.\ (\ref{phiasym}) 
which includes stringy effects at finite curvature of the background.   

Although the normalization fixed by rel.\ \rf{opasym} is the most 
natural one from the point of view of string theory on $H_3^+$ (cf.\ 
\cite{Te2}), we find another set of fields more convenient from the 
mathematical point of view. The new set is introduced by 
$$ \Theta^j(u|z)\ \equiv\  B^{-1}(j)\ \Phi^j(u|z)\ \ . $$ 
This of course amounts to setting the prefactor of the second term 
in rel.\ \rf{opasym} equal to one, 
\begin{equation}\label{Thetaasym}
\Theta^j(u|z) \;\sim \;\; :e^{2j\phi(z)}:
|\ga(z)-u|^{4j}+ B^{-1}(j) :e^{2(-j-1)\phi(z)}:\de^2(\ga(z)-u)\ \ .
\end{equation}
The two-point function of the fields $\Theta^j(u|z)$ on the complex
plane is then given by an expression of the following form 
\begin{equation}\label{twopt}\begin{aligned}
{}  \bigl\bra \Theta^{-j_2-1}& (u_2|z_2) 
\Theta^{j_1}(u_1|z_1)\bigr\ket\,|z_2-z_1|^{4\De_{j_1}}
 = \\[1mm] 
& \hspace*{-1cm} =\;\frac{\pi^3}{4P^2_1}\de(P_2-P_1)\,
\de^{(2)}(u_2-u_1)+\frac{\pi\de(P_2+P_1)}{B\bigl(-\fr{1}{2}-iP_1\bigr)}
|u_2-u_1|^{4j_1}\ \ ,
\end{aligned}
\end{equation}
where $j_i=-\frac{1}{2}+iP_i$, $P_i\in\BR^+$ 
for $i=1,2$. In our analysis we will 
mainly work with the fields $\Theta^j$, but all of our results 
will finally be rewritten in terms of the fields $\Phi^j$ above. The 
translation between the different normalizations is straightforward. 
\medskip

\paragraph{\it Reflection property.} Except from a simple factor
$(2j+1)/\pi$, one can identify the coefficient $B(j)$ with a 
reflection amplitude $R(j)$ for closed strings on $\H3p$.    
Having fixed the normalization of operators by 
\rf{Thetaasym} one may re-express
the reflection of closed strings as a 
linear relation between the operators $\Theta^j(u|w)$ and 
$\Theta^{-j-1}(u|w)$, 
\begin{eqnarray}\label{refrel}
\Theta^{j}(u|z) & = &  -R(-j-1)(\CI_j\Theta^{-j-1})(u'|z)\ \ , 
\\[2mm]
\mbox{ where } \label{reflampl} & & R(j)\ =\ -\nu_b^{2j+1}
\frac{\Ga(1+b^2(2j+1)\bigr)}{\Ga(1-b^2(2j+1)\bigr)}\ \ ,
\end{eqnarray}
and $\CI_j$ is the intertwining operator that establishes
the equivalence of the $\SLC$-representations $P_{-j-1}$ 
and $P_j$,
\begin{equation} \label{CIint} 
(\CI_j\Theta^{-j-1})(u|z)\ =\ \frac{2j+1}{\pi}\int_{\BC}d^2u'\; 
|u-u'|^{4j} \Theta^{-j-1}(u'|z)\ \ .
\end{equation}
The operator $\CI_j$ is normalized such that $\CI_{-j-1}\circ 
\CI_j=\Id$. This normalization ensures its unitarity for $j
\in -\frac{1}{2}+ i\BR$. 

\subsection{Constraints from the gluing condition}

To begin with, let us now analyse the constraints on the form of 
the one-point function that arise from the gluing condition 
(\ref{gluecond}). As usual, the $z$-dependence of the one-point 
functions can be determined from the behavior under conformal 
transformations of the world-sheet theory. This gives  
$$ \bra \Theta^{j}(u|z) \ket_\raa^{}\ = \ \frac{\CA_u(j|\raa)}
{|z-\bz|^{2\De_j}}\ \ . $$  
The dependence w.r.t.\ the variable $u$ is likewise restricted by 
Ward identities associated with the currents $J^a, \bar J^a$. They 
can be derived using the operator product expansions of the chiral 
currents with the primary bulk fields. The resulting differential 
equations for $\CA_u(j|\raa)$ are of the form 
\begin{equation}\label{gluediff}
(\CD^{a}_{j,u}-\CD^{a}_{j,\bu})\CA_u(j|\raa)\;=\;0\ \ .
\end{equation} 
These equations are locally solved by $|u+\bu|^{2j}$. However, 
$u+\bu=0$ is a singular point so that there are two linearly 
independent distributional solutions of eq.\ \rf{gluediff}. Hence, 
we have to consider the two solutions $|u+\bu|^{2j}$ and 
$|u+\bu|^{2j} \sgn(u+\bu)$. This means that the gluing conditions 
\rf{gluediff} restrict the one-point function to be of the form
\begin{equation}\label{oneptansatz}
\bra \Theta^{j}(u|z) \ket_\raa^{} \;=\; \frac{|u+\bu|^{2j}A_\si(j|\raa)}
{|z-\bz|^{2\De_j}}\ \ , 
\end{equation}
where $A_\si(j|\raa)$ still depends on the variable $u$ through the 
function $\si\equiv \sgn(u+\bu)$. 

Let us furthermore note that additional restrictions arise
from the reflection property \rf{refrel} of the bulk primary 
fields. In the Appendix \ref{distr_d} we prove the identity
\begin{equation}\label{distid}\begin{aligned}
\frac{2j+1}{\pi}\int_\BC d^2u \;|u+\bu|^{-2j-2}\sgn^{\ep}(u+\bu) & 
\;|u-\ga|^{4j}=
\\ 
& =\; -(-)^{\ep}|\ga+\bar{\ga}|^{2j}\sgn^{\ep}(\ga+\bar{\ga})\ \ .
\end{aligned}\end{equation}
It implies a nice reflection property for $A_{\si}(j|\raa)$. The latter
is most easily expressed in terms of the coefficients $A^{\ep}(j|\raa)$ 
which appear in the expansion  
$$ A_{\si}(j|\raa)\ \equiv \ A^0(j|\raa) + \si A^1(j|\raa)$$ 
with respect to $\sigma$ (note that $\si^2$ = 1). With the help 
of the property (\ref{refrel}), the coefficients $A^i$ are easily 
shown to satisfy
\begin{equation}\label{refpropA}
A^{\ep}(j|\raa)\;=\;(-)^{\ep}\,R(-j-1)\,A^{\ep}(-j-1|\raa)\ \ .
\end{equation}
The formula (\ref{oneptansatz}) along with the reflection 
property (\ref{refpropA}) encode all the information one 
can extract without considering further constraints.

\subsection{Constraints from two-point functions with degenerate fields}

\paragraph{\it Introductory remarks.} 
As we have mentioned before, the simplest factorization constraints 
are obtained from the two-point functions of the theory. 
This correlation function can be factorized in two different ways:
If one imagines the two bulk fields close to each other it is most natural
to use the bulk operator product expansion to get a factorization 
in the closed string channel, leading to a representation of the 
two-point function as sum over one-point functions. The 
configuration where the two fields are far from each other is 
projectively equivalent to the situation where the fields are 
close to the boundary. In the latter case it is more natural to 
factorize in the open string channel by writing the bulk fields 
as sum over fields localized on the boundary. This yields an 
expression which is bilinear in the corresponding bulk-boundary 
operator product coefficients.

In rational conformal field theories one can exploit the equivalence 
between these two ways of factorizing the two-point function by 
concentrating on the contribution of the identity boundary field
in the open string channel. One thereby gets a powerful quadratic
equation for the one-point functions. In non-compact models, 
however, it is usually not possible to mimic this strategy 
since the identity may not appear in the open string channel 
at all.   

Fortunately, there exists a 
way out. In fact, the fields $\Theta^j$ we have considered so 
far are not the only ones in the theory. They are the fields that 
are in one-to-one correspondence with the normalizable states of 
the model. By analytic continuation in $j$, however, we obtain 
additional fields which are still perfectly well defined even 
though they do not correspond to any normalizable state. For 
certain discrete values of $j$, the new fields are associated 
with degenerate representations of the current algebra. This 
implies that the operator product of these degenerate fields with 
any other field of the theory contains only finitely many blocks
and that the factorization in the open string channel 
includes a contribution from the identity boundary field.

The fact that analytic continuation in $j$ allows to recover 
the degenerate fields is not a priori obvious, though. The power
of the results obtained by {\it assuming} that this is the case, 
cf.\ e.g.\ \cite{Te1,FZZ}, illustrates that one should consider 
it as a rather profound property of the theory. Therefore we 
would like to emphasize that this assumption can now be rigorously 
justified with the help of the results in \cite{Te2,Te3}. 

Here we shall only consider the simplest of the degenerate 
fields, $\Theta^{1/2}$, and study the following two-point 
functions
\begin{equation}\label{twoptdef}
\CG^j_{\raa}\bigl(
\begin{smallmatrix} u_2 & u_1\\ z_2 & z_1 \end{smallmatrix}
\bigr)\;\equiv\;
\bra \Theta^{\frac{1}{2}}(u_2|z_2)\Theta^j(u_1|z_1)\ket_\raa^{}\, .
\end{equation}
The special feature of the degenerate field  $\Theta^{\frac{1}{2}}$ 
is that it satisfies the following differential equations 
\begin{equation} \label{nvd} 
\pa_u^2\Theta^{\frac{1}{2}}(u|z) \ =\ 0  \ \ \ , \ \ \ 
  \pa_{\bu}^2\Theta^{\frac{1}{2}}(u|z) \ = \ 0 \ \ . 
\end{equation}  
They become obvious when we identify $\Theta^{1/2}$   
with the fundamental matrix-valued $h(z)$ through the familiar 
relation  
\begin{equation} \label{1/2fld}   
 \Theta^{\frac{1}{2}}(u|z)\;=\;(-u,1)\cdot h(z)\cdot\left(\begin{matrix}
-\bu\\ 1 \end{matrix}\right)\ \ . 
\end{equation}  
Indeed, the expression is linear in both $u$ and $\bar u$ so that 
the second derivatives $\pl^2_u$ and $\pl^2_{\bar u}$ vanish.   
\medskip
 
\paragraph{\it Differential equations.}
The form of the two-point function \rf{twoptdef} is strongly
constrained by various differential equations which we  now want 
to discuss. 

To begin with, there are six differential equations that arise 
from the symmetry of the theory under the {\sl two actions of $\SLR$}
on the world-sheet coordinates $z_i$ and the parameters $u_i$, 
respectively. The first is generated by the Virasoro modes $L_n, 
|n| \leq 1,$ while the second is associated with the zero modes 
of the currents. The resulting equations imply that the two-point 
function can {\it locally} be represented as  
\begin{equation}\label{SL2symm}\begin{aligned}
\CG^j_{\raa}\bigl(
\begin{smallmatrix} u_2 & u_1\\ z_2 & z_1 \end{smallmatrix}
\bigr)\;=\; \frac{
|z_1-\bz_1|^{2(\De-\De_j)}}{
|z_1-\bz_2 |^{4\De}}  
|u_1+\bu_1|^{2j-1}|u_1+\bu_2|^2 G^j_\raa(u|z)\ \ .
\end{aligned}\end{equation}
Here we have abbreviated $\De\equiv\De_{\frac{1}{2}}$ and 
introduced the cross-ratios $u$ and $z$ as
\begin{equation}
z\;=\;\Bigl|\frac{z_2-z_1}{z_2-\bz_1}\Bigr|^2\ \ ,\qquad
u\;=\;\Bigr|\frac{u_2-u_1}{u_2+\bu_1}\Bigr|^2\ \ .
\end{equation}
Next we are going to exploit the {\sl null vector decoupling}
equations (\ref{nvd}). They were motivated above and express 
the decoupling of the null vector in the Verma module of spin 
$\frac{1}{2}$ of the $\fsl_2$ algebra. From these equations 
we easily conclude that $G^j$ can be expanded in the form  
$$ G^j_\raa(u|z)\ =\ G^j_{\raa,0}(z) +u G^j_{\raa,1}(z) \ \ \ . $$
Finally, we have to take the Knizhnik-Zamolodchikov equations 
into account which can be derived with the help of the Sugawara 
construction. With our choice of the gluing condition (\ref{gluecond}) 
they read 
\begin{eqnarray} 
t z(z-1) \pa_z G^j_\raa(u|z) & = & \bigl[\,  u(u-1)(u-z)\pa_u^2 
         -(2ju(z-1)+(u+z)(u-1))\pa_u \nonumber \\[2mm] 
          & & + u-\fr{z}{2}+j(z-1)\, \big]\,  G^j_\raa(u|z)\ \  .
\label{kzred1} \end{eqnarray}
The null vector decoupling equations furthermore imply that \rf{kzred1} 
reduces to a $2\ti 2$ matrix equation, and that it therefore has a 
two-dimensional space of solutions. Two canonical bases $\CF_{\ep}^{\rm
s}$ and $\CF_{\ep}^{\rm t}$, $\ep=\pm$, for the space of solutions to 
\rf{kzred1} are introduced in Appendix C. 

It is important to note that in our particular case the 
Knizhnik-Zamolodchikov equations are nonsingular for $z\in 
(0,1)$ and $u\in(0,\infty)$.\footnote{This is not true for 
generic values of $j_1,...,j_4$ where one has a singularity at 
$u=z$. This plays an important role e.g.\ in \cite{MaOo3}.} As 
a consequence, the two-point function \rf{twoptdef} can be 
specified uniquely through its asymptotic behavior for $z\ra 0$, 
followed by $u\ra 0$. These asymptotics are what we will  
determine next.  
\medskip

\paragraph{\it Asymptotics $z_2\ra z_1$.} The decomposition of 
$\CG^j_{\raa}$ into conformal blocks can be obtained with the help 
of the operator product expansion 
\begin{equation}\begin{aligned}
\Theta^{\frac{1}{2}}(u_2|z_2)\Theta^j(u_1|z_1)\;\underset{z_2\ra z_1}{\sim}\;
\sum_{\ep=\pm}
\;&|z_2  -z_1|^{2(\De_{j+\frac{\ep}{2}}-\De_j^{}-\De)}
|u_1-u_2|^{1-\ep}\ti\\[1mm]
& \hspace*{-2cm} \ti C_\ep(j)\bigl(\Theta^{j+\frac{\ep}{2}}(u_1|z_1)+
\CO(z_2-z_1)+ \CO(u_2-u_1)\bigr)\ \ .
\end{aligned}\end{equation}
To be specific, let us spell out the explicit expressions for the 
operator product coefficients $C_\ep(j)$, $\ep=\pm$ that appear on the 
right hand side (see \cite{Te1,Te2})
\begin{equation}
C_+(j)\ \equiv\  1\ \ ,\qquad C_-(j)\ =\ \frac{1}{\nu_b}
\frac{\Ga(-b^2(2j+1))\Ga(1+2b^2j)}{\Ga(1+b^2(2j+1))\Ga(-2b^2j)}\ \ .
\end{equation} 
In this way, we have completely determined the expression for 
$\CG^j_{\raa}$ in terms of the coefficients $A_\si(j|\raa)$, 
\begin{equation}\label{CGdecomp}\begin{aligned}
\CG^j_{\raa}\bigl(
\begin{smallmatrix} u_2 & u_1\\ z_2 & z_1 \end{smallmatrix}
\bigr)\;=\; 
|z_1-\bz_1|^{2(\De-\De_j)}
|z_1-\bz_2 & |^{-4\De}  
|u_1+\bu_1|^{2j-1}|u_1+\bu_2|^2\ti\\
& \ti\;  \sum_{s=\pm}\;C_\ep(j)\CF_{\ep}^{\rm s}
(u|z)A_{\si_1}(j+\fr{\ep}{2}|\raa)\ \ .
\end{aligned}\end{equation}
Here, $\CF_\ep^{\rm s}$ is one of 
the bases in the space of solutions 
of eqs.\ \rf{kzred1} that we have mentioned 
before (see Appendix C for details).  
\medskip

\paragraph{\it Asymptotics for $\Im z_2 \ra 0$.} The field 
$\Theta^{\frac{1}{2}}(u|z)$ becomes singular for $\Im z\ra 0$. It 
is possible to describe this singular behavior through an expansion 
into boundary fields. This expansion is restricted by null vector 
decoupling and Knizhnik-Zamolodchikov equations and therefore has
the form  
\begin{equation}\label{blk_bd}\begin{aligned}
\Theta^{\frac{1}{2}}(u|z)\;\underset{\Im{z}\downarrow 0}{\sim}\;& 
\;A(\fr{1}{2},0|\raa)\;(\Im z)^{\frac{3}{2}b^2}\,(u+\bu)\,\id\\[1mm]
+& 
\;A(\fr{1}{2},1|\raa)\;(\Im z)^{-\frac{1}{2}b^2}
\bigl(\Psi_{-}(x)-\fr{1}{2}(u-\bu)
\Psi_{0}(x)-u\bu \Psi_{+}(x)\bigr)\ \ .
\end{aligned}\end{equation}
The three fields $\Psi_m = \Psi^1_m$ are boundary fields which are 
associated with the degenerate spin $j=1$ representation of $\SLR$.
  
$A(\frac{1}{2},0|\raa)$ is a natural device to parametrize the boundary 
conditions in the quantum theory as one can see from the following 
short computation \footnote{It should be possible to calculate 
$A(\frac{1}{2},0|\raa)$ by means of a free field calculation similar 
to what was done in the case of Liouville theory \cite{FZZ}.}
\begin{equation}\label{trbd} \Tr(\om_0 h(z))
\ = \ \, [(\pa_u+\pa_{\bu}) \Theta^{\frac{1}{2}} (u|z)]^{}_{u=0} 
\; \underset{\Im{z}\downarrow 0}{\sim} (2\Im z)^{\frac{3}{2}b^2} 
\;A(\fr{1}{2},0|\raa)
\end{equation}
where we have inserted eq.\ (\ref{1/2fld}) in the first step. 
Equation \rf{trbd} can be regarded as a natural quantum counterpart 
of the equation $\Tr(\om_0 h)=2\sinh \raa$ that defines the boundary 
condition in the classical theory. This motivates us to focus on the 
term proportional to the identity in \rf{blk_bd}. Formally one may 
project out the second term in \rf{blk_bd} by considering
\[ \begin{aligned}
\CP\CG^j_{\raa}\bigl(
\begin{smallmatrix} u_2 & u_1\\ z_2 & z_1 \end{smallmatrix}
\bigr)& \;\equiv\;\fr{1}{2}(u_2+\bu_2)\bigl[(\pa_{u_2}+\pa_{\bu_2})
\CG^j_{\raa}\bigl( \begin{smallmatrix} u_2 & u_1\\ z_2 & z_1 
\end{smallmatrix}\bigr) \bigr]_{u_2=0}^{}\ \ , \\[2mm]
& \;\equiv\; \fr{1}{2}
(u_2+\bu_2)\;\bigl\bra \Tr(\om_0
h(z_2))\Theta^j(u_1|z_1)\bigr\ket_\raa^{}\ \ . 
\end{aligned} \]
The bulk-boundary expansion \rf{blk_bd} then implies that 
the leading asymptotics of $\CP\CG^j_{\raa}$ for $\Im z_2\ra 0$ 
is given by
\begin{equation}\label{bdasym}\begin{aligned}
\CP\CG^j_{\raa}\bigl(
\begin{smallmatrix} u_2 & u_1\\ z_2 & z_1 \end{smallmatrix}
\bigr)\;\underset{\Im z_2\ra 0}{\sim}\; & 
|z_2-\bz_2  |^{\frac{3}{2}b^2}
(u_2+\bu_2)\; A(\fr{1}{2},0|\raa)\;\bra \Theta^j(u_1|z_1)\ket_\raa^{}.
\end{aligned}\end{equation}
\begin{rem}\label{factor_rem}
In rational conformal field theories one can conclude from 
\rf{blk_bd} that $A(\fr{1}{2},0|\raa)$ is proportional to the 
one-point function of the operator $\Theta^{\frac{1}{2}}(u|z)$
and this results in a stronger version of the 
factorization constraint. In  non-rational theories, however, 
one can not expect to find a {\it simple} relation between
$A(\fr{1}{2},0|\raa)$ 
and $\bra\Theta^{\frac{1}{2}}(u|z)\ket_\raa$, as was first 
observed for Liouville theory by Fateev et al.\ \cite{FZZ}, 
cf.\ also our introductory remarks in this Subsection. 
\end{rem}
\medskip

\paragraph{\it Comparison of the asymptotics for $\Im z_2 \ra 0$.}
Finally, we can combine all the information we have collected and 
derive the factorization constraint we were looking for. We achieve
this by comparing eq.\ \rf{bdasym} with the expression \rf{CGdecomp}
for the two-point function. The limit $\Im z_2 \ra 0$ implies $z\ra
1$ so that the asymptotic behavior of the conformal blocks is given 
in terms of the corresponding fusion coefficients $F_{st}(j)$ which 
can be found in Appendix C, 
\begin{equation}\label{bdasym1}\begin{aligned}
\CP\CG^j_{\raa}\bigl(
\begin{smallmatrix} u_2 & u_1\\ z_2 & z_1 \end{smallmatrix}
\bigr)\;\underset{\Im z_2\ra 0}{\sim}\; & 
|z_1-\bz_1|^{2(\De-\De_j)}
|z_1-\bz_2  |^{-4\De}  
|u_1+\bu_1|^{2j-1}|u_1+\bu_2|^2\ti\\[1mm]
& \hspace*{-2cm} \ti \biggl(\frac{4\Im z_2\Im z_1}{|z_2-\bz_1|^2}
  \biggr)^{-2\De} \frac{4\Re u_1\Re u_2}{|u_1+\bu_2|^2}
\sum_{\ep=\pm}\;F_{\ep-}\;C_\ep(j)\;A_{\si_1}(j+\fr{\ep}{2}|\raa)\ \ .
\end{aligned}\end{equation}
In order to simplify this result we introduce some new objects 
$E_{\si}(j|\raa)$ which are related to $A_{\si}(j|\raa)$ by 
\begin{equation} \label{EA} 
 A_{\si}(j|\raa)\ =\ \nu_b^{-j-\frac{1}{2}}\, \Ga(-b^2(2j+1))\, 
E_{\si}(j|\raa)\ \ . 
\end{equation} 
After inserting the explicit expressions for $F_{s-}$ and $C_s(j)$ 
we can now rewrite \rf{bdasym1} in the form  
\begin{equation}\label{bdasym2}\begin{aligned}
\CP & \CG^j_{\raa}\bigl(
\begin{smallmatrix} u_2 & u_1\\ z_2 & z_1 \end{smallmatrix}
\bigr)\;\underset{\Im z_2\ra 0}{\sim}\\[1mm]  
& \underset{\Im z_2\ra 0}{\sim}\; |z_1-\bz_1|^{-2\De_j}
|z_2-\bz_2  |^{-2\De}  
|u_1+\bu_1|^{2j}(u_2+\bu_2)\sgn(u_1+\bu_1) \ti\\[1mm]
& \qquad\ti \nu_b^{-j-1}\Ga(-b^2(2j+1))\frac{\Ga(-2b^2)}{\Ga(-b^2)}
\Bigl[E_{\si}\bigl(j+\fr{1}{2}|\raa)- 
E_{\si}\bigl(j-\fr{1}{2}|\raa)\Bigr]\ \  .
\end{aligned}\end{equation}
We conclude that \rf{bdasym} will hold provided that $E_\si(j|\raa)$ 
satisfies the following functional equation,
\begin{equation}\label{fc:funrel}
 E_{\si}\bigl(j+\fr{1}{2}|\raa\bigr)- 
E_{\si}\bigl(j-\fr{1}{2}|\raa\bigr)\;=\;\si\,
\sqrt{\nu_b}\,\frac{\Ga(-b^2)}{\Ga(-2b^2)}
A(\fr{1}{2},0|\raa)E_{\si}(j|\raa)\ \ \ . 
\end{equation}
This is the factorization constraint for the one-point function 
that we have been looking for. Its solution $E_\si(j|\raa)$ determines 
the coefficients $A_\si(j|\raa)$ of the one-point functions 
\rf{oneptansatz} through eq.\ \rf{EA}. 

\subsection{Solutions of the factorization constraint}

We shall propose the following expression as the relevant solution
of the factorization constraint
\begin{equation}\label{ads2onept}
A_{\si}(j|\raa)\;=\;A_b\,\nu_b^{-j-\frac{1}{2}}\,\Ga(-b^2(2j+1))\, 
e^{-\raa(2j+1)\si}\ \ 
\end{equation}
with some constant $A_b$ that is arbitrary for the moment. The 
parameter $\raa$ is related to $A(\fr{1}{2},0|\raa)$ by 
\begin{equation}\label{onehalf->bd}
A(\fr{1}{2},0|\raa)\;=\;-
\frac{1}{\sqrt{\nu_b}}\frac{\Ga(-2b^2)}{\Ga(-b^2)}\;
2\sinh \raa \ \ .
\end{equation}
This expression can easily be recognized as the most natural quantum 
``deformation'' of the corresponding classical expression that is 
compatible with the factorization constraint and the reflection 
property. If we insert this expression back into eq.\
(\ref{oneptansatz}) and take care of difference between the 
normalizations of $\Theta^j$ and $\Phi^j$, we finally obtain
\begin{equation}\label{res1}  
 \langle \Phi^j(u|z)\rangle_\raa \ = \  |u+\bu|^{2j} \, 
\nu_b^{j+\frac{1}{2}}\,\Ga(1+b^2(2j+1))\, e^{-\raa(2j+1)\si} 
\ \frac{A_b/\pi b^2}{|z-\bz|^{2\De_j}}\ \ .  
\end{equation}   
Up to some overall normalization, this agrees with our expectation 
(\ref{1pfexp1}) in the limit where $b$ is sent to zero.

We should note, however, that \rf{ads2onept} is {\it not} the most 
general solution to \rf{fc:funrel}. By using \cite[Appendix C]{PoTe} 
one may show that the most general solution of \rf{fc:funrel} can 
be written in the following form
\begin{equation}\label{gensol}
E_\si(j|\raa)\;=\; \bigl(e^{0}(j|\raa)+e^{1}(j|\raa)\si\bigr)
e^{-\raa (2j+1)\si}\ \ ,
\end{equation}
where the coefficients $e^{\ep}$ must be periodic in $j$ 
with period $1/2$, i.e.\ they have to satisfy $e^{\ep}(j+1/2|\raa)=
e^{\ep}(j|\raa)$. Since we also want the one-point function 
to obey the reflection property (\ref{refpropA}), we need to  
impose the condition $e^{\ep}(-j-1|\raa)=(-)^{\ep}e^{\ep}
(j|\raa)$. In particular, the latter excludes solutions with $e^1
(j|\raa)$ being constant in $j$.

The freedom that is left by the factorization constraint is 
therefore considerable. Under rather mild assumptions \cite{Sch}, 
however, it is possible to show that the periodicity requirement 
on $e^\ep(j|\raa)$ forbids a nontrivial {\it phase}. One should 
therefore supplement the factorization constraint with conditions 
that restrict the {\it absolute value} of $e^{\ep}(j|\raa)$. 
Such a condition will be provided by the analogue of the Cardy 
condition that we shall discussed in section 5.

\begin{rem}
When appealing to the result of \cite{Sch}, the most important assumption
one has to make is analyticity of $A_{\si}(j|\raa)$ in a strip of width
$1/2$ around the axis $j+\frac{1}{2}\in i\BR$. The necessity of such an 
assumption can be seen by considering a {\it three} point function
in which exactly one of the fields is $\Phi^{\frac{1}{2}}$. By associativity
of the OPE one can get two different representations as sum over
one-point functions, where the contours over which the variable $j$ 
is integrated will differ by shifts of $1/2$. In order to
relate these two representations one will have to shift contours,
which motivates our assumption concerning the analyticity of 
$A_{\si}(j|\raa)$.
\end{rem}

\subsection{The case of the spherical branes}

Let us briefly discuss the modifications of the previous analysis 
that are necessary to treat the spherical branes. The different 
gluing conditions (\ref{glue2}) force us to modify the ansatz 
\rf{oneptansatz} for the one-point function to 
\begin{equation}\label{oneptansatz'}
 \bra \Theta^{j}(u|z)\ket_\ras \;=\;\frac{(1+u\bu)^{2j}A'(j|\ras)}
{|z-\bz|^{2\De_j}}\ \ .
\end{equation}
Note that this time the function $1+ u \bar u$ has no singularities
and hence it dependence on $u$ is entirely fixed by the Ward
identities for currents. In the analysis of the factorization 
constraint one finds instead of \rf{CGdecomp} the expression
\begin{equation}\label{CGdecomp'}\begin{aligned}
\CG^j_{\ras}\bigl(
\begin{smallmatrix} u_2 & u_1\\ z_2 & z_1 \end{smallmatrix}
\bigr)\;=\; 
|z_1 -\bz_1|^{2(\De-\De_j)}
|z_1 &-\bz_2  |^{-4\De}  
(1+|u_1|^2)^{2j-1}|1+u_1\bu_2|^2\ti\\[1mm]
& \ti\;  \sum_{s=\pm}\;(-)^{\frac{1-s}{2}}
C_s(j)\CF_s(u|z)A'\bigl(r|j+\fr{s}{2}\bigr)\ \ ,
\end{aligned}\end{equation}
where $u$ and $1-u$ are now modified to 
\[ u=-\frac{|u_2-u_1|^2}{|1+u_2\bu_1|^2}\ ,\quad\text{and}\quad
1-u=\frac{(1+|u_1|^2)(1+|u_2|^2)}{|1+u_2\bu_1|^2}\ ,  \]  
respectively. Contrary to \rf{bdasym}, we shall now require that 
the two-point functions factorize into the product of two one-point 
functions when we take the arguments of the fields far apart, cf.\ 
remark \ref{factor_rem} below eq.\ \rf{bdasym}. If we introduce 
$E'(j|\ras)$ by 
$$A'(j|\ras)\ = \ \nu_b^{-j}\, \Ga(-b^2(2j+1))E'(j|\ras)\ \ , $$ 
the factorization constraint turns into the following functional 
relation for $E'(j|\ras)$:
\begin{equation}\label{fc:funrel'}
 E'\bigl(j+\fr{1}{2}|\ras\bigr)+ 
E'\bigl(j-\fr{1}{2}|\ras\bigr)\;=\;
\Ga(-b^2)
E'(\fr{1}{2}|\ras)E'(j|\ras) \ \ .
\end{equation}
The relevant solution turns out to be of the form
\begin{equation}
A_{\si}(j|\ras)\;=\;\frac{\Ga(-b^2(2j+1))}{2\,\nu_b^{j}\,\Ga(-b^2)}\,
\frac{\sin \ras(2j+1)}{\sin \ras}\ \ .
\end{equation}
The solution can be inserted back into the ansatz (\ref{oneptansatz'})
and gives the following one-point functions for the primary bulk
fields $\Phi^j$, 
\begin{equation} \label{res1a}  
 \langle \Phi^j(u|z) \rangle_{\ras} \ = \ (1+u \bar u)^{2j} 
\Ga(1+ b^2(2j+1)) \, \frac{\sin \ras (2j+1)}{\sin \ras} 
\frac{-\nu_b^{j+1}}{2\pi \Gamma(1-b^2)} \, \frac{1}{|z-\bz|^{2\De_j}}
\ \ .
\end{equation} 
Upon taking the semi-classical limit $b \rightarrow 0$, this almost 
reduces to the expression we anticipated in eq.\ (\ref{1pfexp2}) except 
that the one has to identify $\Lambda_0$ with $i\ras$. 
It appears that a consistent boundary conformal field theory
only exists for a discrete set of real values for $\ras$, see
also our remarks in subsection \ref{spherCardy}. This would imply 
that the parameter $\Lambda_0$ which controls the radius of the 
spherical branes is imaginary.  

\begin{rem}
The ansatz for the one-point function that was studied in \cite{GKS} 
differs from \rf{oneptansatz'} by replacing the factor $(1+u\bu)^{2j}$ 
with $(u-\bu)^{2j}$, where $(u-\bu)^{2j}$ is defined to be  $\exp (2\pi 
i j)|u-\bu|^{2j}$ whenever $\Im u < 0$. It turns out to be
impossible to satisfy the factorization constraint with this 
ansatz. Thereby we resolve a puzzle pointed out in \cite{GKS}: A
dependence like $(u-\bu)^{2j}$ would create a singularity on  
the boundary of $H_3^+$, which one would not expect to appear
in the case of spherical or instantonic branes. In fact, such 
branes are localized in the interior of $H_3^+$. The problem is 
clearly avoided with the correct $u$ dependence of the form 
$(1+u\bu)^{2j}$.
\end{rem}

\setcounter{equation}{0} 
\section{The open string sector}

The aim of this section will be to study the open string sector 
of the $\H3p$ model, i.e.\ the boundary operators and some simple 
correlations functions thereof. Our main goal is to determine the 
stringy corrections to the semi-classical reflection amplitude we
computed in Section 2. Since there is no such quantity for the 
spherical branes, we shall restrict all the discussion of this 
section to the boundary conditions that preserve an $SL(2,\BR)$ 
subgroup of the $SL(2,\BC)$ symmetry in the bulk, i.e.\ to the 
Euclidean $AdS_2$ brane. 

\subsection{The spectrum of boundary fields} \label{stripspec}

One can rewrite the Lagrangian (\ref{Saction}) for closed strings
on $\H3p$ in a first order formalism (see e.g.\ \cite{Te1}). 
In such a formulation, the interaction terms turn out to vanish 
near the boundary of $\H3p$ so that the theory becomes a free 
field theory in the asymptotic region. This remains true in the 
corresponding boundary problem. In fact, the currents that we 
use to write down the boundary conditions approach the currents
of the asymptotic free field theory and hence the boundary 
conditions become usual Dirichlet/Neumann boundary conditions
when $\phi$ tends to infinity. The boundary conditions therefore
do not introduce any interaction in the asymptotic model. 

Following previous experience from Liouville theory \cite{Te4} 
and the $\H3p$ model with periodic boundary conditions 
\cite{Te1,Te2} one therefore expects that states and the 
corresponding bulk and boundary fields in the $\H3p$ model
on the upper half plane can be characterized by their 
asymptotic behavior at the boundary. We have already used 
this fact for the bulk fields.%
\smallskip%

Near the boundary we may describe the theory in terms of free 
fields $\phi$, $\ga$, $\bar{\ga}$ and their canonical conjugate 
momenta. For gluing conditions for the $AdS_2$ branes imply 
usual Neumann boundary conditions for $\phi$ and $\nu = - \Im 
\ga$, i.e.\ only these two fields possess non-vanishing zero 
modes $\phi_0$ and $\nu_0$, respectively. For the third field
we impose Dirichlet boundary conditions that specify the 
`transverse position' $r$ of the brane. Canonical quantization
of the free field theory then leads to a space of states of the 
form
\begin{equation}\label{FFspst}
\CH^{\rm free}_{}\;=\;L^2(\BR\ti\BR;e^{\phi_0}d\phi_0d\nu)\ot\CF,
\end{equation}
where $\CF$ is a Fock-space that realizes the action of the 
non-zero modes of the three fields (cf.\ \cite{Te1}). As indicated 
in eq.\ \rf{FFspst} we assume the zero modes $\phi_0$ and $\nu$ to 
be realized as multiplication operators. Note that this space does
not depend on the Dirichlet parameter $r$.%
\smallskip%

We want to rewrite the space $\CH^{\rm free}_{}$ as a sum of 
sectors $\CR_j$ for the current algebra $\hfsl_2$. The latter 
is generated by the components $J^a_n, a=0,\pm,$ of the three 
currents in the boundary problem. To begin with, we consider 
the subspace 
$$\CH^{\rm free}_{0}\ \equiv \ 
L^2(\BR\ti\BR;d\phi_0e^{\phi_0}d\nu)\ot\Om\ \ ,$$  
where $\Om$ is the Fock vacuum of $\CF$ which is annihilated by 
the modes $J_n^a$ with $n>0$ of the current algebra. It is not 
difficult to see that the realization of the zero modes $J_0^a$ 
on $\CH^{\rm free}_{0}$ can be decomposed into standard principal 
series representations $\CP_j$ of the $\fsl(2,\BR)$ Lie algebra. 
In fact, the zero mode part in the Hamiltonian of the free theory 
is simply given by $H_0 \sim \hat P^2 - i \hat P$ where $\hat P = 
i \pl_{\phi_0}$ is the momentum conjugate to $\phi_0$.\footnote{
$H_0$ coincides with the quadratic Casimir (\ref{Casads2}) for 
large values of the radial coordinate $\chi$.} Hence, the Fourier 
transformation defined by the basis of plane waves $\exp(j\phi_0)$ 
with $j\in-\frac{1}{2}+i\BR$ provides a decomposition of 
$\CH^{\rm free}_{0}$ into a direct sum of representations 
$\CP_j$ from the principal continuous series of $\SLR$. Each 
representation $\CP_j$ of the zero mode algebra canonically 
extends to a so-called prolongation module $\CR_j$ which is 
generated by acting with the creation operators $ J_n^a$ with
$n<0$. In conclusion we have argued that 
\begin{equation}
\CH^{\rm free}_{}\;=\;\int_{\BR}dP\;\,\CR_{-\frac{1}{2}+iP}\ \ .
\end{equation}
Note that the integral extends over the full real line, i.e.\ 
$ P \in \QR$ runs through all the allowed eigenvalues of the 
momentum operator $\hat P$.%
\smallskip%

So far we have been talking about the spectrum of the free theory
which we obtained by dropping the interactions in the $H_3^+$ model.
The spectrum of the original theory can be embedded into $\CH^{\rm 
free}$ because each state $|\psi\ket$ of the interacting model 
behaves like a state for the free theory when we approach the 
boundary of $\H3p$. In other words, its asymptotic behavior 
assigns a unique element of $\CH^{\rm free}$ to each state 
$|\psi\ket$. In particular, primary states $|j;u\ket_{\raa\raa'}$ 
are represented asymptotically by wave-functions 
\begin{equation}\label{asstate}
\psi^j_{\raa\raa'}
(u|\ga)\;\equiv e^{-(j+1)\phi_0}\de(\ga-u)\;+\;S(j|\raa,\raa')
c^{-1}(j) e^{j\phi_0}|\ga-u|^{2j}\ \ ,
\end{equation}
where the normalizing factor $c(j)$ was defined in \rf{Icdef}
and we are thinking about a theory in which the open strings 
can end on two possibly different $AdS_2$ branes associated 
with the two labels $\raa$ and $\raa'$. 
Of course we do not expect the coefficient $S(j|\raa,\raa')$ 
of the outgoing plane wave to be arbitrary: As there is only 
one asymptotic region in $H_3^+$, the outgoing signal should 
be uniquely determined by the 
incoming signal. Therefore, the quantity $S(j|\raa,\raa')$ 
implicitly describes the (stringy) geometry in the interior 
of $H_3^+$. In particular, it is the first place where the
dependence on the boundary parameters $\raa$ and $\raa'$ 
enters. The relation between incoming and outgoing signals 
implies that the state space of the boundary $H_3^+$ contains
only ``half'' of the representations that we found in 
$\CH^{\rm free}_{}$, i.e.\ 
\begin{equation}
\CH_{\raa\raa'}^{\rm int} \; = \; \int_{\BS}^{\oplus}dj
\;\,\CR^{j}\ , \quad\BS=-\fr{1}{2}+i\BR^+_0\ \ .
\end{equation}
In other words, the integration over the momentum $P$ is now 
restricted to a half line.  
\medskip

\paragraph{\it Normalization.} 
With each of the normalizable states $|j;u\ket_{r_2r_1}^{}$ that 
we discussed above there comes a unique boundary field 
$\Psi^{j}_{r_2r_1}(u|x)$. The normalization of the state and 
the associated field is fixed by the asymptotic behavior 
\rf{asstate}. As in the case of the bulk primary fields, cf.\ 
Subsection \ref{bulk_norm}, we shall find it more convenient 
to work with a different set of boundary fields $\Xi^{j}$ which 
are related to the fields $\Psi^j$ by 
$$ \Xi^j_{r_2r_1}(u|x) \ \equiv \ S^{-1}(j|r_2,r_1)\, c(j)\, 
   \Psi^j_{r_2r_1}(u|x) \ \ . $$    
In comparison to (\ref{asstate}), the second term in an asymptotic 
expansion of $\Xi^j$ has a trivial coordinate independent pre-factor
in front of the second term. On the complex plane, the two-point 
function of the fields $\Xi$ is given by
\begin{equation}\label{norm_states}
\bigl\bra  \, \Xi^{-j_2-1}_{r_1r_2}(u_2|x_2)\,   
 \Xi^{j_1}_{r_2r_1}(u_1|x_1)\, 
\bigr\ket
\;=\;|x_2-x_1|^{-2\De_{j_1}}\; 2\pi\,\frac{\pi\de(P_2-P_1)}
{P_1\tanh\pi P_1}\,\de(u-u')\ \ ,
\end{equation}
with $j_i=-1/2 +iP_i$ and  $P_i\in\BR^+$ as usual. The 
factor on the right hand side of \rf{norm_states} is simply 
$|c(j)|^2$ and it may be identified as the inverse of the 
Plancherel measure for the representations $\CP_j$.

As in \cite{Te2} one can argue that the new choice of 
normalization fixes the leading short-distance behavior in 
the operator product expansions to be 
\begin{equation}\label{OPE-norm}
\Xi^{j_2}_{r_2r_2}(u_2|x_2)
\ \Xi^{j_1}_{r_2r_1}(u_1|x_1)\ \underset{x_2\ra x_1}{\sim} \ 
|x_2-x_1|^{-2b^2(j_1+1)(j_2+1)}\ \Xi^{j_2+j_1}_{r_2,r_1}(u_1|x_1)\ \ .
\end{equation}
Here we assume that $\Re(j_1+j_2+\frac{1}{2})>0$ and there is a 
similar expansion when we multiply the boundary fields with equal 
boundary labels $r$ from the right.  
\medskip

\paragraph{\it Reflection property.}
There is now also an open string version for the reflection
of signals in the interior of the Euclidean $AdS_2$ similar 
to the reflection of closed string modes in the interior of
of $\H3p$. In order to state this more precisely, we extend 
the definition of the states $|j;u\ket$ from the usual range 
of $j$ to $j\in -\fr{1}{2}+i\BR$. The states  $|-j-1;u\ket$ 
are certainly related to $|j;u\ket$ by some linear 
transformation involving $S(j|r_2,r_1)$. For the fields $\Xi^j$ 
such a relation implies  
\begin{equation}\label{refrel1}
\Xi^{j}_{r_2,r_1}(u|x)\;=\;R(j|r_2,r_1)(\CJ^j
\Xi^{-j-1}_{r_2,r_1})(u|x)\ \ \mbox{ with } \ \ 
R(j|r_2,r_1)\ \equiv \ S^{-1}(j|r_2,r_1)
\end{equation}
and $\CJ^j$ is the unitary $SL(2,\BR)$-intertwining operator 
that describes the equivalence of the representations 
$\CP_j$ and $\CP_{-j-1}$.

\subsection{Calculation of the reflection amplitude: Short description}

Sice the details of the analysis that we use to determine the
reflection amplitude are somewhat tedious, we shall now present 
a short summary of the methods we employ and of the the main 
results we shall find. Readers who are not interested in the 
technical details may then skip Subsections 4.3-4.6 in which 
we provide a more thorough discussion. 

In order to calculate $R(j|r_2,r_1)$ we shall make again use 
of a degenerate field much in the same way as in the previous 
section. But this time it is the degenerate boundary field  
$\Xi^1_r(u|x)$ corresponding to the finite dimensional 
representation with spin $j=1$ of the $\fsl(2,\BR)$ zero 
mode algebra. It can be constructed from the degenerate bulk 
field $\Theta^{1/2}$ by  
\begin{equation}
\Xi^1_r(u|\Re z)\;\equiv\; \lim_{\Im z\downarrow 0}
(\Im z)^{\frac{1}{2}b^2} \Theta^{\frac{1}{2}}(u|z) \ \ ,  
\end{equation}
where the limit is assumed to be taken in front of a boundary 
segment with boundary condition labeled by $r$. The properties
of the degenerate field $\Xi^1$ constrain its operator product
expansion with a generic boundary field to be of the form
\begin{equation}\label{1jOPE}
\begin{aligned}
\Xi^1_{r_2}(u_2|x_2)\, & \Xi^{j}_{r_2,r_1}(u_1|x_1)
\;\underset{x_2\ra x_1}{\sim}\;\\[1mm]
& 
\sum_{s=-,0,+}e_s(j|r_2,r_1)
(u_2-u_1)^{1-s}|x_2-x_1|^{\De_{j+s}-\De_j-\De_s}
\ \Xi^{j+s}_{r_2,r_1}(u_1|x_1).\end{aligned}
\end{equation}
A functional equation for $R(j|r_2,r_1)$ is then obtained by 
requiring consistency of the reflection relation \rf{refrel1} 
with the operator product expansion \rf{1jOPE}. This relation
reads  
\begin{equation}\label{funrel}
\frac{R\bigl(j+\fr{1}{2}|r_2,r_1\bigr)}{R
\bigl(j-\fr{1}{2}|r_2,r_1\bigr)}\;=\;
\frac{2j}{2j+1}\,
e_-\bigl(-j-\fr{1}{2}|r_2,r_1\bigr)\ \ .
\end{equation}
Given the coefficient $e_-$, it can be shown under mild assumptions
\cite{Sch} that there is at most one unitary solution to \rf{funrel}. 
The central task is therefore to construct $e_-$. This is done by 
evaluating part of the consistency conditions that were first 
formulated by Cardy and Lewellen \cite{CL}. The dependence on the 
boundary conditions enters the analysis through the particular
bulk-boundary coefficient $A(\frac{1}{2},0|r)$ which determines
the contribution of the identity field in the expansion of 
$\Theta^{1/2}(u|z)$ near the boundary, 
$$ (2 \Im z)^{-\frac{3}{2} b^2} \tr (\omega_0 h) \ 
   \underset{\Im{z}\downarrow 0}{\sim} \ A(\fr{1}{2},0|r) \ \ . $$    
We shall find that $e_-$ is uniquely fixed by the consistency 
conditions. Explicitly, it is given by  
\begin{equation}\label{em_form}\begin{aligned}
e_-^{}(j|r_2,r_1)\;=\;-
\frac{\la_b}{\pi^3} & 
\frac{\Ga(1+b^2(2j-1))\Ga(-b^2(2j+1))}{\sin\pi b^2 2j}\ti\\[1mm]
\ti & \prod_{s=\pm}
\cos\bigl(\pi b^2j+s\fr{i}{2}(r_2+r_1)\bigr)
\sin\bigl(\pi b^2j+s\fr{i}{2}(r_2-r_1)\bigr)\ \ .
\end{aligned}\end{equation}
The constant $\la_b$ shall be spelled out later (see eq.\ \rf{laexp}
below). The parameters $r_i$ we use to label the boundary conditions 
are related to the coefficients $A(\frac{1}{2},0|r)$ via 
\begin{equation}\label{bdparrel}
\sqrt{\la_b}\sinh r\;=\;-\pi
\frac{\Ga(-b^2)}{\Ga(-2b^2)}
\; A(\fr{1}{2},0|r)\ \ .
\end{equation}
Eq. \rf{bdparrel} describes the quantum corrections to the 
classical relation $A(\fr{1}{2},0|r)=2\sinh r$. So far we have 
not made any assumptions about the range of the parameters $r_i$ 
that appear in the expression \rf{em_form} for $e_-$, although
the semi-classical correspondence certainly suggests that the 
$r_i$ should be real. It is quite interesting to note, however, 
that the same conclusion can be reached without any input from  
the expected geometry. In fact, requiring the reflection amplitude
to be unitarity restricts the range for the $r_i$. More 
precisely, it follows from unitarity of $R$ that the l.h.s.\ of 
\rf{funrel} is an absolute square for $j\in\BS$. Therefore, the 
functional equation \rf{funrel} can only have unitary solutions if the 
object on the r.h.s.\ of \rf{funrel} is positive for all $j\in\BS$. 
This can be seen to require not only $r_i\in\BR$ but furthermore 
$r_1=r_2$. We shall comment on the second requirement later. 

The unique solution for $R(j|r)\equiv R(j|r,r)$ is then finally 
given by the expression
\begin{equation} \label{res2} 
R(j|r)\;=\;\biggl(\frac{\la_b}{4\pi^2}\biggr)^{iP} 
\frac{\Ga^2_k(b^{-2}-iP+\frac12)}{\Ga^2_k(b^{-2}+iP+\frac12)}
\frac{\Ga_k^{}(b^{-2}+2iP)}{\Ga_k(b^{-2}-2iP)}
\frac{S_k(2R+P)}{S_k(2R-P)}\ \ ,
\end{equation}
where $R\equiv r/ 2\pi b^2$, $j=-\frac{1}{2}+iP$, and the special function 
$S_k(x)$ and $\Gamma_k$ are defined through, 
\begin{eqnarray} \label{Sk}
\log S_k(x) & = & i\int\limits_{0}^{\infty}\frac{dt}{t}
\Biggl(\frac{\sin 2tb^2x}{2\sinh b^2t\sinh t}-\frac{x}{t}
\Biggr) \ \ , \\[2mm] 
\label{Gk} 
\Ga_k(x) & = & b^{b^2x(x-b^{-2})}(2\pi)^{\frac{x}{2}}
\Ga_2^{-1}(x|1,b^{-2}) \ \ .
\end{eqnarray}
Here, $\Ga_2(x|\omega_1,\omega_2)$ denotes Barnes Double Gamma 
function \cite{Bar}.     

\begin{rem}
Another outcome of the analysis seems to be worth noting: The 
operator product coefficients $e_s$, $s=-,0,+$ turn out to be 
related to fusion matrices of particular conformal blocks, 
similar to what was found in rational CFT (see 
\cite{Ru,BPPZ,AlReSc1,FFFS2}) and in Liouville theory \cite{PoTe}. 
We find that there exists a change of normalization of the 
boundary operators such that the re-normalized boundary 
fields have operator products coefficients 
\begin{equation}\label{E-Fus}
E_s(j|r_2,r_1)\;\equiv\;\Fus{j(r_2)\ }{j+s}
{1}{j} {j(r_2)}{j(r_1)}\ \ .
\end{equation}
instead of $e_s$. The spins $j=j(r)$ that one has to insert into 
the fusing matrix on the right hand side depend on the 
parameter $r$ which labels the boundary conditions,  
\begin{equation} \label{shift} 
j(r)\;=\;-\frac{1}{2}-\frac{1}{4b^2}+i\frac{r}{2\pi b^2} \ \ .
\end{equation}
We find it remarkable that the range of the representation 
labels $j$ one uses in this eq.\ (\ref{E-Fus}) for a real 
boundary parameter $r$ is not identical to the spectrum 
$\BS$. This represents a marked difference to previous 
cases where relations similar to \rf{E-Fus} were found.  
\end{rem}

\subsection{The basic bulk field} \label{htobd} 

To obtain the results we have sketched above, we consider a 
setup on the upper half plane in which the boundary condition
$r_1$ is imposed along the negative real line, while we impose 
a possibly different boundary conditions $r_2$ on the other 
side of the boundary. As in the previous section, we shall 
exploit the special properties of the primary bulk field 
$\Theta^{\frac{1}{2}}(x|z)$ but now in this more general 
situation where the boundary condition jumps at $x=0$. We 
find it useful to decompose this field into chiral vertex 
operators, 
\begin{equation}\label{hexp}\begin{aligned}
\Theta^{\frac{1}{2}}(x|z)\ =\  & V_+^{\frac{1}{2}}(x|z)
V_+^{\frac{1}{2}}(-\bx|\bz)\, a(\hat{\jmath}|r)+
V_+^{\frac{1}{2}}(x|z)
V_-^{\frac{1}{2}}(-\bx|\bz)\, b(\hat{\jmath}|r) \\[2mm]
+\  & V_-^{\frac{1}{2}}(x|z)
V_+^{\frac{1}{2}}(-\bx|\bz)\, c(\hat{\jmath}|r)+
V_-^{\frac{1}{2}}(x|z)
V_-^{\frac{1}{2}}(-\bx|\bz)\, d(\hat{\jmath}|r)\ \ .
\end{aligned}
\end{equation}
\newcommand{\hj}{\hat{\jmath}}
The operator $\hj$ is defined by $\hj\CP_j=j\CP_j$. 
Equation \rf{hexp} can be read as an expansion into a complete set of 
solutions to the Knizhnik-Zamolodchikov equations. It also takes the 
null vector decoupling equations (\ref{nvd}) and the gluing conditions 
(\ref{gluecond}) between left and right currents into account. The 
coefficients $a,b,c,d$ that are introduced through \rf{hexp} will now 
be determined by some of the consistency conditions that were 
formulated in \cite{CL}. To this end we study the behavior of the 
degenerate field with $j = 1/2$ near the boundary $\Im z=0$. 

As a preparation let us note that the chiral vertex operators 
$V_{s}^{\frac{1}{2}}(x|z)$, $s=+,-$ satisfy the following operator 
product expansion, 
\begin{equation}\label{CVOPE1}
\begin{aligned}
V_{r}^{\frac{1}{2}}(x'|z')\, 
V_{s}^{\frac{1}{2}}(x|z)\;=\;(z'-z)^{\De_1-2\De}
V_{r+s}^{1}\bigl(\fr{x+x'}{2}|\fr{z+z'}{2}\bigr)\ ,
\end{aligned}
\end{equation}
in the case where $r+s=\pm 1$, whereas for $r+s=0$ one finds an 
expansion of the form
\begin{eqnarray*}
V_{-s}^{\frac{1}{2}} (u_2 |z_2)\, 
V_{s}^{\frac{1}{2}}(u_1|z_1) & = &  f_{s,-}\;(z_2-z_1)^{-2\De}
 \;(u_2-u_1) \\[2mm]
  & & \hspace*{-4.5cm} + \; f_{s,+}\;(z_2-z_1)^{\De_1-2\De}\;
  \bigl(V^1_{0,-}(z_1)-\fr{1}{2}(u_2+u_1)
  V^1_{0,0}(z_1)+u_2u_1 V^1_{0,+} (z_1)\bigr)\ \ . 
\end{eqnarray*}
Here, the $f_{st}$, $s,t\in\{+,-\}$ are the special elements of the 
fusing matrix given by (see Appendix C.2 for explicit expressions) 
\begin{equation}
f_{st}(j)\;\equiv\;\Fus{j+\frac{s}{2},}{\frac{1}{2}+\frac{t}{2}}
{\frac{1}{2}}{\frac{1}{2}}
{j}{j}\ \ . 
\end{equation}

By means of these operator product expansions one indeed finds a 
singular behavior of the expected form 
\begin{equation}\label{blk-bd2}\begin{aligned}
\Theta^{\frac{1}{2}}(u|z)\;\underset{\Im{z}\downarrow 0}{\sim}\;& 
\;A(\fr{1}{2},0|r_\nu)\;(\Im z)^{\frac{3}{2}b^2}\,(u+\bu)\,\id\\
+& 
\;A(\fr{1}{2},1|r_\nu)\;(\Im z)^{-\frac{1}{2}b^2}
\bigl(\Xi_{-}(x)-\fr{1}{2}(u-\bu)\, 
\Xi_{0}(x)-u\bu \, \Xi_{+}(x)\bigr)\ \ 
\end{aligned}\end{equation}
where $\nu=1$ for $\Re z < 0$ and $\nu = 2$ otherwise. Let us note that 
we certainly are want the $\H3p$ model to be local. This implies 
that the coefficients $A(\fr{1}{2},i|r)$, $i=0,1$ depend only on 
the boundary condition $r$ assigned to the segment of the boundary 
that our bulk field approaches. 

Keeping this in mind, let us now determine $A(\fr{1}{2},0|r_\nu)$ 
from our general ansatz (\ref{hexp}) by sending the degenerate bulk 
field to the boundary once along the positive real axis and then 
again somewhere along the negative half line. This will provide
expressions for $A(\fr{1}{2},0|r_\nu)$ in terms of the $b$, $c$
which we can solve for the latter. If we let the bulk field 
approach the real line with $\Re z >0$ we obtain the first 
formula to be given by 
\begin{equation}
A(\fr{1}{2},0|r_2)\;=\;
e^{\frac{3}{4}\pi i b^2}
\bigl(
 f_{--}\,b\,+ f_{+-}\,c\,\bigr)\ \  .
\end{equation}
In order to treat the case where $\Re z <0$, it is useful to 
note that the degenerate chiral vertex operators with $j = 1/2$ 
satisfy the braid relation
\begin{equation}
V^{\frac{1}{2}}_{r}(x|e^{\pi i}z)\;=\;
e^{\pi i(\De_{j+\frac{r}{2}}-\De_j-\De)}
\, V^{\frac{1}{2}}_{r}(x|z)\ \ .
\end{equation}
With the help of this relation it is then obvious that the operator 
product coefficient $A(\fr{1}{2},0|r_1)$ must be of the form
\begin{equation}
A(\fr{1}{2},0|r_1)\;=\;
e^{\frac{3}{4}\pi i b^2} \bigl(
e^{-\pi i b^2(2j+2)} f_{--}\,b\,+\,
e^{+\pi i b^2 2j}f_{+-}\,c\, \bigr)\ \ .
\end{equation}
The two expressions for the coefficients of the bulk boundary 
expansion can now be solved for $b$, $c$, 
\begin{equation}\label{bc-sol}\begin{aligned}
b(j|r_1,r_2)\;=\;&+e^{\frac{1}{4}\pi i b^2}f_{--}^{-1}
\frac{A_1
e^{\pi i b^2 2j}- A_2}{2i\sin(\pi i b^2(2j+1))}\\[2mm]
c(j|r_1,r_2)\;=\;& -e^{\frac{1}{4}\pi i b^2}f_{+-}^{-1}\frac{A_1
e^{-\pi i b^2(2j+2)}- A_2}{2i\sin(\pi i b^2(2j+1))}\ \ . 
\end{aligned}\end{equation}
Here, we have use the shorthand notation $A_\nu\equiv 
A(\frac{1}{2},0|r_\nu)$ and explicit formulas for $f_{rs}$ 
are spelled out in Appendix C.2. 

\subsection{The basic boundary field}

Up to now we have focused on the contribution proportional to the
identity in the bulk-to-boundary OPE \rf{blk-bd2}. But there exists
one non-trivial boundary field in eq.\ \rf{blk-bd2} which is 
extracted from the expansion of the degenerate bulk field by 
\begin{equation}
\Xi^1_{r_\nu}(u|\Re z)\; \equiv\;  \lim_{\Im z\downarrow 0}
(\Im z)^{\frac{1}{2}b^2} \Theta^{\frac{1}{2}}(u|z)\ \ , 
\end{equation}
where the value of $\nu\in\{1,2\}$ depends on whether $z$ approaches
the left or right real half-axes.
The formula defines a boundary operator $\Xi^1_{r}(u|x)$ that transforms
in the degenerate $j=1$ representation of $\SLR$ and it possesses an 
expansion into chiral vertex operators $V^1_s(u|x)$, $s=-,0,+$ that 
takes the following form, 
\begin{equation}  \label{degbfld} 
\Xi^1_{r}(u|x)\;=\;\sum_{s=-1}^{+1}\;e_{s}(\hj|r_1,r_2)\,
V^1_s(u|x)\ \ .
\end{equation}
It is rather easy to express the coefficient $e_{0}(j|r_1,r_2)$ 
in terms of the coefficients $b,c$ which we determined in eqs.\  
 \rf{bc-sol} of the previous subsection, 
\begin{eqnarray}
e_0(j|r_1,r_2) & = &  e^{-\frac{\pi i}{4}b^2}\bigl( 
     b(j|r_1,r_2)f_{-+}+c(j|r_1,r_2)f_{++}\bigr) \nonumber\\[2mm] 
    \label{e0expr1} 
       & = & \frac{\Ga(-b^2)}{\Ga(-2b^2)}
\frac{\Ga(1+2b^2j)\Ga(-b^2(2j+2))}{2i\sin(\pi b^2(2j+1))}\ti\\[2mm]
& & \ti\Bigl(A_2\bigl(\sin\pi b^2(2j+2)+\sin\pi b^2 (2j)\bigr)-
A_1\sin\pi b^2(4j+2)\Bigr)\ \ . \nonumber
\end{eqnarray}
We have also inserted formulas for the elements of the fusing matrix 
from the Appendix C.2. Let us furthermore note that the normalization 
\rf{OPE-norm} implies $e_+\equiv 1$. It therefore remains to determine
$e_-$. We now want to show how $e_-$ can be computed from $e_0, e_1$
and elements of the fusing matrix, even though the procedure will be 
rather complicated at first.   

\begin{claim}\label{uniclaim}
The conditions of associativity of the operator product expansion
yield an equation that expresses $e_-(j|r_1,r_2)$ 
algebraically in terms of $e_s(j|r_1,r_2)$, $s=+,0$ and certain
fusion coefficients.
\end{claim}

To prove the claim, let us study the constraints from 
associativity of the boundary operator product expansion. It 
suffices to analyze the following special four-point functions 
in which only two fields are non-degenerate,  
\begin{eqnarray}\label{fourpt-1}
\bra \, \Xi^{-j-1}_{r_1,r_2}(u_4|\infty)\, \Xi^1_{r_2}(u_3|x)\, 
 \Xi^1_{r_2}(u_2|x')\,  \Xi^{j}_{r_2,r_1}(u_1|0)\, \ket & = & \\[2mm] 
& & \hspace*{-5cm}  = \ \bra j,u_4|\, \Xi^1(u_3|x)\, 
 \Xi^1(u_2|x')\, |j,u_1 \ket\ \  . 
\nonumber \end{eqnarray}
Since the boundary field $\Xi^1_r(u|x)$ is degenerate with $j=1$, it 
satisfies the null vector decoupling equation $\pa_u^3\, \Xi^1_r(u|x)=0$. 
This restricts its operator products with other boundary fields to 
have the form 
\begin{eqnarray}\label{1jOPE'}
\Xi^1_{r_2}(u_2|x_2)\, \Xi^{j}_{r_2,r_1}(u_1|x_1)
\; & \underset{x_2\ra x_1}{\sim}\; & \\[2mm]
& & \hspace*{-4cm} 
\sum_{s=-1}^{+1} \ e_s(j|r_1,r_2)
(u_2-u_1)^{1-s}|x_2-x_1|^{\De_{j+s}-\De_j-\De_s}\, 
\Xi^{j+s}_{r_2,r_1}(u_1|x_1)\ \ .\nonumber
\end{eqnarray}
In the case $j=1$ one finds the degenerate boundary fields with 
$j=0,1,2$ on the right hand side eq.\ (\ref{1jOPE'}) in which  
$\Xi^0$ is the identity field. If the product of two fields with 
$j=1$ is inserted into the correlation function \rf{fourpt-1} each of 
the three contributions to the operator product yields one 
equation. With $s=-1$ we obtain 
\begin{equation}\label{asso1}
e_+(j-1)e_-(j)F^1_{--} +(e_0(j))^2F^1_{0-}+e_-(j+1)e_+(j)
F^1_{+-}\;=\;
e_-(1)\ \ ,
\end{equation}
where we have used the following abbreviations   
\begin{equation}
e_s(j) \ \equiv \ e_s(j|r_1,r_2) \ \ \ \ , \ \ \ \ 
F_{st}^1\ \equiv\ F_{st}^1(j)\ \equiv\ \Fus{j+s\ }{1+t} {1}{1}
{j}{j}\ \ .
\end{equation}
Explicit expressions for the fusion coefficients $F_{st}^1(j)$ 
can be found in the Appendix C.2. From the case $s=0$ we infer
\begin{equation}\label{asso2}
e_+(j-1)e_-(j)F^1_{-0} +(e_0(j))^2F^1_{00}+e_-(j+1)e_+(j)F^1_{+0}\;=\;
e_0(1)e_0(j)\ \ .
\end{equation}
By imposing $e_+(j)\equiv 1$ and combining \rf{asso1} and \rf{asso2} 
we get an algebraic equation involving $e_-(j)$ and $e_-(1)$. When 
this is specialized to $j=1$ it yields an equation that expresses 
$e_-(1)$ algebraically in terms of $e_0(1)$ and the elements 
$F_{st}^1|_{j=1}$ of the fusing matrix. The expression for $e_-(1)$ 
can now be inserted into the equation for $e_-(j)$ and this then 
leads to a formula which determines $e_-(j)$ in terms of $e_0(j)$ 
and the fusion coefficients $F_{st}^1(j)$. This proves our 
Claim \ref{uniclaim}.

\subsection{Relation with fusion coefficients}

At the end of the last subsection we sketched a procedure that 
allows to determine the last coefficient $e_-(j)$ we are missing 
for our construction of the degenerate boundary field
(\ref{degbfld}). The recipe looks rather complicated but it 
turns out that there is a way of solving the conditions on 
$e_-(j)$ directly which is also conceptually very interesting. 
It is based on the observation \cite{Ru} that the conditions 
\rf{asso1}\rf{asso2} are satisfied if we insert the following 
quantity for $e_s(j)$, 
\newcommand{\te}{E}
\newcommand{\RN}{{\rm N}}
\begin{equation}
\te_s(j|\rho_2,\rho_1)\;\equiv\;\Fus{\rho_2,}{j+s}
{1}{j}
{\rho_2}{\rho_1}\ \ .
\end{equation}
Validity of eqs.\ \rf{asso1}\rf{asso2} with $e_s(j)$ replaced by 
$\te_s(j)$ is assured by the pentagon identity for the fusion 
coefficients \cite{MS}. Let us stress that the equations hold 
true for any choice of the parameters $\rho_2$, $\rho_1$. This
freedom is related to the choice of boundary conditions $r_i$ 
but the precise relation between $\rho_i$ and $r_i$ will turn 
out to be non-trivial, unlike in the rational case. 
While $\te_s(j)$ solve eqs.\ \rf{asso1}\rf{asso2}, it is not 
normalized in the same way as $e_s(j)$, i.e.\ in particular 
$\te_+(j) \neq 1$. To achieve proper normalization it is 
useful to observe that there is a whole family of solutions to  
eqs.\ \rf{asso1} \rf{asso2} which is generated by setting
\begin{equation}
\te_s^{\rm N}(j|\rho_2,\rho_1)\;\equiv\ 
\te_s(j|\rho_2,\rho_1)\frac{\RN(j|\rho_2,\rho_1)
\RN(1|\rho_2,\rho_2)}{\RN(j+s|\rho_2,\rho_1)}\ \ .
\end{equation}
As we shall show now, one can find functions $\RN(j,\rho_2,\rho_1)$
and $\rho_i = \rho_i(r_i)$ such that $\te_s^N$ is normalized in the 
same way as $e_s$. We can then use this $\te^N_s$ to construct the 
degenerate boundary field $\Xi^1$, i.e.\ we can set 
\begin{eqnarray}
e_-(j|r_2,r_1) & \equiv & 
\te_0^{\rm N}\bigl(j|\rho_2(r_2),\rho_1(r_1)\bigr)\ = \ 
\Fus{\rho_2,}{j-1} {1}{j} {\rho_2}{\rho_1}
\frac{\RN(j|\rho_2,\rho_1)\RN(1|\rho_2,\rho_2)}
{\RN(j-1|\rho_2,\rho_1)}
\nonumber  \\[2mm] 
& = &\Fus{\rho_2,}{j-1}{1}{j}{\rho_2}{\rho_1}
\Fus{\rho_2,}{j} {1}{j-1} {\rho_2}{\rho_1}
\RN^2(1|\rho_2,\rho_2)\ \ .
\end{eqnarray}
A precise formula for the proper choice of $\RN(j,\rho_2,\rho_1)$ 
is spelled out in the process of proving the following claim. 

\begin{claim}\label{gaugeclaim}
There exists a function $\RN(j|\rho_2,\rho_1)$ together with a
choice for the parameters $\rho_2=\rho_2(r_2)$, $\rho_1=
\rho_1(r_1)$ such that
\begin{align}
\te_0^{\rm N}\bigl(j|\rho_2(r_2),\rho_1(r_1)\bigr)\;= & \;
e_0\bigl(j|r_2,r_1\bigr)\label{gauge1}\\[2mm] 
\te_+^{\rm N}(j|\rho_2(r_2),\rho_1(r_1))\;= &\; 1 \ \ .
\label{gauge2}\end{align}
More precisely, these properties are ensured if we construct  
$\RN(j|\rho_2,\rho_1)$ by eq.\ (\ref{Neq}) and choose $\rho_i$ 
such that eqs.\ (\ref{bdparrel'}) are satisfied (see below).  
\end{claim}

In order to prove our claim, we start by solving equation
\rf{gauge2} which is equivalent to the functional equation
\begin{equation}\label{funeq:N}
1\;\equiv\;
\Fus{\rho_2,}{j+1}
{1}{j}
{\rho_2}{\rho_1}\ \frac{\RN(j|\rho_2,\rho_1)\RN(1|\rho_2,\rho_2)}
{\RN(j+1|\rho_2,\rho_1)}\ \ .
\end{equation}
It is not difficult to see that the functional equation \rf{funeq:N} 
can be solved in terms of a special function $\Ga_k(x)$ satisfying 
the functional equation $\Ga_k(x+1)=\Ga(b^2x)\Ga_k(x)$. With this 
functional equation and the explicit expression for the relevant 
fusion coefficient given in the Appendix C one can easily verify 
that the expression
\begin{equation}\begin{aligned} \label{Neq} 
\RN(j|\rho_2,\rho_1)\;=\; \la_b^{\frac{j}{2}} &\frac{
\Ga_k(2+2\rho_2+b^{-2})\Ga_k(-2\rho_2)}
{\Ga_k(j+\rho_2+\rho_1+2+b^{-2})\Ga_k(j-\rho_2-\rho_1)
} \ti \\[1mm] 
& \ti \frac{\Ga_k(1+b^{-2})\Ga_k(2j+1+b^{-2})}
{\Ga_k(j+\rho_2-\rho_1+1+b^{-2})\Ga_k(j+\rho_1-\rho_2+1+b^{-2})}
\end{aligned}\end{equation}
does the job for some positive constant $\la_b$. Now we can 
compute $\te_0^N$ using this choice of the function $\RN(j|
\rho_2,\rho_1)$,  
\begin{equation}\nonumber \label{e0expr2}\begin{aligned}
\te_0 & (j|\rho_2,\rho_1)\;=\; \Fus{\rho_2,}{j}
{1}{j}{\rho_2}{\rho_1}\RN(1|\rho_2,\rho_2) \;=\;
\frac{\sqrt{\la_b}}{2\pi} \frac{\Ga(1+2b^2j)\Ga(-b^2(2j+2))}
{\sin(\pi b^2(2j+1))} \times \\[1mm]  
& \hspace*{-8mm} 
\Bigl(\cos\pi b^2(2\rho_1+1) \bigl(\sin\pi b^2(2j+2)+\sin\pi b^2 
(2j)\bigr) -\cos\pi b^2(2\rho_2+1)\sin\pi b^2(4j+2)\Bigr)\ . 
\end{aligned}\end{equation}
Comparing the expression for $\te_0$ to the one for 
$e_0$ given in eq.\ \rf{e0expr1} it becomes obvious that the 
normalization \rf{gauge1} can be satisfied by a suitable choice 
of $\rho_2=\rho_2(r_2)$, $\rho_1=\rho_1(r_1)$, 
\begin{equation}\label{bdparrel'}
\frac{\sqrt{\la_b}}{2\pi}\cos\pi b^2(2\rho_i+1)\;=\;\frac{1}{2i}
\frac{\Ga(-b^2)}{\Ga(-2b^2)}
\; A(\fr{1}{2},0|r_i)\ \ .
\end{equation} 
This proves the claim. Let us note briefly that in this way we have 
obtained an explicit formula for the missing coefficient $e_-$,  
\begin{equation}\begin{aligned}
e_-^{}(j|r_2,r_1)\;=\;
\frac{\la_b}{\pi^3} & 
\frac{\Ga(1+b^2(2j-1))\Ga(-b^2(2j+1))}{\sin\pi b^2 2j}\ti\\
\ti & \prod_{s=\pm}
\sin\pi b^2(j+s(\rho_2+\rho_1+1))\sin\pi b^2(j+s(\rho_2-\rho_1))\;\, .
\end{aligned}\end{equation}
Along with the normalization $e_+ = 1$ and our formula 
(\ref{e0expr1}) this completely determines the boundary field $\Xi^1$. 

\begin{rem}
It is clear that eq.\ \rf{bdparrel'} determines the boundary parameters
$\rho_i$ only up to $\rho_i\ra\rho_i+b^{-2}$. We will find restrictions
that eliminate that ambiguity in the next subsection.
\end{rem}

\subsection{Reflection amplitude}

With an explicit formula for basic boundary field $\Xi^1$, we 
can now turn to the construction of the reflection amplitude 
$R(j|r_2,r_1)$. A functional equation for this quantity is 
obtained by requiring consistency of the reflection relation 
\rf{refrel1} with the OPE \rf{1jOPE}. We can either use 
\rf{refrel1} first followed by \rf{1jOPE} or do it the other 
way around and then compare the leading asymptotics for $x_2
\ra x_1$ followed by $u_2\ra u_1$ . This gives 
\begin{equation}\label{funrelrep}
\frac{R\bigl(j+\fr{1}{2}|\rho_2,\rho_1\bigr)}{R
\bigl(j-\fr{1}{2}|\rho_2,\rho_1\bigr)}\;=\;
\frac{2j}{2j+1}\,
e_-\bigl(-j-\fr{1}{2}|\rho_2,\rho_1\bigr)\ \ .
\end{equation}
By definition, the reflection amplitude $R(j|r_2,r_1)$ must 
also satisfy the simple condition 
$$R(j|r_2,r_1)R(-j-1|r_2,r_1)\ = \ 1\ \ . $$
If we finally take into account that $R(j|r_2,r_1)$ should 
be unitary, i.e.\ $|R(j|r_2,r_1)|=1$ for all $j\in\BS$, then
we arrive at important restrictions on the allowed values of 
$\rho_2$, $\rho_1$. In fact, the unitarity of $R(j|r_2,r_1)$ 
allows to rewrite eq.\ \rf{funrelrep} as 
\begin{equation}\label{funrel'}
|R\bigl(j+\fr{1}{2}|\rho_2,\rho_1\bigr)|^2\;=\;
\frac{2j}{2j+1}\,
e_-\bigl(-j-\fr{1}{2}|\rho_2,\rho_1\bigr)\ \ ,
\end{equation}
which implies that the right hand side of eq.\ \rf{funrel'} 
has to be positive for all $j\in\BS$. To analyze the resulting 
restrictions, we introduce $j=-\frac{1}{2}+iP$, $q_{\pm}=
-i(\rho_1+\frac{1}{2}\pm \rho_2+\frac{1}{2})$, and then 
re-write eq.\ \rf{funrel'} in the form
\begin{equation}\begin{aligned}
|R\bigl(j+\fr{1}{2}|\rho_2,\rho_1\bigr)|^2\;=\;
\frac{\la_b}{\pi^4}&
|\Ga(1-b^2+2ib^2P)\Ga(2ib^2P)|^2\ti\\[1mm]
 & \hspace*{-2cm} \ti (\cosh 2\pi b^2P-\cosh 2\pi b^2q_+)
(\cosh 2\pi b^2P-\cosh 2\pi b^2q_-)\ \ . 
\end{aligned}\end{equation}
The right hand side of this relation can only be positive if either 
\begin{itemize}
\item[i)] the $q_{\pm}$ are of the form 
$q_{\pm}=q'_{\pm}+ik_{\pm}/2b^2$, $q_\pm\in\BR$, $k_\pm\in 1+2\BZ$, or
\item[ii)] $q_{s}$ vanishes for one $s\in\{+,-\}$, 
$q_{-s}=q'_{-s}+ik_{-s}/2b^2$, $q'_{-s}\in\BR$ and $k_{-s}\in 1+2\BZ$.
\end{itemize}
Let us note that the first of these cases does not seem to be of 
physical interest. One would have parameters $k_i$ $i=1,2$ defined 
by $k_\pm=k_1\pm k_2$ labeling boundary conditions to the 
left and the right of $x=0$. However, $k_\pm\in 1+2\BZ$ implies 
that $k_1\neq k_2$. As we will be interested in the possibility 
to put the same boundary condition along the whole real line, 
we are forced to discard possibility i). 

Hence, we are left with possibility ii). Since the two sub-cases 
$q_{\pm}=0$ are very similar, we shall focus on the case $q_-=0$. 
This means that $\rho_i=-1/2- 1/4b^2+iR$ and the functional 
equation \rf{funrelrep} then takes the form
\begin{equation}\begin{aligned}
\frac{R\bigl(j+\fr{1}{2}|\rho,\rho\bigr)}{R
\bigl(j-\fr{1}{2}|\rho,\rho\bigr)}\;=\;
\frac{\la_b}{(2\pi)^2}& \biggl|
\frac{\Ga(1-b^2+2ib^2P)\Ga(1+2ib^2P)}{\Ga^2(1+ib^2P)}
\biggr|^2\\[1mm] 
&\qquad \ti \cosh b^2(2R+P)
\cosh b^2(2R-P)>0 \ \ .
\end{aligned}\end{equation}
A unitary solution to this equation can be constructed in terms
of the special functions $\Gamma_k$ and $S_k$ which were defined 
in eqs.\ (\ref{Gk},\ref{Sk}) above. These functions possess a 
number of nice properties which are relevant for us, 
\begin{equation}
\Ga_k(x+1)\ =\ \Ga(b^2x)\Ga_k(x)\ , \quad 
\frac{S_k\bigl(x-\frac{i}{2}\bigr)}{S_k\bigl(x+\frac{i}{2}\bigr)}
\ =\ 2\cosh\pi b^2 x\ ,
\quad |S_k(x)|^2\ =\ 1\ .
\end{equation}
They are used to verify that the following expression represents 
a unitary solution of (\ref{funrelrep})
\begin{equation}
R(j|\rho,\rho)\;=\;\biggl(\frac{\la_b}{4\pi^2}\biggr)^{iP} 
\frac{\Ga^2_k(b^{-2}-iP + \frac12)}{\Ga^2_k(b^{-2}+iP + \frac12)}
\frac{\Ga_k(b^{-2}+2iP)}{\Ga_k(b^{-2}-2iP)}
\frac{S_k(2R+P)}{S_k(2R-P)} \ \ .
\end{equation}
This concludes the construction of the unitary reflection 
amplitude $R(j|\rho,\rho)$ for the special choice $\rho = 
\rho_1 = \rho_2$. There is no physical unitary solution 
for $\rho_1 \neq \rho_2$.   

\begin{rem}
We have obtained two expressions for $A(\frac{1}{2},0|r)$, eq.\ 
\rf{onehalf->bd} and eq.\ \rf{bdparrel'}. Comparing these two 
expressions finally allows us to determine $\la_b$,  
\begin{equation}\label{laexp}
\la_b\;=\;(2\pi)^2
\frac{\Ga(1+b^2)}{\Ga(1-b^2)}\ \  .
\end{equation} 
\end{rem}

\setcounter{equation}{0} 
\section{Analogs of the Cardy condition}

At this point we have found the full string corrections for 
the semi-classical closed string couplings and the reflection 
amplitude. The latter determines the relative open string 
spectral density as a logarithmic derivative of the ratio 
of two reflection amplitudes (see Appendix B). 
In the point particle limit of our theory, there is no relation 
between the bulk couplings to the brane and the spectral 
density. But the situation is entirely different within 
string theory: here, the closed string couplings determine
the annulus amplitude which is related to the open string 
partition function by world-sheet duality. 
The partition function involves the 
open string spectral density. For rational boundary conformal 
field theory, the analysis of world-sheet duality is a crucial 
ingredient in obtaining exact solutions \cite{Car}. Similarly, 
world-sheet duality gives rise to an important consistency 
condition also in the case of non-rational models \cite{TB}. Our aim 
here is to explore this ``Cardy condition''.  

\subsection{Boundary state}

We shall now introduce boundary states for our $AdS_2$-branes.  
To this end, we re-interpret the one-point function of $\Theta^j$ 
as a linear form ${}_{\rm B}^{}\bra \raa|$ on the space of closed 
string states. More precisely, we define   
\begin{equation}
{}^{}_{\rm B}\bra \raa|j;u\ket\; := \;{}_{\rm B}^{}
\bra \Theta^{j}(u|\fr{i}{2})\ket_\raa \ \ . 
\end{equation} 
We have placed a subscript $B$ on the boundary state to distinguish 
it from the usual closed string states. For the following it will be 
useful to perform a Fourier-transformation over the variable $u$,
i.e.\ to evaluate the boundary state in a new basis of closed string 
states $|j;n,p\ket$,  
\begin{equation}
{}^{}_{\rm B}\bra \raa|j;n,p\ket\;=
\;\int_{\BC} d^2u \;e^{-in\arg(u)}|u|^{-2j-2-ip}
{}^{}_{\rm B}\bra \raa|j;u\ket \ \ .
\end{equation}
The integral can be performed (see subsection \ref{distr_d} in
Appendix A) and yields ${}^{}_{\rm B}\bra \raa|j;n,p\ket=
2\pi\de(p)A(j,n|\raa)$, where
\begin{equation}\begin{aligned}\label{bstads}
A(j,n|\raa)\;=\;2\pi & 
\, A_b \,\Ga(-b^2(2j+1))\; d^j_n \ti\\[1mm]
\ti & \bigl( \pi^0_n\cosh r(2j+1)-\pi^1_n\sinh r(2j+1)\bigr),
\end{aligned} 
\end{equation}
where $d^{j}_n$ and $\pi^{\ep}_n$, $\ep=0,1$ are defined by
\begin{equation} 
d^{j}_n\;=\;\frac{\Ga(2j+1)}{\Ga(1+j+\frac{n}{2})
\Ga(1+j-\frac{n}{2})}, \quad
\pi^{\ep}_n\;=\;\left\{ \begin{aligned}
1-\ep \;\;& {\rm if}\;\;n\;\;{\rm even}\\
\ep\;\;& {\rm if}\;\;n\;\;{\rm odd.}
\end{aligned}\right.
\end{equation}
Note that in the new basis, $n$ runs through integers and $p$ is a 
real number. 

\subsection{The annulus amplitude} 

Following Cardy \cite{Car}, we would like to consider the quantity
${}^{}_{\rm B}\bra \raa|\tilde{q}^{\frac{1}{2}\SH^{\rm cyl}}|\raa
\ket_{\rm B}^{}$, where $\tilde{q}=\exp(-2\pi i/\tau)$ and 
$\SH^{\rm cyl}$ is the Hamiltonian for the theory with periodic 
boundary conditions, i.e.\ $\SH^{\rm cyl}=\SL_0 +\bL_0-c/12$. In 
our case, this expression is ill-defined due to various 
divergencies. The strategy will be to first cut off all 
potentially divergent summations by introducing a regularized 
boundary state. After identifying similar divergencies in the 
partition function on the strip in the next subsection, we shall 
discuss the interpretation of these divergencies and the relations 
between the respective cut-offs in Subsection \ref{compcut-off}. 
This naturally leads to the identification of physically 
meaningful quantities that one can compare between the open 
and closed string channels.

The "regularized boundary state" we want to work with automatically 
removes the divergent contributions coming from the summations over 
$j=-\frac{1}{2}+iP$, $n$ and $p$. It is defined by 
\begin{eqnarray}\label{regbdst}
{}^{}_{\rm B,reg}\bra \raa|j;n,p\ket & := &  
\Theta(\de-P)\,\Theta_\lambda(n)\, 2\pi\de_T(p)\;
A(j,n|\raa) \ \ , \\[2mm] 
\mbox{where} \ \ \Theta_\lambda (n)  & = & \left\{ \begin{array}{ll} 
1 & \mbox{ for } \ n\in\{-\lambda+1,-\lambda+2,\dots,\lambda\}\\
0 & \mbox{ otherwise} \end{array} \right. \ \ , \nonumber
\end{eqnarray}   
$\Theta$ is the usual step function, i.e.\  $\Theta(x) = 1$ for $x>0$
and it vanishes for negative arguments, and $2\pi\de_T(p)= \frac{2}{p}
\sin Tp$. Having introduced the cut-offs $\de,\lambda,T$, we may now 
study the regularized annulus amplitude 
\begin{equation}\begin{aligned}
{}^{}_{\rm B,reg} & \bra
\raa|\tilde{q}^{\frac{1}{2}\SH_{\rm p}}|\raa\ket_{\rm B,reg}^{}\;
\equiv\\[2mm]
& \hspace*{-8mm} \equiv\;
-\int_{\BS}\frac{dj}{\pi^3}\,(2j+1)^2\;\frac{1}{2\pi}\sum_{n\in\BZ}
\;\int_{\BR}\frac{dp}{2\pi}
\;\chi^j(\tilde{q})\;{}^{}_{\rm B,reg}\bra
\raa|j;n,p\ket\bra j;n,p|\raa\ket^{}_{\rm B,reg} \ \ .
\end{aligned}\end{equation}
It involves the characters $\chi^j(q)=q^{b^2P^2}\eta^{-3}(\tau)$ 
with  $q=\exp({2\pi i \tau})$. The Dedekind eta-function  $\eta
(\tau)$ is familiar from flat space theories and it appears here 
because there are no singular vectors in the modules 
of the current algebra when $j = - 1/2 + iP$. After inserting 
\rf{regbdst} and performing some straightforward calculations, 
one can bring the regularized annulus amplitude into the following 
form, 
\begin{equation}\label{annamp}
{}^{}_{\rm B,reg} \bra
\raa|\tilde{q}^{\frac{1}{2}\SH_{\rm p}}|\raa\ket_{\rm B,reg}^{}\;=\;
 \int_{\de}^{\infty}
dP\,\chi^j(\tilde{q})\,N_{\rm ann}(P|\raa)\ \ .  \end{equation}
We have changed to an integration over the variable $P$ using the
standard relation $j=-\frac{1}{2}+iP$. The density $N_{\rm ann}(P|\raa)$ 
is given by
\begin{equation}
N_{\rm ann}(P|\raa)\;\equiv\;
2\lambda\;4\pi T\;\frac{|A_b|^2}{\pi^3 b^2}
\frac{\cos^2 2\raa P \cosh^2\pi P +\sin^2 2\raa P \sinh^2\pi P}
 {\sinh 2\pi P\sinh
 2\pi b^2P}\ \ .
\end{equation}
The  previous two equations provide an explicit formula for the 
regularized annulus amplitude that comes with the brane $\raa$.

\subsection{The open string partition function} \label{pf_strip}

We now want to discuss the second important quantity that enters 
Cardy's consistency condition, namely the partition function on 
the strip with the boundary condition $\raa$ imposed along both 
boundaries. The naive definition of the partition function on 
the strip would be to take the trace of $ \exp(2\pi i \tau\SH^{\rm 
strip}_\raa)$ over the space $\CH^{\rm strip}$. This trace turn out 
to be divergent. There are two sources of divergencies: One is 
the unboundedness of the spectrum for the $SL(2,\BR)$-generators 
in the principal series representations. One may e.g. consider
a basis for the principal series representation $\CP_j$ that
consists of vectors $|j;m\ket$ which diagonalize the generator
of the compact subgroup of $\SLR$. The variable $m$ will then take 
on any integer value. We will regularize this 
divergence by considering traces that are performed over subspaces 
$\CH^{\rm strip}_{\kappa} \subset \CH^{\rm strip}$ where  $\kappa
\in\BZ^{>0}$. They are generated by acting with current algebra
generators on the states $|j;m\ket$, $m\in\{-\kappa+1,-\kappa +2,
\dots,\kappa\}$.

More interesting is the divergence of the open string partition 
function which comes from the infinite volume of the radial coordinate 
on $H_3^+$. We shall imagine having restricted the coordinate $\phi$ 
in $H_3^+$ to be smaller than some cut-off $L$ so that $\H3p$ gets 
effectively replaced by a compact space. Standard arguments suggest 
(see Appendix B) that the leading behavior of the spectral 
density $N_{\rm str}(P|\raa)$ for $L\ra\infty$ is of the form 
\begin{equation}\label{divL}
N_{\rm str}(P|\raa) \;\underset{L\ra\infty}{\sim}\frac{L}{\pi}+
\frac{1}{2\pi i}\frac{\pa}{\pa P}\log R\bigl(-\fr{1}{2}+iP|\raa\bigr).
\end{equation}
There are two ways of dealing with such a divergence: One possibility 
is to simply divide by the ``volume'' $L$ of the radial coordinate.
We are thus lead to an object $Z_{\kappa}(q)$ which does not depend 
at all on the boundary condition $\raa$. But there is a second more 
interesting possibility which exploits that the divergence for $L
\rightarrow \infty$ is universal. This suggests to focus attention on 
the sub-leading term in eq.\ \rf{divL} by subtracting the spectral 
density $N_{\rm str}(P|\raa_\ast)$ for a fixed reference boundary 
condition $\raa_\ast$. The corresponding {\it relative partition 
function}
\begin{equation}
Z^{}_{\rm rel}(q|\raa;\raa_\ast)\;\equiv\;
Tr_{\CH^{\rm strip}_{\kappa}}\bigl(q^{\SH^{\rm strip}_\raa}-q^{\SH^{\rm
strip}_{\raa_\ast}}\bigr)
\end{equation}
is indeed well-defined and given by the expression
\begin{equation}\label{Zrel}  
Z^{}_{\rm rel}(q|\raa;\raa_\ast)\;\equiv\;
\int\limits_{0}^{\infty}dP \;\frac{1}{2\pi i}\frac{\pa}{\pa P}\log
\frac{R\bigl(-\frac{1}{2}+iP|\raa\bigr)}{R\bigl(-\frac{1}{2}+iP|\raa_\ast\bigr)}\;
\chi^j_{\kappa}(q)\ \ .
\end{equation}
We have used the notation $\chi^j_{\kappa}(q)$ for the following 
regularized characters,  
\begin{equation}
\chi^j_{\kappa}(q)\;=\;2\kappa \;q^{b^2P^2}\eta^{-3}(\tau)\ \ .
\end{equation}
The previous two formulas provide an explicit expression for the 
regularized partition function. We claim that it is related with  
the annulus amplitude by world-sheet duality. In order to make 
this more precise, we need to compare the various cut-offs we 
have introduced so far.

\subsection{Comparison of cut-offs}\label{compcut-off}

First, we observe that the cut-off $T$ does not have a direct 
counterpart in our discussion of the partition function on the 
strip. $T$ was introduced to regularize the factor $\de(p)$ in 
the boundary state. The variable $p$ is related via Fourier 
transformation to the time-variable $t$ on the cylinder that 
describes the boundary of $AdS_3$. Hence, we can interpret $2T$ 
as the length of a time-interval to which one has restricted 
the path-integral of the model. This interpretation suggests 
to consider the transition amplitude per unit of time which 
is obtained by simply dropping the factor $\de_T(p)$ in 
\rf{regbdst}. Let us note that an analogous prescription is 
implicit in the identification of the partition function on 
the strip with a trace over the space of states. 

Furthermore, the constants $\lambda$ and $\kappa$ can both be 
considered as ultraviolet cut-offs for the periodic angular 
variable $\theta$ of the asymptotic cylinder. In fact, the 
variable $n$ that is cut off by $\lambda$ is related to $\theta$ 
by Fourier transformation. To find the precise relation between 
$\lambda$ and $\kappa$, let us note that the annulus amplitude can 
be naturally interpreted in terms of the radial evolution 
that is generated by $\SH^{\rm cyl}$ in an annulus
bounded by two concentric circles in the complex plane.
One must take the conformal transformation 
of the currents that is induced by the change of coordinates
from the strip to the complex plane into account.
For definiteness, let us assume that 
the annulus in the complex plane is bounded by two concentric circles 
around the origin with inner and outer radii $R_1$ and $R_2$,
respectively. The parameter $\tau$ is related to $R_1$ and $R_2$ 
through 
\[ 
\log\frac{R_2}{R_1}\;=\;\frac{\pi i}{\tau}\ \ .
\]
$\lambda$ is cutting off the spectrum of the zero mode $J^3_0$ for 
the current $J^3(z)$. Here $z$ is the usual coordinate on the 
complex plane. In order to relate that current to the one used 
in a Hamiltonian formulation of the theory on the strip, we 
perform the canonical mapping from the annulus to the strip,
\[
z\;=\;R_1e^{aw}, \quad a\equiv
\frac{i}{\pi}\log\frac{R_2}{R_1}\;=\;\frac{i}{\tau}\ \ . 
\]
The current $J^3(z)$ on the annulus is then related to its 
counterpart $J^3(w)$ on the strip via $zJ^3(z)=aJ^3(w)$. This 
implies that the cut-offs $\lambda$ and $\kappa$ should obey 
\footnote{The fact that $\lambda$ 
and $\kappa$ were originally defined as integers, whereas $a$ is not
integer in general will not matter when taking $\lambda$, $\kappa$ 
to infinity.}
\begin{equation} \label{uv-cuts} 
 -i \tau \lambda\;=\;  \kappa \ \ . 
\end{equation} 
We finally need to relate the infra-red cut-off $\de$ to the 
``volume'' cut-off $L$. Standard arguments from Fourier analysis
lead to the relation 
\begin{equation}\label{infra-cuts}
L\;=\;(2\pi\de)^{-1}\ \ .
\end{equation}
The two relations (\ref{uv-cuts},\ref{infra-cuts}) must be taken 
into account when we compare the regularized annulus amplitude
with the regularized partition function.

\subsection{Analogs of the Cardy condition}

We are now ready to formulate and verify analogues of the Cardy condition:
One possibility is to simply divide the expression (\ref{annamp}) by
$2 \lambda\, 2T\,\de^{-1}$, and compare the resulting quantity to 
$Z_{\kappa}(q)/2 \kappa$, as defined in subsection \ref{pf_strip}. 
Using the modular transformation of the characters, 
\begin{equation}\label{mod_char}
\chi_P\bigl(-\fr{1}{\tau}\bigr)\;=\;
2\sqrt{2}b\frac{i}{\tau}\int\limits_{0}^{\infty}dP'\;\cos(4\pi b^2 PP')
\chi_{P'}(\tau)\ \ ,
\end{equation}
and the identifications (\ref{uv-cuts},\ref{infra-cuts}) between the 
different cut-offs, it becomes trivial to verify that the two 
expressions agree provided that $A_b$ satisfies 
\begin{equation}\label{Ab_constr}
2\sqrt 2|A_b|^2 \;=\;\pi^2b^3\ \ .
\end{equation}
 
More interesting is the comparison of the sub-leading terms in the 
divergence for $L\ra\infty$. We shall say that the {\it relative} 
Cardy condition is fulfilled if 
\begin{equation}\label{Cardy}\begin{aligned}
0\;=\lim_{\kappa\ra\infty}
\biggl( & \frac{1}{2\kappa}Z_{\rm rel}^{}(q|\raa;\raa_\ast)
\;-\; \\[2mm]  
& \hspace*{-1cm} -\lim_{\de\ra 0}\;\frac{1}{2\lambda\,2T}\Bigl(
{}^{}_{\rm B,reg}  \bra
\raa|\tilde{q}^{\frac{1}{2}\SH_{\rm p}}|\raa\ket_{\rm B,reg}^{}-
{}^{}_{\rm B,reg} 
\bra \raa_\ast|\tilde{q}^{\frac{1}{2}\SH_{\rm p}}|\raa_\ast\ket_{\rm B,reg}^{}
\Bigr)_{\lambda=a\kappa}\biggr)\ \ .
\end{aligned}\end{equation}
In order to verify that our expression for the boundary state 
indeed satisfies this condition, let us start by considering 
the second term in \rf{Cardy}. It follows from \rf{annamp} and
the simple identity $\sin^2 a-\sin^2b=\cos^2 b-\cos^2a$ that 
\begin{equation}
\begin{aligned}\lim_{\de\ra 0}\frac{1}{4\lambda T}\Bigl(
{}^{}_{\rm B,reg} \bra
\raa|\tilde{q}^{\frac{1}{2}\SH_{\rm p}} & |\raa\ket_{\rm B,reg}^{}-
{}^{}_{\rm B,reg} 
\bra \raa_\ast|\tilde{q}^{\frac{1}{2}\SH_{\rm p}}|\raa_\ast\ket_{\rm B,reg}^{}
\Bigr)\;=\;\\
& =\;2\pi\;\frac{|A_b|^2}{\pi^3 b^2} \;\int\limits_{0}^{\infty}
dP\,\chi^j(\tilde{q}) \frac{\cos^2 (2\raa P)-\cos^2
(2 \raa_\ast P)}
{\sinh (2\pi P)\sinh (2\pi b^2P)}\ \ . \end{aligned}\end{equation}
Inserting the modular transformation law for the characters, equation
\rf{mod_char}, leads to an expression of the desired form
$\frac{i}{\tau}\int_{0}^{\infty}dP'\; N(P'|\raa;\raa_\ast)\;\chi_{P'}(\tau)$,
with $N(P'|\raa;\raa_\ast)$ given by
\begin{equation}\begin{aligned}
N & (P|\raa;\raa_\ast)\;=\;\\
= & 
\frac{\sqrt{2}|A_b|^2}
{\pi^3b^3}\frac{\pa}{\pa P}\int\limits_{0}^{\infty}\frac{dt}{t}\;
\frac{\sin 2tb^2\bigl(P+\frac{\raa}{\pi b^2}\bigr)
+\sin 2tb^2\bigl(P-\frac{\raa}{\pi b^2}\bigr)-(\raa\leftrightarrow \raa_\ast)}
{2 \sinh t\sinh b^2 t}\ \ .
\end{aligned}\end{equation}
With the help of the special function $S_k(x)$ that we defined 
in eq.\ (\ref{Sk}) one finally rewrites the spectral density in 
the following form
\begin{equation}
N(P|\raa;\raa_\ast)\;=\;
\frac{\sqrt{2}|A_b|^2}{\pi^3b^3}\,\frac{1}{i}\frac{\pa}{\pa P}\,\log
\frac{S_k(P+2R)S_k(P-2R)}
{S_k(P+2R_0)S_k(P-2R_0)}\ \ ,
\end{equation}
where $R\equiv \frac{\raa}{2\pi b^2}$, $R_0\equiv\frac{\raa_\ast}{2\pi b^2}$.
This expression should be compared to \rf{Zrel}. We find agreement 
provided that $A_b$ satisfies \rf{Ab_constr}.

\subsection{The spherical branes}\label{spherCardy}

Towards the end of this section let us briefly discuss the
verification of Cardy's condition for the spherical branes. We 
will not find any divergence in this case, reflecting the fact 
that the spherical or instantonic branes are compact and do 
not to extend to the boundary of $H_3^+$. 

We start by introducing the boundary state ${}^{}_{\rm B}\bra \ras|$
of the spherical branes through the standard prescription 
${}^{}_{\rm B}\bra \ras|j;u\ket\;\equiv\;{}_{\rm B}^{}\bra 
\Theta^{j}(u|\fr{i}{2})\ket_{\ras}$. Since there is no 
divergence, we can now study the usual expression for the 
annulus amplitude, 
\begin{equation}\begin{aligned}
{}^{}_{\rm B}  \bra 
\ras'|& \tilde{q}^{\frac{1}{2}\SH_{\rm p}}|\ras\ket_{\rm B}^{}\;\equiv\\
& \equiv\;
-\int_{\BS}\frac{dj}{\pi^3}\,(2j+1)^2\;\int_{\BC}d^2u
\;\chi^j(\tilde{q})\;{}^{}_{\rm B,reg}\bra
\ras'|j;u\ket\bra j;u|\ras\ket^{}_{\rm B,reg}\ \ .
\end{aligned}\end{equation}
The integral over $u$ is trivial to carry out and it  
leads us to an expression
\begin{equation}\label{sphCardy}
{}^{}_{\rm B}  \bra 
\ras'| \tilde{q}^{\frac{1}{2}\SH_{\rm p}}|\ras\ket_{\rm B}^{}\;\propto\;
\frac{2}{\sin \ras'\sin \ras}\int_0^{\infty}dP\,P\, 
\frac{\sinh 2\ras' P\sinh 2\ras P}{\sinh2 \pi
b^2P}\,\chi_P^{}(\tilde{q})\ \ .
\end{equation}
If one now restricts the values of $\ras$, $\ras'$ to be 
elements of the following discrete set 
$$\BS_{\rm deg}\equiv
\bigl\{ \pi b^2(2J+1);J=0,\fr{1}{2},1,\dots\bigr\}\ \ ,$$ 
then one can proceed with the verification of the Cardy 
condition along the lines of \cite[Section 4]{GKS},
see also \cite{ZZ2} for a very similar previous discussion 
for the case of Liouville theory. The 
crucial ingredient of their discussion is the identity
\begin{equation}\label{sinh_id}
\frac{\sinh2\pi n b^2 P \sinh 2\pi m b^2 P}{\sinh 2\pi b^2P}
\;=\; \sum_{l=0}^{\min(n,m)-1}\sinh 2\pi b^2(n+m-2l-1)P\ \ .
\end{equation}
This allows them to rewrite the r.h.s.\ of eq.\ \rf{sphCardy}
as a sum of terms which can be directly identified as characters 
for the theory on the strip by using a close relative of the 
modular transformation formula \rf{mod_char}. After a short 
computation one obtains the open string partition function 
\begin{equation} \label{res3a} 
Z(q|\ras,\ras') \ =\ \sum_{J=|R'_1 - R'_2|}^{R'_1 + R'_2 - 1}
\ \chi^J(q) \ \ .    
\end{equation} 
Here $\chi^J(q)$ are characters for the sectors which are 
generated by the current algebra from ground states with 
a $2J+1$-fold degeneracy. This is somewhat similar to the 
spectrum of maximally symmetric D-branes on a 3-sphere with 
infinite radius \cite{AlReSc1}.   

\begin{rem}
For labels of the boundary parameters outside of $\BS_{\rm deg}$, 
it does not seem possible to satisfy the Cardy condition. This is 
supported by the evaluation of the factorization constraint for a 
second degenerate field in \cite{GKS,PaSa}. We have also argued 
for the discreteness of the parameter $\ras$ in our semi-classical 
analysis of the open string spectrum. 
\end{rem}

\setcounter{equation}{0} 
\section{Conclusions and open problems} 

In this work we have proposed an exact solution for all the 
maximally symmetric branes in the Euclidean $AdS_3$. They 
fall into one of two classes depending on whether they 
preserve a $\SLR$ or a $\SU$ subgroup of $\SLC$. Branes 
within each class are related by symmetries of the 
background. The most interesting representative of the 
former class are branes localized along a Euclidean $AdS_2 
\subset AdS_3$. The boundary states for these branes are 
given by formula (\ref{bstads}). Branes preserving an 
SU(2) symmetry are point-like or spherical (with imaginary 
radius) and their exact solution is encoded in the 
one-point function (\ref{res1a}). 
\smallskip

Our strategy in solving these theories was to look at  
factorization constraints on the one-point functions 
in the boundary conformal field theory. The latter arise 
from considering two-point functions of bulk fields in 
the model. We have evaluated one such constraint 
explicitly that uses the degenerate field $\Phi^{1/2}$. 
It would certainly be interesting to check our solution 
against factorization constraints involving other
degenerate fields like the one with label $j = 1/2b^2$
that was used in \cite{Te1}. 
This would further restrict the freedom left by the first 
factorization constraint (see discussion in Subsection 3.4). 
\smallskip
     
The analysis we have carried out also provides important 
information on the open string sector of the model. In 
particular, we have derived the open string spectral density in 
two different ways. For the $AdS_2$ branes the answer is 
given in eq.\ (\ref{Zrel}) whereas the partition function for
the spherical branes has been spelled out in eq.\ (\ref{res3a}). 
The latter is discrete and agrees with the findings of \cite{GKS}.
Let us note, however, that there is more information in the open 
string sector than just the spectrum. It would certainly be 
interesting to compute scattering amplitudes of open string 
states. The latter require to derive and solve the factorization 
constraints on the three-point functions of the boundary fields, 
similar to what was done in \cite{PoTe} for the case of Liouville theory.
So far we have only considered a boundary field that 
corresponds to a degenerate representation of the current algebra.
The results of Section 4 imply that the three point functions
involving this degenerate boundary field can be expressed in terms  
of special fusion coefficients (see \ref{E-Fus}). 
The only difference is that there appears an interesting 
shift (\ref{shift}) in the identification of boundary 
conditions and sectors of the model. This observation supports our 
expectation that 
a calculation of the three point function of generic boundary fields
should be possible along similar lines as in \cite{PoTe}.
\smallskip  
  
The boundary two-point functions are non-trivial 
in the case of non-compact branes and it allow to read off the
reflection amplitude for open strings. We have computed the 
latter for open strings ending on the same brane and our 
expression is manifestly unitary. If one considers open 
strings which have one end on a $AdS_2$ brane $\raa$ and 
the other on $\raa' \neq \raa$, however, a unitary 
reflection amplitude does not exist. This might be related 
to the fact that all the $AdS_2$-branes from the family 
parametrized by $\raa$ end along the same lines on the 
boundary of the Euclidean $AdS_3$. The geometry suggests 
that unitarity can be retained by considering {\it all} the open 
string modes that that exist in the brane configuration 
consisting of two branes with parameters $\raa$ and 
$\raa'$ (and similarly for any larger number of branes).       
\smallskip

As we have remarked in the introduction, one of the most 
interesting applications of our results concerns the coset 
$\H3p/\QR_\tau$ which describes a 2D Euclidean black hole 
geometry. In this construction, $\QR_\tau$ acts by constant 
shifts of the Euclidean time. Since our $AdS_2$ branes are 
left unchanged by translations 
along the time direction, they descend trivially to 
the black hole geometry. In particular, their boundary state
is simply given by omitting the factor $\delta(p)$ from the 
boundary states of the $AdS_2$ brane, i.e.\ 
\begin{equation}\begin{aligned}\label{bstBH}
{}^{}_{\rm B}\bra \raa|j;n\ket^{BH} \;=\; (2\pi)^2 & 
\, C_b \,\Ga(-b^2(2j+1))\,d^j_n \ti\\[1mm]
\ti & \bigl( \pi^0_n\cosh \raa (2j+1)-\pi^1_n\sinh 
\raa(2j+1)\bigr)\ \ .
\end{aligned} 
\end{equation}
The definition of the constants $d^j_n$ and $\pi^0_n$ can 
be found after eq.\ (\ref{bstads}) and the constant $C_b$ can 
be determined by a Cardy type computation as above. The
expression (\ref{bstBH}) describes the coupling of closed 
string modes that do not wind around the semi-infinite 
cigar as $\phi \rightarrow \infty$. 

Geometrically, the boundary state (\ref{bstBH}) corresponds 
to a one-dimensional brane that is stretched in between two
antipodal points on the asymptotic circle at the boundary of 
the black hole geometry and extends into 2-dimensional 
space. It crosses the tip of the semi-infinite cigar for 
$\raa = 0$. In Section 2 we have remarked that the bulk 
and boundary theories for $\H3p$ do not possess a positive 
definite state space because of the imaginary B-field. This 
problem disappears in the coset model for purely 
dimensional reasons both for the closed and the open 
string sector. It would also be interesting to study the 
holographic dual of the branes (\ref{bstBH}) in a black hole 
geometry using the matrix model description that was proposed 
in \cite{KKK}. Following \cite{DeWFrOo,BdBDO}, these would 
show up as point-like defects on the asymptotic circle. We 
hope to return to all these issues in the near future.  
\bigskip

{\bf Acknowledgments:} It is a pleasure to thank C.\ Bachas, 
V.\ Kazakov, I.\ Kostov, P.\ Lee, H.\ Ooguri, J.\ Park, 
M.\ Petropoulos, S.\ Ribault, M.\ Rozali, A.\ Schwimmer for 
correspondence and interesting discussions. The work of B.P.\ 
was supported by the  Gif Project Nr. I. 645 130-14|1999 and 
by the EU contract HPRN-CT-2000-00122.

\newpage

\appendix

\setcounter{equation}{0} 
\section{Analysis on the Euclidean $AdS_2$ and $AdS_3$} 

\subsection{Some integrals for Subsection 2.3}

Our aim here is to compute the integrals (\ref{aux1}). 
To this end let us introduce the following distributions $D^j_{\ep}$ 
on $\CS(\H3p)$, 
\begin{equation}\label{Djepdef}
D^j_{\ep}(f)\;\equiv\;
\int_{\BC} d^2u \;|u+\bu|^{2j} \sgn^{\ep}(u+\bu)
F^j_u[f] \ \ . 
\end{equation}
Here, $F^j_u$ denotes the generalized Fourier transformation on $\H3p$ 
that is defined by
\begin{equation}
F^j_u[f]\;=\;\int_{H_3^+}dh\;\bigl(\Phi^j(u|h)\bigr)^*\,f(h).
\end{equation}
\begin{prop}\label{distclaim} The distributions $D^j_{\ep}$ can be 
represented in the following simple form
\begin{eqnarray}
D^j_{\ep}(f) \ = \ 
\int_{\H3p}dh\;D^j_{\ep}(h)f(h)\ & &  \ \mbox{ where } \\[2mm]
D^j_{0}(h)\;=\;\frac{\cosh\psi(2j+1)}{\cosh\psi}\qquad
& & \quad D^j_{1}(h)\;=\;\frac{\sinh\psi(2j+1)}{\cosh\psi}.
\end{eqnarray}  
Here, $h$ is an element of $\H3p$ which is parametrized by $(\psi, 
\chi,\nu)$ that we introduced in equation\ (\ref{param1}).  
\end{prop}

\begin{proof} Let us begin by proving the following simple lemma
about the transformation properties of $D^j_\e$ under the action 
of $\SLR$. 
\begin{lem}\label{invlem}
The functionals $D^j_{\ep}$ are invariant under the subgroup of 
matrices $g\in \SLC$ that satisfy $\om(g^{\dagger})=g^{-1}$, 
\begin{equation}
D^j_{\ep}(T_g f)\;=\;D^j_{\ep}(f) \ \ . 
\end{equation}
Here, $\SLC$ acts on functions $f\in L^2(\H3p)$ according to     
\ $(T_g f)(h)\, = \, f\bigl(g^{-1}h(g^{-1})^\dagger\bigr)$. 
\end{lem}
\begin{proof}
We begin by noting that the generalized Fourier transformation 
$F^j_{u}(f)$ satisfies the following intertwining property,
\begin{equation}
F^j_u[T_g f]\;=\; |\be u+\de|^{-4j-4} F^j_{g\cdot u}[f]
\;\;{\rm where} \;\;
g=\bigl(
\begin{smallmatrix}
\al & \be \\ \ga & \de\end{smallmatrix}\bigr)\;\;{\rm and}\;\; 
g\cdot u\equiv\frac{\al u+\ga}{\be u +\de}.
\nonumber \end{equation} 
This follows easily from $\SLC$-invariance of the measure $dh$
on $\H3p$ and the identity (\ref{trafo1}). One may then substitute 
the $u$-integration by an integration over $u'\equiv g\cdot u$. It 
remains to observe that 
\begin{equation} \label{trfrealpart}
u+\bu\;=\;2\Re\biggl(\frac{-\de\bar{\be}u'\bu'
+u'(\de\bar{\al}+\be\bar{\ga})-\ga\bar{\al}}{|-\be u'+\al|^{2}}\biggr)
\;=\;  \frac{2 (u'+\bu')}{|-\be u'+\al|^{2}}\;\, ,
\end{equation}
where the second equality holds true for all $g$ which satisfy  
$\om(g^{\dagger})=g^{-1}$. This completes the proof of the lemma.  
\end{proof}
We can use the invariance of the functionals $D^j_{\ep}$ 
to express them as averages over the Euclidean $AdS_2$ 
branes $\CC_{\psi}$,   
\begin{equation}
D^j_{\ep}(f)\;=\; E^j_{\ep}(\CA f)\ , \quad (\CA f)(\psi)\equiv
\int_{\CC}dc\;(T_cf)(h_\psi)\ \ , 
\end{equation}
where $h_{\psi}$ was defined in eq.\ (\ref{param1}, the measure 
$dc$ is given by $dc=dtd\chi \cosh \chi$, and $E^j_{\ep}$ is a 
distribution on functions $f(\psi)$ of a single real variable. 
The Casimir $Q$ acts diagonally on the functions $E^j_{\ep}$, 
i.e.\ $E^j_{\ep}(Qf)=j(j+1)E^j_{\ep}(f)$. For functions which 
are constant in $\chi,\nu$, the Casimir takes the simple form 
\begin{equation}
Q_{\psi}\;=\;\frac{1}{4}\pa_{\psi}^2+\frac{1}{2}
\frac{\sinh\psi}{\cosh\psi}\pa_{\psi}^{}\;=\;
\frac{1}{4}\frac{1}{\cosh\psi}\bigl(\pa_{\psi}^2-1\bigr)\cosh\psi.
\end{equation} 
By inserting this into the eigenvalue equation for the functions 
$E^j_\e$ we can now easily conclude that 
\begin{equation}
E^j_{\ep}(f)\;=\; \int_{\BR}d\psi\cosh^{-1} \psi 
\;\bigl(K_+e^{(2j+1)\psi}+K_-e^{-(2j+1)\psi}\bigr)f \ \ .
\end{equation}
In order to fix the two unknown coefficients $K_{\pm}$ one may 
consider $D^j_{\ep}(f)$ for functions $f$ that are supported near the 
boundary of $\H3p$. More precisely, let us consider the set of 
functions $f_{\raa}$ that vanish if $|\psi-\raa|>\de$ for 
sufficiently small $\de>0$. We will be interested in the asymptotic 
behavior of $D^j_{\ep}(f_{\raa})$ for large values of $|\raa|$. 
From the asymptotic behavior (\ref{phiasym}) of the functions 
$\Phi^j_u$ we obtain
\begin{equation}
F^j_u[f]\;\underset{|\raa|\ra\infty}{\sim}\;
\int\limits_{\H3p}dh \;f(h)\biggl(e^{2j\phi}\de(\ga-u)+
\frac{2j+1}{\pi}e^{-2(j+1)\phi}|\ga-u|^{-4j-4}\biggr)
\nonumber \end{equation}
up to terms of $\CO(e^{-\raa})$. Inserting this expression into 
\rf{Djepdef} of $D^j_{\ep}(f)$ and using \rf{distid} yields 
\begin{eqnarray}\label{Djepas}
D^j_{\ep}(f)\; & \underset{|\raa|\ra\infty}{\sim} & 
  \int\limits_{\H3p}dh \;f(h)\sgn^{\ep}(\ga+\bar{\ga})
\left(\bigl(e^{\phi}|\ga+\bar{\ga}|\bigr)^{2j} + 
(-)^{\ep} \bigl(e^{\phi}|\ga+\bar{\ga}|\bigr)^{-2j-2}
\right)\ .
\nonumber \end{eqnarray}
We finally note that $(\ga+\bar{\ga})\exp\phi=2\sinh\psi
\sim \exp{|\psi|}$ in order to rewrite the previous property 
as  
\begin{equation}
D^j_{\ep}(f)\;  \underset{|\raa|\ra\infty}{\sim}  \;  
\int_{\H3p}dh \;f(h)e^{-|\psi|}
\Bigl( e^{(2j+1)\psi} + (-)^{\ep}e^{-(2j+1)\psi}\Bigr)\ \ . 
\end{equation}
We can finally conclude from here that $K_+ = 1/2$ and 
$K_- \ (-1)^\e /2$. This completes the proof of Proposition 
\ref{distclaim}.
\end{proof}

Let us finally state without proof the corresponding result 
for the spherical branes. To this end we introduce  
\begin{equation}\label{Djepdef1}
\tilde D^j (f)\;\equiv\;
\int\limits_{\BC} d^2u \;|u\bu+ 1|^{2j} F^j_u[f] \ \ . 
\end{equation}
Using the same ideas as in the proof of the previous 
proposition one can establish the following result. 

\begin{prop}\label{distclaim1} The distributions $\tilde D^j$ 
can be represented in the following simple form
\begin{equation}
\tilde D^j (f) \ = \ 
\int_{\H3p} dh\; \frac{\sinh \Lambda (2j+1)}{\sinh \Lambda}  
 f(h)\ \ .  
\end{equation}  
Here, $h$ is an element of $\H3p$ which is parametrized by $(\psi, 
\chi,\nu)$ that we introduced in equation\ (\ref{param2}).  
\end{prop}

\subsection{Harmonic analysis on the $AdS_2$ branes}

As we discussed in some detail above, the space of wave functions
on the Euclidean $AdS_2$-brane carries an action of $\SLR$. Let us 
denote the surface on which the brane is localized by $\CC_{\psi}$
and recall that it comes equipped with a measure $dc = 2 e^\chi 
d\chi d \nu$. The action of $\SLR$ on the Hilbert space $L^2(
\CC_{\psi},dc)$ is easily shown to be unitary. Our claim is 
that $L^2(\CC_{\psi},dc)$ decomposes into irreducible 
representations $\CP_j, j = -\frac12 + ip,$ from the principal 
series of $\SLR$.   

\begin{thm} 
There exists an isomorphism between the following representations
of  $\SLR$  
\begin{equation}
L^2(\CC_{\psi},dc)\;\simeq\; \int_{\BS}^{\oplus} d\mu_{\rm P}^{}(j)
\;\CP_j\ \ ,
\quad \ \ \  d\mu_{\rm P}(j)\ =\ \frac{2j+1}{2\pi}\coth\pi j \ dj
\ \ .
\end{equation}
This isomorphism can be realized explicitly by a generalized Fourier 
transform. It implies that any function $f \in L^2(\CC_{\psi},dc)$ 
may be decomposed in the form 
\begin{eqnarray}\label{ads2ft}
f(c) & = &  \int_{\BS}d\mu_{\rm P}^{}(j)\int_{\BR}du 
\ \Xi^j(u|\psi;c)\ \CF^j_u[f]\ \\[2mm] \ \ \mbox{where}
& & \ \ \Xi^j(u|\psi; c) \ = \ \left( \, v'_u\  h\  {v'_u}^\dagger \, 
\right)^j 
\end{eqnarray} 
with $v'_u = (iu,1)$ and the functions $\CF^j_u[f]$ are the generalized 
Fourier coefficients of $f$, i.e.\  
\begin{equation}  
\CF^j_u[f]\ =\ \int_{\CC_{\psi}}dc\;
\bigl(\Xi^j(u|\psi;c)\bigr)^* \ f(c)\ \ .
\end{equation}
The Fourier transformation $\CF$ diagonalizes the action of $\SLR$ on 
$\CC_{\psi}$ in the sense that 
\begin{equation}\label{ads2ft:int}
\begin{aligned}
\CF^j[T_g f]\;=\; P_g^j\CF^j[f]\ \ \ \ \mbox{ where }& \ \ \ 
T_g f(c) \ = \ f (g^{-1} c (g^{-1})^\dagger)\;\, ,\\
\mbox{ and }& \ \ \ P_g^j h(u) \ = \ |\beta u + \delta|^{2j}\, h(g\cdot u)
\end{aligned}\end{equation}
for all $g \in \SLC$ of the form 
$g=(\begin{smallmatrix} \de & -i\be \\ i\ga & \de 
\end{smallmatrix})$, $\al,\be,\ga,\de\in\BR$. 
Moreover, the Fourier transform satisfies a reflection 
property which is related to the equivalence $\CP_j\simeq
\CP_{-j-1}$,  
\begin{equation}\label{ads2ft:refl}
\CF^{-j-1}_u[f]\;=\;-(\cosh\psi)^{2j+1}\int_{\BR}du' \; J^j(u,u')
\ \CF^j_u[f]\ \ ,
\end{equation}
where $J^j$ is the integral kernel that we have defined in  eq.\ 
(\ref{psiasym}) above.  
\end{thm}

{\it Sketch of proof.} To begin with let us note that the action 
of the Lie algebra of $\SLR$ on the surface $\CC_\psi$ is 
represented by the differential operators 
\begin{equation}
\begin{aligned}
\CD^{+}\;=\;&-\pa_{\nu}
\\ 
\CD^{-}\;=\;& -(\nu^2-e^{-2\chi})\pa_{\nu}+2\nu\pa_{\chi}
\end{aligned}\qquad
\CD^0\;=\; -\nu\pa_{\nu}+\pa_{\chi}\ \ .
\end{equation}
Here we use the coordinates from eq.\ (\ref{param2}). The expression 
the Laplace operator on $\CC_\psi$ is 
\begin{equation}
Q\;=\;\pa_{\chi}^2+\pa_{\chi}^{}+e^{-2\chi}\pa_{\nu}^2\ \ .
\end{equation}
It is easy to determine the common spectral decomposition of 
$Q$ and $\CD^+$. The latter is diagonalized by the Fourier 
transformation w.r.t.\ $\nu$ and on the eigen-spaces of 
$\CD^+$ the operator $Q$ gets represented by $Q_k\equiv
\pa_{\chi}^2+\pa_{\chi}^{}-k^2\exp{(-2\chi)}$ where $ik$ is the 
eigen-value of $\CD^+$. By a simple re-definition of the 
eigen-functions one can see that the spectral problem for 
$Q_k$ in $L^2(\BR,d\chi e^{\chi})$ is equivalent to 
the spectral problem for the Hamilton-operator $H_k=
-\pa_{\chi}^2+k^2\exp{(-2\chi)}$, for which the solution 
is well-known \cite{Le}. One thereby finds that the set of 
functions
\begin{equation}
\Psi^j_k(\chi,\nu)\; := \;
\frac{2\bigl(\frac{|k|}{2}\bigr)^{-j-\frac{1}{2}}}{\Ga\bigl(-j-\frac{1}{2}
\bigr)}
e^{-ik\nu}e^{-\frac{1}{2}\chi}K_{j+\frac{1}{2}}(|k|e^{-\chi}),\quad
j\in\BS,\;\,k\in\BR
\end{equation}
forms a basis for $L^2(\BR^2,d\nu d\chi e^{\chi})$ 
with normalization given by
\begin{equation}
\int_\BR d\nu \int_{\BR}d\chi \,e^{\chi}\;
\Psi^{-\frac{1}{2}-ip}_{-k}(\chi,\nu)\,
\Psi^{-\frac{1}{2}+ip'}_{k'}(\chi,\nu)\,=\,(2\pi)^2
\de(p-p')\de(k-k').
\end{equation}
Let us compare this with the Fourier transform of $\Xi^j(u|\psi;c)$
\begin{eqnarray} 
\hat \Xi^j(k|\psi;x) & = & 
\int_\BR du \; e^{-iku} \;\Xi^j(u|\psi;c) \nonumber \\[2mm] 
& = & \cosh^j\psi\ \frac{\left(\frac{|k|}{4}
\right)^{-j-\frac12}}{\Ga(-j-\frac12)}\ e^{-ik\nu} 
e^{-\frac12 \chi } K_{j+\frac{1}{2}}(|k|e^{-\chi})\ \ . \nonumber 
\end{eqnarray} 
The second line follows easily from one of the standard integral 
representation for the Bessel function $K_{\rho}$. Hence we 
conclude from the completeness and orthogonality of the 
$\Psi^j_k(\chi,\nu)$ that eq.\ \rf{ads2ft} defines a 
generalized Fourier transformation on $L^2(\CC_{\psi},dc)$.

The intertwining property \rf{ads2ft:int} is easily verified by a 
direct calculation. To finally verify the reflection property
\rf{ads2ft:refl} one may note that $K_\rho=K_{-\rho}$ implies 
a simple reflection property for the $\Xi^j(k|\psi;c)$. This is
translated into \rf{ads2ft:int} by means of the following formula
\cite{GS} for the Fourier transformation of the distribution 
$|x|^{2j}$, 
\begin{equation}\label{FTform}
\int_{\BR}dx\; e^{ikx} |x|^{2j}\;=\;\sqrt{\pi}\;\Bigl|\frac{2}{k}
\Bigr|^{2j+1} \frac{\Ga\bigl(j+\frac{1}{2}\bigr)}{\Ga(-j)}\ \ .
\end{equation}
Thereby we did prove all the assertions of our theorem describing
the harmonic analysis of the Euclidean $AdS_2$ branes.

\subsection{The distributions $d^{j}_{\ep}$}
\label{distr_d}

Our aim here is to prove a reflection property for the tempered 
distributions $d^{j}_{\ep}$ that are defined by 
\begin{equation}\label{djepdef}
d^{j}_{\ep}(f)\;\equiv\; \int_{\BC}d^2u \;|u+\bu|^{2j}\sgn^{\ep}(u+\bu)\, 
f(u).
\end{equation}
\begin{prop}\label{reflprop}
The tempered distributions $d^{j}_{\ep}$ satisfy the following 
reflection property 
\begin{equation}
d^{j}_{\ep}(f)\ =\ -(-)^{\ep}\, d^j_{\ep}(\CI_jf)\ \ , 
\end{equation}
where $\CI_j$ is the intertwining operator that establishes
the equivalence of the $SL(2,\BC)$-representations $P_{-j-1}$ 
and $P_j$ (see eq.\ (\ref{CIint}))
\end{prop}
\begin{rem}
Proposition \ref{reflprop} can be re-interpreted as the following 
identity between the corresponding distributions,  
\begin{equation}\label{distidapp}\begin{aligned}
\frac{2j+1}{\pi}\int_\BC d^2u \;|u+\bu|^{2j}\sgn^{\ep}(u+\bu) & 
\;|u-\ga|^{-4j-4}=
\\ 
& =\; (-)^{\ep}|\ga+\bar{\ga}|^{-2j-2}\sgn^{\ep}(\ga+\bar{\ga})\ . 
\end{aligned}\end{equation}
\end{rem}

\begin{proof}
It will be convenient to use a kind of Fourier-transformed version 
of the definition \rf{djepdef}. The function $f(u)$ may be represented 
as 
\begin{equation}\label{invFT}
f(u)\;=\;(2\pi)^{-2}\sum_{n\in\BZ}\int_{\BR} dp\;e^{in\arg(u)}|u|^{-2j-2+ip}\,
F^{j}_{np}(f)\ \ , 
\end{equation}
where the Fourier-transform $F^{j}_{np}(f)$ of $f$ is defined by
\begin{equation}F^{j}_{np}(f)\;=\;
\int_{\BC} d^2u \;e^{-in\arg(u)}|u|^{2j-ip}\, f(u)\ \ .
\end{equation}
\begin{lem}
The distribution \, $d^{j}_{\ep}(f)$\,  can be 
represented in terms of \, $F^{j}_{np}(f)$\,  as follows
\begin{equation}\label{FTdjep}
d^{j}_{\ep}(f)\;=\;\sum_{n\in\BZ}\; d^{j,\ep}_n\,F^j_{n0}(f)\ \ , 
\end{equation}
where the coefficients $d^{j,\ep}_n$ are given by the expression 
$d^{j,\ep}_n=d^{j}_n\pi^{\ep}_n$ with 
\begin{equation}
d^{j}_n\;=\;\frac{\Ga(2j+1)}{\Ga(1+j+\frac{n}{2})
\Ga(1+j-\frac{n}{2})}, \quad
\pi^{\ep}_n\;=\;\left\{ \begin{aligned}
1-\ep \;\;& {\rm if}\;\;n\;\;{\rm even}\\
\ep\;\;& {\rm if}\;\;n\;\;{\rm odd\ \ .}
\end{aligned}\right.
\end{equation}
\end{lem}
\begin{proof}
Inserting eq.\ \rf{invFT} into eq.\ \rf{djepdef} immediately leads 
to a representation of the form \rf{FTdjep} with $d^{j,\ep}_n$ 
given through 
\begin{equation}
d^{j,\ep}_n\de(p)\;=\;(2\pi)^{-2}\int_{\BC}d^2u\;
e^{in\arg(u)}|u|^{-2j-2+ip}|u+\bu|^{2j}\sgn^{\ep}(u+\bu)\ \ .
\end{equation}
It is straightforward to reduce the resulting integral for 
$d^{j,\ep}_n$ to the form
\begin{equation}
d^{j,\ep}_n\;=\;e^{-in\frac{\pi}{2}}\int_0^{\pi}d\vf\;
(e^{in\vf}+(-)^{\ep}e^{-in\vf})(2\sin\vf)^{2j}\ \ .
\end{equation}
By studying the behavior of the integrand under $\vf \ra \pi-\vf$ 
one may verify that $d^{j,\ep}_n=\pi^{\ep}_n d^j_n$.
The integral for $d^j_n$ can be found e.g.\ on
p. 427 of \cite{GGV}.
\end{proof}
\begin{lem}\label{FTrefl}
The Fourier-transform $F^{j}_{np}(f)$ satisfies the following
reflection property $F^{j}_{np}(f)=-r^j_{np}F^{-j-1}_{np}(\CI_jf)$, 
where
\begin{equation}
r^j_{np}\equiv\frac{\Ga(-2j-1)}{\Ga(2j+1)}
\frac{\Ga\bigl(1+j-\frac{1}{2}(n+ip)\bigr)\Ga\bigl(1+j-\frac{1}{2}(n-ip)\bigr)}
{\Ga\bigl(-j-\frac{1}{2}(n-ip)\bigr)\Ga\bigl(-j-\frac{1}{2}(n+ip)
\bigr)}\ \ .
\end{equation}\end{lem}
\begin{proof}
The claim follows easily with the help of
\begin{equation}\begin{aligned}
\int_{\BC}d^2x'\;|x & -x'|^{-4j-4}  x'{}^{j-m}\bx'{}^{j-\bm}
\;=\;
\\
=\; & \pi \,\frac{\Ga(1+j-m)\Ga(1+j+\bm)}
{\Ga(-j-m)\Ga(-j+\bm)}\frac{\Ga(-2j-1)}{\Ga(2j+2)}\,x ^{-j-1-m}
\bx^{-j-1-\bm}\ \ . \end{aligned}\end{equation}
This can be obtained by a minor generalization from an integral 
calculated in \cite{Do}. 
\end{proof}
To complete the proof, one may note that the functional relation for 
the Gamma function implies $r^{j}_{n0}d^j_n/d^{-j-1}_n=(-)^n$. From 
this we conclude 
\begin{equation}\label{refldjep}
r^{j}_{n0}\,d^{j,\ep}_n\;=\;(-)^{\ep}d^{-j-1,\ep}_n\ .
\end{equation}
Inserting the result of Lemma \ref{FTrefl} into eq.\ \rf{FTdjep} 
yields
\begin{equation}
d^{j}_{\ep}(f)\;=\;\sum_{n\in\BZ}\; d^{j,\ep}_n(-r^j_{n0})
F^{-j-1}_{n0}(\CI_jf)\ .
\end{equation}
If we finally take eq.\ \rf{refldjep} into account and use eq.\  
\rf{FTdjep} again we can establish the Proposition \ref{reflprop}.
\end{proof}

\setcounter{equation}{0} 
\section{Relative partition functions}

In this appendix we are going to review the relation between 
reflection amplitudes and spectral densities in a quantum 
mechanical setting. This is certainly well-known to many
people, but may not be familiar to all potential readers.
For a mathematical rigorous treatment
the reader may consult e.g. \cite{Y}.

Assume we are given a quantum mechanical system with a 
Hamiltonian $H=p^2+V(q)$, where the potential 
$V(q)$ rapidly approaches zero for $q\ra -\infty$, but 
diverges for $q\ra \infty$. The vanishing of the potential for 
$q\ra -\infty$ implies that eigen-functions of the Hamiltonian 
can be specified by their asymptotic behavior in this region, 
\begin{equation}
\Xi_E(q)\ \sim \ A_p \;e^{ipq}\;+\;B_p\; e^{-ipq}\ \ , \qquad 
  p\ =\ \sqrt{E} \ \ . 
\end{equation}
In general, the eigenvalue equation $H\psi=E\psi$ has two linearly 
independent solutions. But due to the divergence of the potential 
for $q\ra \infty$, a generic eigen-function will have a similar 
divergence and there can be at most one solution which is 
well-behaved for $q\ra \infty$. From this solution we may read 
off the ratio $R(p)=B_p/A_p$. This is the quantum mechanical 
characteristics of a totally reflecting potential: An incoming 
plane wave $e^{ipq}$ is reflected\footnote{Of course 
the qualification ``incoming'' resp. ``outgoing'' requires 
consideration of the problem of asymptotic time-evolution of 
wave-packets that in the asymptotic past approximate an 
incoming plane wave. The relation between time-asymptotics of 
the scattering problem and space-asymptotics of the 
eigen-functions follows by applying stationary phase methods 
to the eigenfunction expansion of the time-dependent 
wave-function.} 
into an outgoing wave $e^{-ipq}$ times the {\it reflection 
amplitude} $R(p)$. The reflection amplitude is a functional 
of the potential. 

Having introduced the reflection amplitude $R(p)$ we want to 
analyse how it is related with the partition function of the 
system. The definition of partition functions gets subtle in 
the case of systems with continuous spectrum. One might hope 
that partition functions could be represented in the following 
form
\begin{equation} \label{trace}
\Tr\bigl(\CO(H)e^{-\be H}\bigr)\ =\ 
\int dE \;\rho(E) \;\CO(E)e^{-\be E}\ ,
\end{equation}
where $\rho(E)$ is some spectral density. Intuitively one would 
consider $\rho(E)$ to represent the ``density of eigenvalues''. 

The most naive version of this idea does not quite work: 
It is instructive to consider the system obtained by putting a 
perfectly reflecting wall at $q=-L$, with large positive $L$, and taking
the limit $L\ra \infty$. For any finite value of $L$ one then has
a system with discrete spectrum, but when $L$ increases, the number
$N(E)$ of eigenvalues corresponding to energies $E'<E$ will
likewise increase. The average density of eigenvalues in an 
interval $[E-\de,E]$ is 
\[ \rho_{\de}(E) \ =\ \frac{N(E)-N(E-\de)}{\de}\ \ . \]
Now one needs to consider the behavior of such quantities for
$L\ra \infty$. Quantization of the energy eigenvalues for finite
$L$ is a consequence of the boundary condition 
\begin{equation}\label{quantcond}
\psi_E(q)\bigr|_{q=L}\ =\ 0\ \ .
\end{equation}
If $L$ is large enough one may approximate $\psi_E$ by its asymptotic
behavior
\[
\psi_E(q)\ \sim\  e^{ipq}+R(p)e^{-ipq}\ .
\]
This is the basic source for the relation between reflection
amplitude and spectral density: The quantization condition \rf{quantcond}
turns into an equation that determines the possible eigenvalues from
the reflection amplitude
\begin{equation}
R(p)\ =\ -e^{-2ipL} \quad\text{for any eigenvalue $E=p^2$.}
\end{equation}

By introducing the function $\De_L(p)=p-\frac{i}{2L}\ln(R(p))$ one may 
express the positions $p_n$ of eigenvalues in terms of the inverse function
$\De_L^{-1}$ as 
\[
p_n=\De_L^{-1}\biggl(\frac{2n+1}{2L}\pi\biggr)\ \ .
\]
One will have to consider the eigenvalues near a fixed value $E_n=E(p_n)$.
When taking $L\ra \infty$ one obviously needs to consider values 
of the eigen-value label $n$ of the same order as $L$. The spacing 
$\de p\equiv p_{n+1}-p_{n}$ of two momenta can then be estimated as 
\begin{equation}\begin{aligned}
\de p\equiv p_{n+1}-p_{n} \ =\ & \;\De_L^{-1}\biggl(\frac{2n+3}{2L}\pi\biggr)-
\De_L^{-1}\biggl(\frac{2n+1}{2L}\pi\biggr)\\[2mm]
\ \sim\  & \;\frac{\pi}{L}\frac{\pa}{\pa y}\De_L^{-1}(y)
\bigr|_{y=\frac{2n+1}{2L}}\\[2mm]
\ =\ & \;\frac{\pi}{L} \frac{1}{\De'_L(p_n)}\ \ .
\end{aligned}
\end{equation}
The average density $\rho_{\de}$ of eigenvalues is therefore 
approximately
\begin{equation}\label{dos_asym}
\rho_{\de}(p)\ \sim\ \;\frac{L}{\pi} \frac{\pa}{\pa p} \De_L(p)\;=\;
\frac{L}{\pi}+\frac{1}{2\pi i}\frac{\pa}{\pa p}\ln R(p)\ \ .
\end{equation} 
This quantity diverges for $L\ra \infty$. It follows that traces like
\rf{trace} do not make sense in a theory with continuous spectrum. 
However, one may note that this divergence is universal, i.e. 
to a large extend independent of the interaction $V(q)$. Interesting 
objects to study are therefore the {\it relative} partition functions, 
which compare the spectrum in the potential $V$ to the spectrum of a 
fixed reference Hamiltonian $H_\ast=p^2+V_\ast$
\begin{equation} \label{reltrace}
\Tr_{\text{\tiny rel }}\bigl(\CO(H)\bigr)_{H_\ast}\equiv
\Tr\bigl(\CO(H)-\CO(H_\ast)\bigr)\ \ .
\end{equation}
We assume that $V_\ast(q)$ belongs to the same class of potentials 
as $V(q)$, i.e. it also rapidly approaches zero for $q\ra -\infty$ 
and diverges for $q\ra \infty$. The spectra of $H$ and $H_\ast$ will 
therefore have the same continuous part. If we denote the reflection 
amplitude for $V_\ast$ by $R_\ast(p)$, we immediately get the 
following {\it relative trace formula} from eq.\ \rf{dos_asym} 
\begin{equation}
\Tr_{\text{\tiny rel }}\bigl(\CO(H)\bigr)_{H_\ast}\;=\;
\int d\mu(E) \;\rho_{\text{\tiny rel }}(E) \;\CO(E)\ \ ,
\end{equation}
where the relative eigenvalue density $\rho_{\text{\tiny rel }}(E)$
is given by the expression
\begin{equation}
\rho_{\text{\tiny rel }}(E)\;=\;
\frac{1}{2\pi i}\frac{\pa}{\pa p}\ln 
 \frac{R(p)}{R_\ast(p)}\bigg|_{p=\sqrt{E}}\ \ .
\end{equation}

\setcounter{equation}{0} 
\section{Conformal blocks}
 
For the convenience of the reader, we want to gather some basic 
definitions and results concerning chiral vertex operators, 
conformal blocks and the associated fusion matrices. These 
results are mostly well-known, but it may be helpful to 
list the required formulae in a uniform notation. The only 
slightly unusual point comes from the definition \rf{cfbldef} 
of the conformal blocks by means of the invariant bilinear 
form \rf{bilin} on the representations $\CP_j$, which accounts 
for some absolute value signs in the formulae below.

\subsection{Chiral vertex operators and conformal blocks}

We only need to consider one rather special class of chiral 
vertex operators. Consider operators $\SV_{m}^{j}(u|z):
\CP_j\ra\CP_{j+m}$ with $j=\frac{1}{2},1,\dots$ and $m=-j,-j+1,
\dots,j$ that are uniquely defined by the properties
\begin{equation}\begin{aligned}
\text{(i)} & \;\; J_n^{a}\SV_{m}^{j}(u|z)-
\SV_{m}^{j}(u|z)J_n^{a}
=z^n\CD_{j,u}^a\SV_{m}^{j}(u|z),\\
\text{(ii)} & \;\;\SV_{m_2}^{j_2}(u_2|z)|j_1;u_1\ket=z^{\De_{j_1+m_2}-\De_{j_2}
-\De_{j_1}}(u_2-u_1)^{j_2-m_2}\ti\\& 
\hspace{5cm}\ti\bigl(|j_1+m_2;u_1\ket +\CO(z)+\CO(u_2-u_1)\bigr)\ \ .
\end{aligned}\end{equation}
In this particular case the dependence of $\SV_{m}^{j}(u|z)$
on its variable $u$ happens to be polynomial, which means that 
$\SV_{m}^{j}(u|z)$ satisfies an equation of the form 
$\pa_u^{2j+1}\SV_{m}^{j}(u|z)\equiv 0$.

Conformal blocks can be defined with the help of the invariant 
bilinear form on $\CP_j$ which can be 
described using the following object 
\begin{equation}\label{bilin}
B\bigl(\,|j,u_1\ket\, , |j,u_2\ket\,\bigr)\;=\;|u_2-u_1|^{2j}\ \ .
\end{equation}
One can then introduce a class of conformal blocks (``s-channel'')
by  
\begin{equation}\label{cfbldef}\begin{aligned}
\CF^{(s)}_{j_1+m_1}\bigl[\begin{smallmatrix} j_3 & j_2 \\
j_4 & j_1 \end{smallmatrix}\bigr] & (u_4,\dots,u_1|z_3,z_2)\;\equiv\;\\
& \equiv\;B\bigl(\,|j_4,u_4\ket\, , \,
\SV_{m_2}^{j_3}(u_3|z_3)
\SV_{m_1}^{j_2}(u_2|z_2)
|j_1;u_1\ket\,
\bigr) 
\end{aligned}\end{equation}
where $j_4=j_1+m_2+m_1$. As usual, one finds that $\CF^{(s)}$ can be
expressed in terms of a function of the cross-ratios $u,z$, 
\begin{eqnarray}
\CF^{(s)}_{j_1+m_1}\bigl[\begin{smallmatrix} j_3 & j_2 \\
j_4 & j_1 \end{smallmatrix}\bigr] (u_4,\dots,u_1|z_3,z_2) & = & 
|u_{4}-u_{1}|^{j_4+j_1-j_2-j_3}
(u_{4}-u_{3})^{j_4+j_3-j_2-j_1} \times \nonumber \\[2mm] 
& & \hspace*{-3cm} \;(u_{4}-u_{2})^{2j_2}\, 
(u_{3}-u_{1})^{j_1+j_2+j_3-j_4}F^{(s)}_{j_1+m_1}
\bigl[\begin{smallmatrix} j_3 & j_2 \\
j_4 & j_1 \end{smallmatrix}\bigr] (u|z) \\[3mm] 
\mbox{ where } & & u \ = \
\frac{(u_4-u_3)(u_2-u_1)}{(u_4-u_2)(u_3-u_1)}\ \ \ , \ \ \ 
z \ = \ \frac{z_2}{z_3}\nonumber  
\end{eqnarray}
The Knizhnik-Zamolodchikov (KZ-) equations follow in the usual manner 
from the Sugawara-construction. The resulting equation for the ``reduced''
conformal blocks $F(u|z)$ takes the form \cite{FZ} 
$$tz(z-1)\pa_z F \ = \  \CD_u^{(2)}F\ \ , $$  
where the differential operator $\CD_u^{(2)}$ is given by the expression
\begin{eqnarray}
\CD_u^{(2)} & = & u(u-1)(u-z)\pa_u^2 +2j_2\kappa(u-z)+2j_1j_2(z-1)+
                  2j_2j_3z \nonumber \\[2mm] 
            & & -\left( (\kappa-1)(u^2-2zu+z)+2j_1u(z-1)+2j_2u(u-1)+ 
                   2j_3z(u-1)\right)\pa_u \nonumber
\end{eqnarray}
We have used the abbreviation $\kappa=j_1+j_2+j_3-j_4$. In the presently
considered case one has {\it polynomial} dependence on $u$, so that the 
KZ-equations have a finite dimensional space of solutions. Two
canonical bases for the space of solutions (``s- and t-channel conformal
blocks'') can be defined by the asymptotics
\begin{equation}\begin{aligned}
\CF^{\rm s}_{j_{21}}(u|z)\underset{z\ra 0}{\sim} & 
 z^{\De_{j_{21}}-\De_{j_2}-\De_{j_1}}x^{j_1+j_2-j_{21}}(1+\CO(x)+\CO(z)), \\
\CF^{\rm t}_{j_{32}}(u|z)\underset{z\ra 1}{\sim} & 
 (1-z)^{\De_{j_{32}}-\De_{j_3}-\De_{j_2}}(1-x)^{j_2+j_3-j_{32}}
(1+\CO(1-x)+\CO(1-z)),
\end{aligned}\end{equation}
where it is understood that the limits are first taken in the $z$-variable.

For our purposes it suffices to write down the solutions in the special
case $j_1=\frac{1}{2}$. To this end let us introduce the notation
\begin{equation}
\begin{aligned}
u=&-b^2(j_1+j_3+j_4+\fr{3}{2})-1,\\
v=&-b^2(j_1+j_3-j_4+\fr{1}{2}),
\end{aligned}\qquad w=-b^2(2j_1+1)\ \ .
\end{equation}
A set of normalized solutions for the s- and t-channel is then given by 
\[\begin{aligned} 
\CF_{+}^{\rm s} \ =\ & z^{b^2j_1}(1-z)^{b^2j_3}\Bigl( F(u+1,v,w;z)
                      -x\frac{v}{w}F(u+1,v+1,w+1;z)\Bigr) \\[2mm] 
\CF_{-}^{\rm s} \ =\ & z^{-b^2(j_1+1)}(1-z)^{b^2j_3} 
\Big(  xF(u-w+1,v-w+1,1-w;z)- \\
    & \hspace{1.5cm}  -z\frac{u-w+1}{1-w}F(u-w+2,v-w+1,2-w;z)\Big), \\[2mm]
\CF_{+}^{\rm t}\ =\ & 
              (1-z)^{b^2j_3}z^{b^2j_1} \Big(  F(u+1,v,u+v-w+1;1-z)+ \\
  &  \hspace{1.5cm} +(1-x)\frac{v}{u+v-w+1}F(u+1,v+1,u+v-w+2;1-z)\Big) \\[2mm]
\CF_{-}^{\rm t}\ =\ & 
(1-z)^{-b^2(j_3+1)}z^{b^2j_1}\Big(  (1-x)F(w-u,w-v,w-u-v;1-z)- \\
      & \hspace{1.5cm}
-(1-z)\frac{w-v}{w-u-v}F(w-u,w-v+1,w-u-v+1;1-z)\Big)\ .
\end{aligned}\]

\subsection{Fusion matrices}

The fusion matrices used in the main text can all be obtained from the 
following basic example
\begin{equation}
F_{st}(j|\rho_2,\rho_1)\;\equiv\;
\Fus{\rho_2+\frac{s}{2}\ }{j+\frac{t}{2}}{\frac{1}{2}}{j}{\rho_2}{\rho_1}
\;\equiv\;
\Fus{\rho_2+\frac{s}{2}\
}{j+\frac{t}{2}}{j}{\frac{1}{2}}{\rho_1}{\rho_2}\ ,
\end{equation}
where $s,t$ take the values $\pm 1$. The matrix elements $F_{st}$ 
have the following expressions
\begin{equation}\begin{aligned}
F_{++}\ =\ & \;\frac{\Ga(-b^2(2\rho_2+1))\Ga(1+b^2(2j+1))}
{\Ga\big(1+b^2(j+\rho_1-\rho_2+\frac{1}{2})\big)
 \Ga\big(b^2(j-\rho_1-\rho_2-\frac{1}{2})\big)}\\[1mm]
F_{+-}\ =\ & \;\frac{\Ga(-b^2(2\rho_2+1))\Ga(-b^2(2j+1))}
{\Ga\big(-b^2(j+\rho_1+\rho_2+\frac{3}{2})\big)
 \Ga\big(-b^2(j-\rho_1+\rho_2+\frac{1}{2})\big)}\\[1mm]
F_{-+}\ =\ & \;\frac{\Ga(1+b^2(2\rho_2+1))\Ga(1+b^2(2j+1))}
{\Ga\big(1+b^2(j+\rho_1+\rho_2+\frac{3}{2})\big)
 \Ga\big(1+b^2(j-\rho_1+\rho_2+\frac{1}{2})\big)}\\[1mm]
F_{--}\ =\ & -\frac{\Ga(1+b^2(2\rho_2+1))\Ga(-b^2(2j+1))}
{\Ga\big(-b^2(j+\rho_1-\rho_2+\frac{1}{2})\big)
 \Ga\big(1-b^2(j-\rho_1-\rho_2-\frac{1}{2})\big)}\ . 
\end{aligned}\end{equation}
In Subsection \ref{htobd} we need the following special case of these
formulae
\begin{equation}
f_{st}(j)\;\equiv\;
\Fus{j+\frac{s}{2}\ }{\frac{1}{2}+\frac{t}{2}}
{\frac{1}{2}}{\frac{1}{2}}{j}{j}\ \ .
\end{equation}
The previously given expressions simplify to
\begin{equation}\begin{aligned}
f_{++}\ =\ & \frac{\Ga(1+2b^2)}{\Ga(1+b^2)}\frac{\Ga(-b^2(2j+1))}
      {\Ga(-2b^2j)}\\[1mm]
f_{-+}\ =\ & \frac{\Ga(1+2b^2)}{\Ga(1+b^2)}
         \frac{\Ga(1+b^2(2j+1))}{\Ga(1+2b^2(j+1))}
\end{aligned}\qquad
\begin{aligned}
f_{+-}\ =\ & \frac{\Ga(-2b^2)}{\Ga(-b^2)}\frac{\Ga(-b^2(2j+1))}
{\Ga(-2b^2(j+1))}\\[1mm] 
f_{--}\ =\ & -\frac{\Ga(-2b^2)}{\Ga(-b^2)}
    \frac{\Ga(1+b^2(2j+1))}{\Ga(1+2b^2j)}\ \ . 
\end{aligned} \nonumber \end{equation}
We finally need 
\begin{equation}
F^1_{s}(j|\rho_2,\rho_1)\;\equiv\;\Fus{\rho_2\ }{j+s}
{1}{j}
{\rho_2}{\rho_1} \ \ . 
\end{equation}
These elements of the fusing matrix can be calculated in terms of 
$F_{st}$ by means of the pentagon identity \cite{MS}, 
\begin{equation}
\Fus{\rho_2\ }{j+s}{1}{j}
{\rho_2}{\rho_1}\;=\;
\sum_{t=\pm}\;\frac{F_{t+}\big(\fr{1}{2}|j,j+s\big)}
{F_{++}\big(\frac{1}{2}|\rho_2,\rho_2\big)}\;
F_{-t}\big(j|\rho_2+\fr{1}{2},\rho_1\big)
F_{+,s-t}\big(j+\fr{t}{2}|\rho_2,\rho_1\big)\ \ .
\nonumber 
\end{equation}
Explicit expressions for $F^1_{s}$ are given by (with $g(b)=
\frac{\Ga(1+b^2)}{\Ga(1+2b^2)}$)   
\begin{equation} \nonumber 
\begin{aligned}
F^1_{+}\ =\ &\;
\frac{g(b)\Ga(1+b^2(2\rho_2+2))\Ga(-2b^2\rho_2)\Ga(1+b^2(2j+2))
\Ga(1+b^2(2j+1))}
{\prod_s^\pm \Ga\big(\frac{1}{2}+s(\frac{1}{2}+b^2(\rho_1+\rho_2+1))
+b^2(j+1)\big)\Ga\big(1+b^2(j+s(\rho_1-\rho_2)+1)\big)},\\[2mm]
F^1_{0}\ =\ &\; 
\frac{\Ga(1+b^2(2\rho_2+2))\Ga(-2b^2\rho_2)\Ga(1+2b^2j)\Ga(-b^2(2j+2))}
{2\pi\sin\pi b^2(2j+1)}\ti\\[1mm] 
&\hspace{2cm}\ti\Bigl(\cos\pi b^2(2\rho_1+1)
\bigl(\sin\pi b^2(2j+2)+\sin\pi b^2 (2j)\bigr)\\[1mm]
& \hspace{6.5cm} -\cos\pi b^2(2\rho_2+1)\sin\pi b^2(4j+2)\Bigr)\ ,
\\[2mm] 
F^1_{-}\ =\ &\;
\frac{g(b)\Ga(1+b^2(2\rho_2+2))\Ga(-2b^2\rho_2)\Ga(-2b^2j)\Ga(-b^2(2j+1))}
{\prod_{s=\pm}\;
 \Ga\big(\frac{1}{2}-s(\frac{1}{2}+b^2(\rho_1+\rho_2+1))-b^2j\big)
 \Ga\big(-b^2(j+s(\rho_1-\rho_2))\big)}\ . 
\end{aligned}
\end{equation}

\end{document}